\documentclass[aps]{revtex4}
\usepackage{eurosym}
\usepackage{amsfonts}
\usepackage{amsmath}
\usepackage{amssymb,epsf}
\usepackage{color}
\usepackage{xcolor}
\usepackage{graphicx}
\usepackage{epstopdf}
\usepackage{float}
\usepackage{caption}
\usepackage{subfig}
\usepackage{hyperref}
\usepackage{multirow}

\begin{document}
	
\title{Some perspective of thermodynamical and optical properties of black holes in Maxwell-dilaton-dRGT-like massive gravity}
\author{B. Eslam Panah$^{1}$\footnote{email address: eslampanah@umz.ac.ir}}
\author{N. Heidari$^{1}$\footnote{email address: heidari.n@gmail.com}}
\author{M. Soleimani$^{1}$}
\affiliation{$^{1}$ Department of Theoretical Physics, Faculty of Basic Sciences, University of Mazandaran, P. O. Box 47416-95447, Babolsar, Iran}
	
\begin{abstract}
Motivated by integrating the dilaton field (as a UV correction) with dRGT-like massive gravity (as an IR correction) into Einstein gravity, we investigate the thermodynamic and optical properties of black holes within this gravitational framework. We begin by reviewing the black hole solutions in Maxwell-dilaton-dRGT-like massive gravity, followed by an analysis of how various parameters influence on the asymptotical behavior of the spacetime and the event horizon of these black holes. In the subsequent section, we examine the conserved and thermodynamic quantities associated with these black holes, paying particular attention to the effects of parameters like $\beta$, $\alpha$, and the massive parameters ($\eta_{1}$ and $\eta_{2}$) on their local stability by simultaneously evaluating the heat capacity and temperature. We also adopt an alternative method to study phase transitions using geometrothermodynamics. Furthermore, we explore how the parameters of Maxwell-dilaton-dRGT-like massive gravity impacts the optical characteristics and radiative behavior of black holes. In particular, we analyze the effects of the dilaton coupling constant ($\alpha$), charge ($q$), the massive gravity parameter ($\eta_1$), and the graviton mass ($m_g$) on the radius of the photon sphere and the resulting black hole shadow. Moreover, the theoretical shadow radius is compared to the observational data from $Sgr A^*$.
Additionally, we investigate the energy emission rate of these black holes, revealing that these parameters substantially influence the emission peak.
\end{abstract}

\maketitle
	
\section{introduction}
\label{sec1}
Based on observational data such as supernovae \cite{Acc1,Acc2} and cosmic microwave background (CMB) radiation \cite{CMB1,CMB2}, our
universe is undergoing an accelerating expansion. General relativity (GR)
cannot describe this acceleration, which motivates the study of alternative
theories of gravity. Massive gravity is one of the interesting alternative
theories of gravity, which can explain the late-time acceleration without
	considering dark energy \cite{Accmass1,Accmass2,Accmass3,Accmass4,Accmass5}.
	It is notable that in massive gravity, the graviton includes a mass ($m_{g}$%
	), which reduces to GR when $m_{g}\longrightarrow 0$. Furthermore, the
	gravitational field can be understood as a theory of a spin$-2$ graviton in
	modern particle physics \cite{Modphy1,Modphy2}. Historically, Fierz and
	Pauli introduced the first version of massive gravity in 1939. This theory
	was a linear theory of massive gravity \cite{FP}. However, this linear
	theory of massive gravity has a fundamental flaw. Specifically, the
	Fierz-Pauli (FP) massive gravity does not account for the observational
	evidence from the solar system when $m_{g}=0$. To address this issue,
	Vainshtein extended the FP theory to include a nonlinear framework \cite%
	{Vainshtein1972}. Subsequently, Boulware and Deser discovered the presence
	of a ghost in this nonlinear theory, now known as the Boulware-Deser (BD)
	ghost \cite{BD}. In response to this challenge, several attempts have been
	made to eliminate the ghost, including the development of new massive
	gravity in three-dimensional spacetime \cite{NewMass}. In recent years, a
	ghost-free massive theory was introduced by de Rham-Gabadadze-Tolley, which
	is known as dRGT-massive gravity \cite{dRGT1,dRGT2}. Furthermore,
	dRGT-massive gravity removes the BD ghost in arbitrary dimensions of
	spacetime \cite{Do1,Do2}.
	
	Black holes provide intriguing opportunities to test the theoretical and
	phenomenological aspects of modified theories of gravity, such as
	dRGT-massive gravity. Research on black holes in the context of dRGT-massive gravity has been conducted in Refs. \cite%
	{dRGTBH2,dRGTBH3,dRGTBH4,dRGTBH5,dRGTBH6,dRGTBH7,dRGTBH8,dRGTBH9,dRGTBH10,dRGTBH11,dRGTBH12,dRGTBH13,dRGTBH14,dRGTBH15,dRGTBH16,dRGTBH17}%
	. Considering black holes in massive gravity, it has been indicated that the
	final stage of Hawking evaporation could lead to the formation of a black
	hole remnant, potentially resolving the information paradox \cite{dRGTBH14}.
	On the other hand, the presence of additional terms in black hole solutions
	may provide explanations for dark matter and dark energy within the
	framework of massive gravity. Additionally, it has been established that the massive graviton can serve as a substitute for the cosmological constant
	over cosmic distances \cite{LDEMMass1,LDEMMass2,LDEMMass3,LDEMMass4}.
	
	A noteworthy study suggests that using a dynamic reference metric, rather
	than a static one, allows for the neglect of the BD-ghost in a nonlinear
	bimetric theory involving a massless spin$-2$ field \cite{HassanR2012}. In a
	different approach, Vegh introduced an alternative branch of dRGT-massive
	gravity by applying holographic principles and utilizing a singular
	reference metric \cite{Vegh}. This framework is recognized as a ghost-free
	theory of massive gravity. Building on Vegh's work, numerous studies have
	examined various aspects of this theory, including black hole solutions and
	their thermodynamic properties, as detailed in Refs. \cite%
	{VeghBH1,VeghBH2,VeghBH3,VeghBH4,VeghBH5,VeghBH6,VeghBH7,VeghBH8,VeghBH9,VeghBH10,VeghBH11,VeghBH12,VeghBH13,VeghBH14,VeghBH15}%
	. An interesting study \cite{Zhang2018} discusses that the mass of a
	graviton ($m_{g}$) is typically very small in weak gravitational
	environments but can be significantly larger in strong gravity regimes, such
	as near black holes and other compact objects. Recent observations from the
	advanced LIGO/Virgo collaborations suggest there may be a tight constraint
	on the graviton mass \cite{LIGO1,LIGO2}. Additionally, various empirical and
	theoretical limits on the graviton's mass have been established \cite%
	{TheExp1,TheExp2,TheExp3,TheExp4}. From an astrophysical perspective, the
	properties of compact objects in the context of Vegh massive gravity have
	been explored in the literature, indicating that these compact objects (such
	as massive neutron and quark stars, white dwarfs, and dark energy stars)
	could be potential candidates for regions of mass gap \cite%
	{comp1,comp2,comp3,comp4,comp5,comp6,comp7}. Notably, a correspondence has
	been identified between the black hole solutions in conformal gravity and
	Vegh massive gravity theory \cite{EslamPanah2019}. It was found that Vegh's
	massive gravity can exhibit new phase transitions for topological black
	holes due to the presence of the graviton mass \cite{TopologyBHPhase}. These
	issues prompted us to investigate this type of massive gravity (known as
	dRGT-like massive gravity) to explore some thermodynamic and optical
	properties of black holes.
	
	The emergence of a scalar field in the low-energy limit of string theory has
	led many researchers to explore dilaton gravity from various perspectives.
	The interaction between the dilaton and other gauge fields significantly
	affects the resulting solutions \cite%
	{gaugeD1,gaugeD2,gaugeD3,gaugeD4,gaugeD5,gaugeD6}. Notably, it has been
	shown that the dilaton field can alter the asymptotic behavior of spacetime.
	Specifically, in the presence of one or two Liouville-type dilaton
	potentials, black hole spacetimes are found to be neither asymptotically
	flat nor (anti)-de Sitter ((A)dS) \cite%
	{AdSD1,AdSD2,AdSD3,AdSD4,AdSD5,AdSD6,AdSD7,AdSD8}. This occurs because the
	dilaton field does not approach zero as $r \to \infty$. Additionally,
	integrating three Liouville-type dilaton potentials allows for the
	construction of dilatonic black hole solutions within the (A)dS spacetime
	framework \cite{AdSThree1,AdSThree2,AdSThree3,AdSThree4}. Research has also
	examined neutron stars in the context of dilaton gravity \cite%
	{CompactD1,CompactD2}, as well as combining dilaton gravity with other
	modified theories to derive black holes \cite%
	{DilatonMod1,DilatonMod2,DilatonMod3,DilatonMod4,DilatonMod5,DilatonMod6,DilatonMod7,DilatonMod8,DilatonMod9,DilatonMod10}%
	.
	
	Combining a scalar field, such as the dilaton field, with the massive theory
	of gravity known as dilaton-massive gravity poses challenges due to the
	complexity of the field equations. It is notable that although dRGT massive
	gravity can explain the accelerated expansion of the Universe without dark
	energy, it is only valid for an open Friedman-Lemaitre-Robertson-Walker
	(FLRW) solution and lacks stable solutions for a homogeneous and isotropic
	Universe \cite{Dilatonmassive1}. Additionally, the scalar and vector
	perturbations in this massive theory of gravity face issues due to a strong
	coupling problem and a nonlinear ghost instability \cite{Dilatonmassive2}.
	To address these challenges, considering additional degrees of freedom, such
	as an extra scalar field, has proven to be a fruitful approach. For example,
	the quasi-dilaton massive theory of gravity successfully explains the
	accelerated expansion of the Universe in FLRW cosmology \cite%
	{Dilatonmassive3}. However, this theory also encounters perturbation
	instability, leading to the development of various extensions \cite%
	{Dilatonmassive4,Dilatonmassive5,Dilatonmassive6}, see Refs. \cite%
	{Dilatonmassive8,Dilatonmassive10,Dilatonmassive11,Dilatonmassive12,Dilatonmassive13,Dilatonmassive14}%
	, for more details about the importance of the study of quasi-dilaton
	massive gravity in cosmology. In this regard, a recent study has
	successfully extracted a black hole solution within dilaton-dRGT-like
	massive gravity \cite{Liu2014}. Notably, dilaton-dRGT-like massive gravity
	arises from the coupling of the dilaton field to the terms involving the
	dRGT-like massive graviton. Generalizing to include the Maxwell field
	results in charged black holes within the framework of dilaton-dRGT-like
	massive gravity, or equivalently, black holes in Maxwell-dilaton-dRGT-like
	massive gravity \cite{Yue2024}. Building on these recent advancements, our
	goal is to further explore and reveal additional properties of charged black
	hole solutions in dilaton-dRGT-like massive gravity.
	
	\section{Black hole solutions}\label{Sec:fr}
	
	The action of Maxwell-dilaton-dRGT-like massive gravity is given by \cite%
	{Liu2014,Yue2024} 
	\begin{equation}
		\mathcal{I}=\frac{1}{16\pi }\int_{\partial \mathcal{M}}d^{4}x\sqrt{-g}\left[ 
		\mathcal{R}-F_{\mu \nu }F^{\mu \nu }-2\left( \nabla \varphi \right)
		^{2}-V\left( \varphi \right) +e^{-2\beta \varphi }m_{g}^{2}{%
			\sum\limits_{i}^{4}\eta _{i}\mathit{u_{i}(g,h)}}\right] ,  \label{action}
	\end{equation}%
	where $\varphi \left( r\right) $ is the dilaton field, and $V\left( \varphi
	\right) $ is a potential for $\varphi \left( r\right) $. Furthermore, $%
	F_{\mu \nu }=\partial _{\mu }A_{\nu }-\partial _{\nu }A_{\mu }$ is the
	electromagnetic tensor field, and $A_{\mu }$ is the gauge potential.
	Notably, we set $G=c=1$ in the action (\ref{action}), where $c$ is the speed
	of light and $G$ is the gravitational constant. Additionally, $g$ refers to
	the determinant of the metric tensor $g_{\mu \nu }$, i.e., $g=\det (g_{\mu
		\nu })$. Also, $\mathcal{R}$ and $m_{g}$ denote the Ricci scalar and the
	graviton mass, respectively. The exponential factor of the last term denotes
	the nonminimal coupling of the scalar dilaton field to the massive graviton
	with coupling constant $\beta $. The constants ${\eta }${$_{i}$ serve as
		free parameters of the action. Indeed, }${\eta }${$_{i}$'s are arbitrary
		constants whose values can be determined according to observational or
		theoretical considerations. }The quantities ${\mathit{u_{i}}}$ are
	introduced as symmetric polynomials of the eigenvalues of the matrix $%
	K_{~~\nu }^{\mu }=\sqrt{g^{\mu \sigma }h_{\sigma \nu }}$. Here, $g_{\mu \nu
	} $ is the dynamical metric tensor, and $h_{\mu \nu }$ is the reference
	metric.
	
	In the action (\ref{action}), $u_{i}$ are given in the following form 
	\begin{equation}
		{\mathit{u}}_{i}=\sum_{y=1}^{i}\left( -1\right) ^{y+1}\frac{\left(
			i-1\right) !}{\left( i-y\right) !}{\mathit{u}}_{i-y}\left[ K^{y}\right] ,
		\label{U}
	\end{equation}%
	where ${\mathit{u}}_{i-y}=1$, when $i=y$. In addition, $[K]=K_{a}^{a}$ and $%
	[K^{n}]=(K^{n})_{a}^{a}$.
	
	By varying the action with respect to the metric tensor $g_{\mu \nu }$ and
	the gauge potential $A_{\mu }$, and the dilaton field $\varphi$, the
	equations of motion are given by \cite{Liu2014,Yue2024}
	
	\begin{eqnarray}
		G_{\mu \nu } &=&2\partial _{\mu }\varphi \partial _{\nu }\varphi -\frac{1}{2}%
		\left( V\left( \varphi \right) +2\partial ^{\sigma }\varphi \partial
		_{\sigma }\varphi \right) g_{\mu \nu }+2\left( F_{\mu \sigma }F_{~\nu
		}^{\sigma }-\frac{1}{4}F^{2}g_{\mu \nu }\right) +e^{-2\beta \varphi
		}m_{g}^{2}\chi _{\mu \nu },  \label{EM1} \\
		&&  \notag \\
		\nabla ^{2}\varphi &=&\frac{1}{4}\left( \frac{\partial V\left( \varphi
			\right) }{\partial \varphi }+2\beta e^{-2\beta \varphi }m_{g}^{2}{%
			\sum\limits_{i}^{4}\eta _{i}\mathit{u_{i}(g,h)}}\right) ,  \label{EM2} \\
		&&  \notag \\
		\nabla _{\mu }F^{\mu \nu } &=&0,  \label{EM3}
	\end{eqnarray}%
	where $G_{\mu \nu }$ is the Einstein tensor. $\chi _{\mu \nu }$ is referred
	to as the massive tensor in the following form 
	\begin{equation}
		\chi _{\mu \nu }=-\sum_{i=1}^{d-2}\frac{\eta _{i}}{2}\left[ {\mathit{u}}%
		_{i}g_{\mu \nu }+\sum_{y=1}^{i}\frac{\left( -1\right) ^{y}i!}{\left(
			i-y\right) !}{\mathit{u}}_{i-y}\left[ K_{\mu \nu }^{y}\right] \right] ,
		\label{chi}
	\end{equation}%
	where $d$ is related to the dimensions of spacetime. We work in a $4-$%
	dimensional spacetime, so $d=4$.
	
	We consider a $4-$dimensional static spacetime with the following form 
	\begin{equation}
		ds^{2}=-f(r)dt^{2}+\frac{dr^{2}}{f(r)}+r^{2}R^{2}\left( r\right) \left(
		d\theta ^{2}+\sin ^{2}\theta d\varphi ^{2}\right) ,  \label{Metric}
	\end{equation}%
	where $f(r)$ is the metric function.
	
	Considering the spatial reference metric (or spatial fiducial metric) in the
	following form 
	\begin{equation}
		h_{\mu \nu }=diag\left( 0,0,{c}_{0}^{2},{c}_{0}^{2}\sin ^{2}\theta \right) ,
		\label{reference metric}
	\end{equation}%
	where $c_{0}$ is a positive constant, the only non-zero components of $u_{i}$
	are $u_{1}$ and $u_{2}$, and obtained as \cite{Liu2014,Yue2024}%
	\begin{eqnarray}
		{\mathit{u}}_{1} &=&\dfrac{2}{rR\left( r\right) },~~~\&~~~{\mathit{u}}_{2}=%
		\dfrac{2}{r^{2}R^{2}\left( r\right) },  \notag \\
		{\mathit{u}}_{i} &=&0,\text{ \ \ when \ \ }i>2.  \label{ui}
	\end{eqnarray}
	
	Using $A_{\mu }=\left( A_{t},0,0,0\right) $, and Eqs. (\ref{EM3}), and (\ref%
	{Metric}), the electromagnetic tensor field is given by $F_{tr}=\frac{q}{%
		r^{2}R^{2}\left( r\right) }$. Applying Eqs. (\ref{EM1}), (\ref{EM2}), and
	the $4-$dimensional static spacetime (\ref{Metric}), the metric function is
	obtained as \cite{Liu2014,Yue2024} 
	\begin{eqnarray}\label{metric}
		f(r) &=&\frac{\gamma _{1,1}}{\gamma _{-1,1}}\left( \frac{b}{r}\right) ^{%
			\frac{-2\alpha ^{2}}{\gamma _{1,1}}}-m_{0}r^{\frac{\gamma _{1,-1}}{\gamma
				_{1,1}}}+\frac{\gamma _{1,1}^{2}}{\gamma _{1,-3}}\Lambda r^{2}\left( \frac{b%
		}{r}\right) ^{\frac{2\alpha ^{2}}{\gamma _{1,1}}}  \notag \\
		&&  \notag \\
		&&-\frac{\gamma _{1,1}^{2}c_{0}\eta _{1}m_{g}^{2}r\left( \frac{b}{r}\right)
			^{\frac{-\xi _{1,2,0}}{\gamma _{1,1}}}}{\xi _{1,2,2}\xi _{2,2,-1}}-\frac{%
			\gamma _{1,1}^{2}c_{0}^{2}\eta _{2}m_{g}^{2}\left( \frac{b}{r}\right) ^{%
				\frac{-\xi _{2,2,0}}{\gamma _{1,1}}}}{\xi _{1,1,-1}\xi _{1,2,-1}}  \notag \\
		&&  \notag \\
		&&-\frac{2q^{2}\gamma _{1,1}^{2}}{\gamma _{1,-1}\gamma _{1,-2}r^{2}}\left( 
		\frac{b}{r}\right) ^{\frac{-4\alpha ^{2}}{\gamma _{1,1}}},  \label{f(r)}
	\end{eqnarray}%
	Also, $\varphi \left( r\right) $, $R(r)$ and $V\left( \varphi \right) $ are
	extracted in the following forms \cite{Liu2014,Yue2024}%
	\begin{eqnarray}
		R\left( r\right) &=&e^{\alpha \varphi \left( r\right) },  \label{R} \\
		&&  \notag \\
		\varphi \left( r\right) &=&\frac{\alpha }{\gamma _{1,1}}\ln \left( \frac{b}{r%
		}\right) ,  \label{Phi} \\
		&&  \notag \\
		V\left( \varphi \right) &=&2\Lambda e^{2\alpha \varphi }+\frac{2\alpha
			^{2}e^{\frac{2\varphi }{\alpha }}}{b^{2}\gamma _{1,-1}}-\frac{2\alpha
			^{2}q^{2}e^{\frac{4\varphi }{\alpha }}}{b^{4}\gamma _{1,-2}}  \notag \\
		&&  \notag \\
		&&+\frac{4\xi _{1,1,0}c_{0}\eta _{1}m_{g}^{2}e^{\frac{\varphi \xi _{0,-2,1}}{%
					\alpha }}}{b\xi _{2,2,-1}}+\frac{2\xi _{1,1,0}c_{0}^{2}\eta _{2}m_{g}^{2}e^{%
				\frac{2\varphi \xi _{0,-1,1}}{\alpha }}}{b^{2}\xi _{1,1,-1}},  \label{V}
	\end{eqnarray}%
	where $\gamma _{i,j}=i\alpha ^{2}+j$, and $\xi _{i,j,k}=i\alpha ^{2}+j\alpha
	\beta +k$, where they are dimensionless quantities. In addition, $	\alpha $ and $b$ are constants. The dimension of $b$ is \texttt{Length}, i.e., $[b]=L$, however $\alpha$ is dimensionless parameter. Notably, $\alpha $ is the parameter of the dilaton field.
	
Here, we are going to investigate the massive coefficients via dimensional analysis. In general, all terms of Eq. (\ref{f(r)}) must be dimensionless. On the other hand, in dimensional analysis we know that $[m_{g}]=[r]=[b]=L$ and $[\alpha ]=[\gamma _{i,j}]=[\xi _{i,j,k}]=1$. Indeed, $\alpha $, $\gamma _{i,j}$, and $\xi_{i,j,k}$, are dimensionless. Also, the dimension $m_{0}$ is $[m_{0}]=L^{\frac{\gamma _{-1,1}}{\gamma_{1,1}}}$. Therefore, the dimensional interpretation of massive terms are
	\begin{eqnarray}
		\lbrack \eta _{1}c_{0}] &=&L^{-3},  \label{et1} \\
		\lbrack \eta _{2}c_{0}^{2}] &=&L^{-2},  \label{et2}
	\end{eqnarray}
	
	Using Eqs. (\ref{et1}), and (\ref{et2}), one can show that massive coefficients are, dimensionally,
	\begin{equation}
		\lbrack c_{0}]=L\text{ \ \ \ \ \ \ \&\ \ \ \ \ \ \ \ }[\eta _{i}]=L^{-4},\ \
		\ \ i=1,2
	\end{equation}
	Now, we can calculate the Ricci scalar by using the $4-$dimensional
	spacetime in Eq. (\ref{Metric}), and the metric function (\ref{f(r)}). After
	some complete calculations, we can obtain the Ricci scalar in the following
	form 
	\begin{eqnarray}
		R &=&\frac{6\Lambda \gamma _{1,-2}}{\gamma _{1,-3}}\left( \frac{b}{r}\right)
		^{\frac{2\alpha ^{2}}{\gamma _{1,1}}}-\frac{2m_{0}\alpha }{\gamma
			_{1,1}^{2}r^{\frac{\gamma _{1,3}}{\gamma _{1,1}}}}  \notag \\
		&&  \notag \\
		&&+\frac{2\xi _{1,-1,1}\xi _{1,1,1}\left( \frac{b}{r}\right) ^{\frac{%
					-2\alpha ^{2}}{\gamma _{1,1}}}}{\gamma _{1,1}\gamma _{1,-1}r^{2}}+\frac{%
			4\alpha ^{2}q^{2}}{\gamma _{1,-1}\left( \frac{b}{r}\right) ^{\frac{4\alpha
					^{2}}{\gamma _{1,1}}}r^{4}}  \notag \\
		&&  \notag \\
		&&+\frac{c_{0}\eta _{1}m_{g}^{2}\xi _{2,2,3}}{\xi _{2,2,-1}\left( \frac{b}{r}%
			\right) ^{\frac{\xi _{1,2,0}}{\gamma _{1,1}}}r}+\frac{2c_{0}^{2}\eta
			_{2}m_{g}^{2}\xi _{1,1,1}\xi _{1,2,1}}{\xi _{1,2,-1}\xi _{1,1,-1}\left( 
			\frac{b}{r}\right) ^{\frac{\xi _{2,2,0}}{\gamma _{1,1}}}r^{2}},
		\label{Ricci}
	\end{eqnarray}%
	where indicates that the Ricci scalar is not constant, and it depends on all
	of parameters of charged black holes in dillaton-dRGT-like massive gravity.
	Furthermore, the Ricci scalar diverge at $r=0$ (i.e. $\lim_{r\longrightarrow
		0}R\longrightarrow \infty $).
	
	Using the spacetime (\ref{Metric}), we obtain the Kretschmann scalar in the
	following form
	
	\begin{eqnarray}
		R_{\alpha \beta \gamma \delta }R^{\alpha \beta \gamma \delta } &=&\frac{%
			8R^{^{\prime \prime ^{2}}}f^{2}}{R^{2}}+\frac{8R^{^{\prime }}R^{^{\prime
					\prime }}\left( 4f+rf^{^{\prime }}\right) f}{rR^{2}}+\frac{8R^{^{\prime
					\prime }}Rff^{^{\prime }}}{rR^{2}}+f^{^{\prime \prime ^{2}}}  \notag \\
		&&  \notag \\
		&&+\frac{4R^{^{\prime ^{4}}}f^{2}}{R^{4}}+\frac{16R^{^{\prime ^{3}}}f^{2}}{%
			rR^{3}}+\frac{4f^{^{\prime ^{2}}}}{r^{2}}+\frac{4\left( f-1\right) ^{2}}{%
			r^{4}}  \notag \\
		&&  \notag \\
		&&+\frac{4R^{^{\prime ^{2}}}}{r^{2}R^{4}}\left[ 14R^{2}f^{2}+4rR^{2}ff^{^{%
				\prime }}+r^{2}R^{2}f^{^{\prime ^{2}}}-2f\right]  \notag \\
		&&  \notag \\
		&&+\frac{8R^{^{\prime }}}{r^{3}R^{3}}\left[ 2R^{2}f^{2}+r^{2}R^{2}f^{^{%
				\prime ^{2}}}+2rR^{2}ff^{^{\prime }}-2f\right] ,  \label{Kresch}
	\end{eqnarray}%
where $f=f\left( r\right) $, and $R=R\left( r\right) $
. By substituting the metric function (Eq. (\ref{f(r)})) and Eq. (\ref{R}) into the Kretschmann scalar (Eq. (\ref{Kresch})), we find that the Kretschmann scalar diverges at $r=0$. This indicates the presence of a curvature singularity at $r=0$.

The investigation of the asymptotic behavior of the metric function is
intriguing, as it may depend on all the parameters of the theory. Our
analysis indicates that the parameters $\alpha$ and $\beta$ play an
important role in determining the asymptotic behavior of spacetime. We
categorize the effects of various parameters on the asymptotic behavior of spacetime into three parts by adjusting the values of $\alpha$ and $\beta$, which are:
	
i) For all values of $\alpha$, the asymptotic behavior of the spacetime is determined solely by the following term: 
\begin{equation}
		\lim_{r\longrightarrow \infty }f(r)\longrightarrow \frac{\gamma _{1,1}^{2}}{%
			\gamma _{1,-3}}\Lambda r^{2}\left( \frac{b}{r}\right) ^{\frac{2\alpha ^{2}}{%
				\gamma _{1,1}}},
\end{equation}%
which shows the spacetime will not be asymptotically (A)dS, because it depends on both $\Lambda $ and $\alpha $.
	
ii) Considering $\beta >\frac{-\gamma _{2,-1}}{2\alpha }$, the asymptotical behavior of the spacetime is given by 
\begin{equation}
		\lim_{r\longrightarrow \infty }f(r)\longrightarrow \frac{\gamma _{1,1}^{2}}{%
			\gamma _{1,-3}}\Lambda r^{2}\left( \frac{b}{r}\right) ^{\frac{2\alpha ^{2}}{%
				\gamma _{1,1}}}-\frac{\gamma _{1,1}^{2}c_{0}\eta _{1}m_{g}^{2}r\left( \frac{b%
			}{r}\right) ^{\frac{-\xi _{1,2,0}}{\gamma _{1,1}}}}{\xi _{1,2,2}\xi _{2,2,-1}%
		},
\end{equation}%
which reveals that the asymptotic behavior of the spacetime depends on
the cosmological constant ($\Lambda$), the parameters of the dilaton field ($\alpha$), the parameter of the reference metric ($c_{0}$), massive gravity ($\eta_{1}$), and also the graviton mass ($m_{g}$).
	
iii) For $\alpha >-\beta $, we find that 
\begin{equation}
		\lim_{r\longrightarrow \infty }f(r)\longrightarrow \frac{\gamma _{1,1}^{2}}{%
			\gamma _{1,-3}}\Lambda r^{2}\left( \frac{b}{r}\right) ^{\frac{2\alpha ^{2}}{%
				\gamma _{1,1}}}-\frac{\gamma _{1,1}^{2}c_{0}\eta _{1}m_{g}^{2}r\left( \frac{b%
			}{r}\right) ^{\frac{-\xi _{1,2,0}}{\gamma _{1,1}}}}{\xi _{1,2,2}\xi _{2,2,-1}%
		}-\frac{\gamma _{1,1}^{2}c_{0}^{2}\eta _{2}m_{g}^{2}\left( \frac{b}{r}%
			\right) ^{\frac{-\xi _{2,2,0}}{\gamma _{1,1}}}}{\xi _{1,1,-1}\xi _{1,2,-1}},
\end{equation}%
	where it is clear that the asymptotic behavior of the spacetime is
	determined by all of the parameters of the system, such as $\Lambda$ , $\alpha$, $\beta$, $c_{0}$, $\eta_{1}$, $\eta_{2}$, and $m_{g}$.
	
\begin{figure}[h]
		\centering
		\includegraphics[width=80mm]{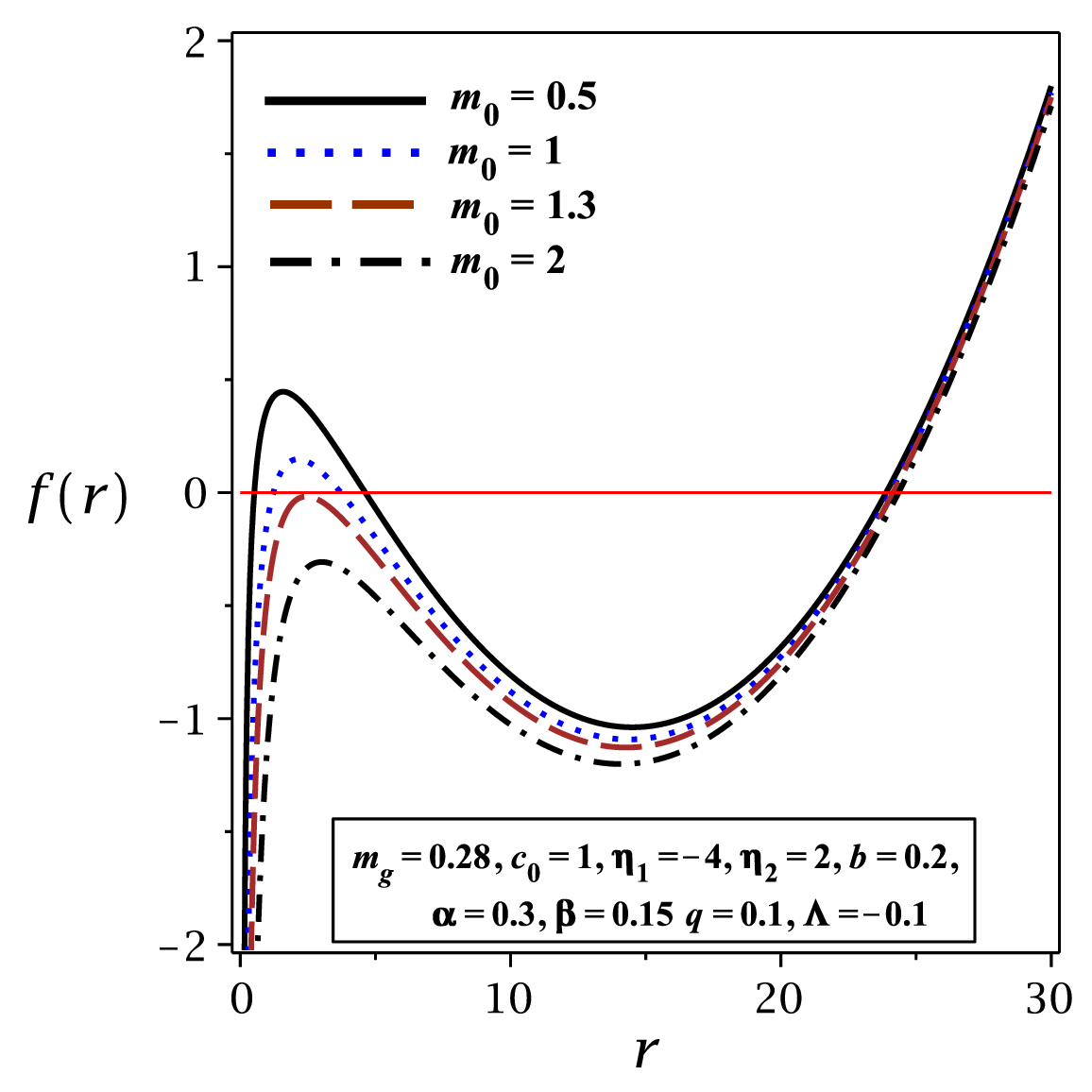} \includegraphics[width=80mm]{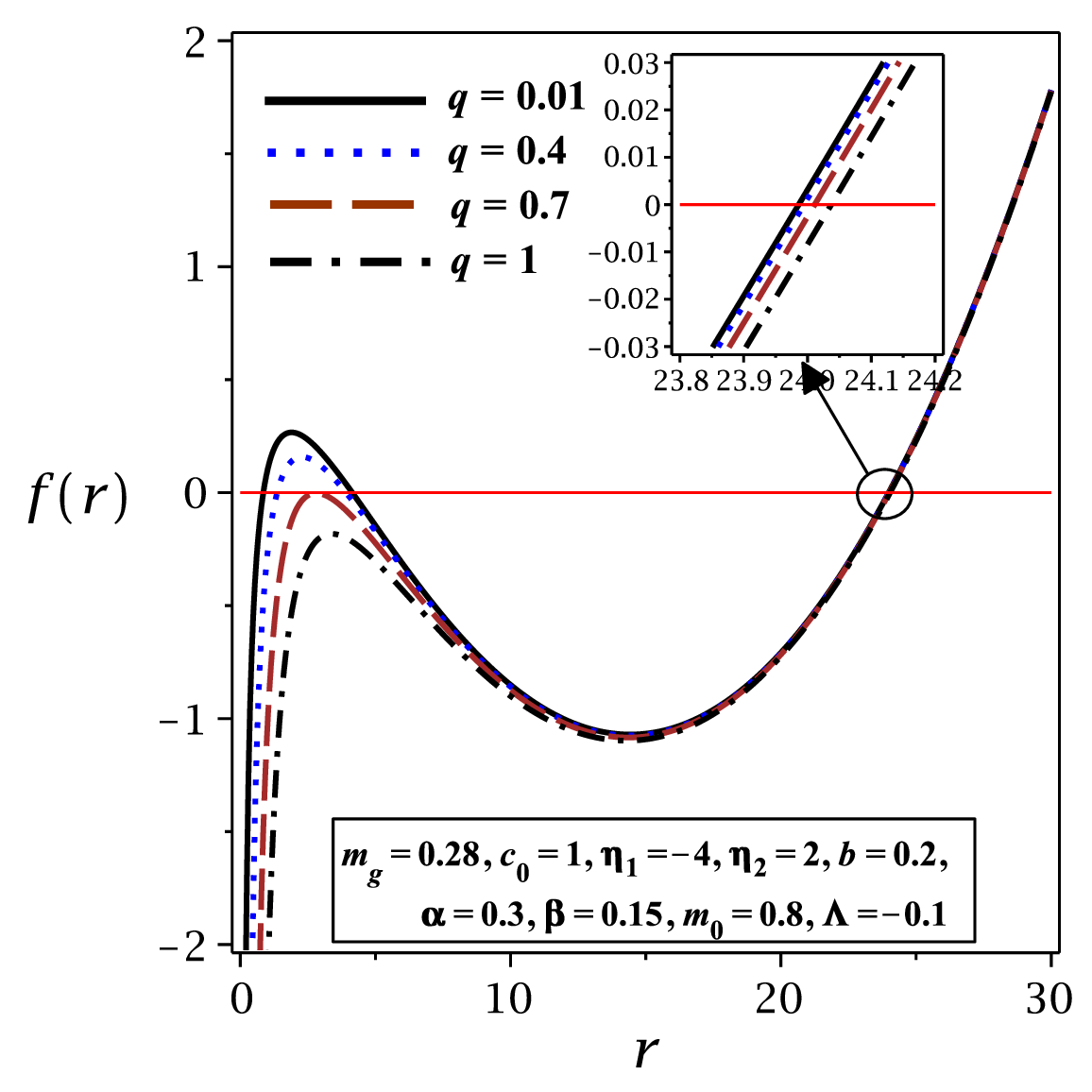} 
		\newline
		\includegraphics[width=80mm]{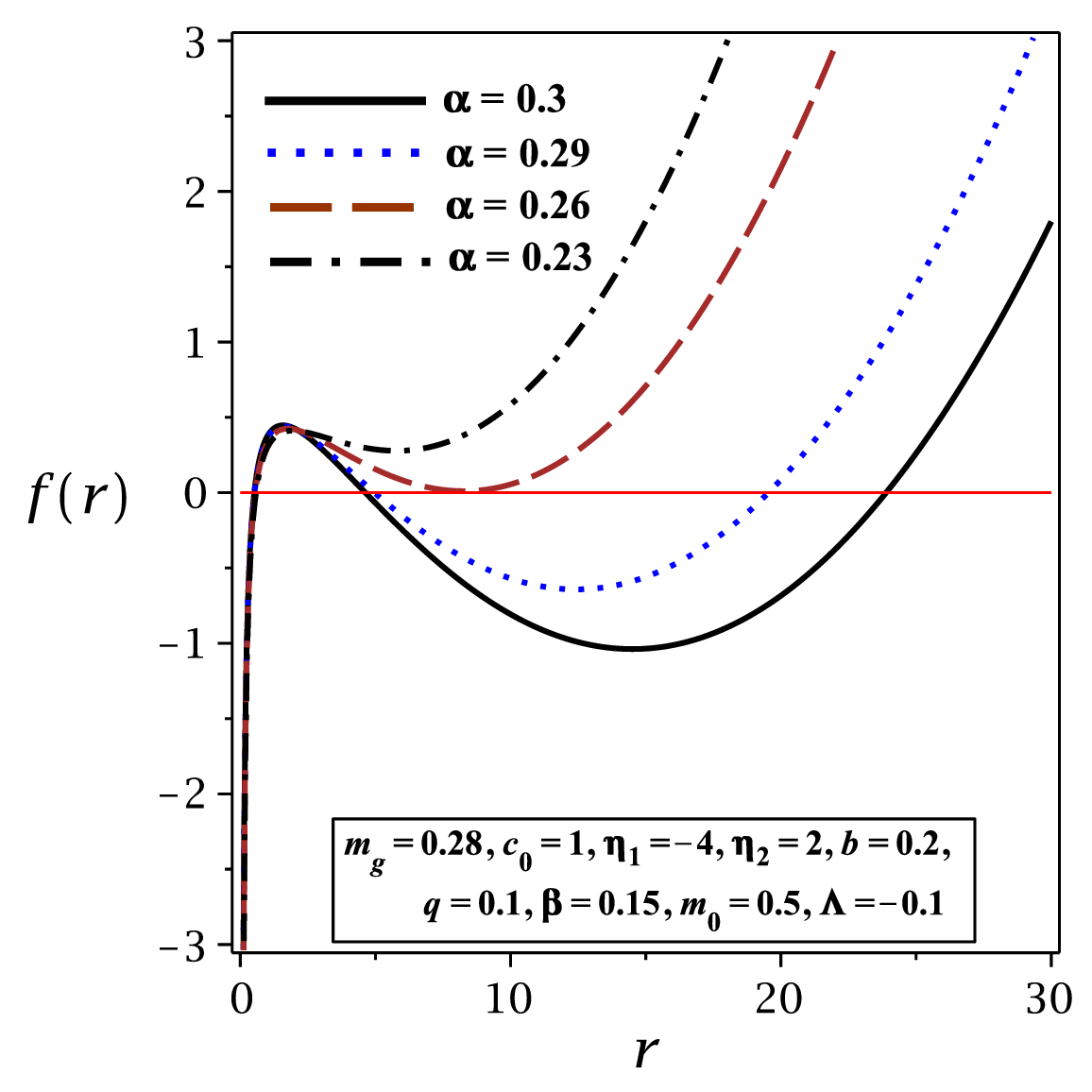} \includegraphics[width=80mm]{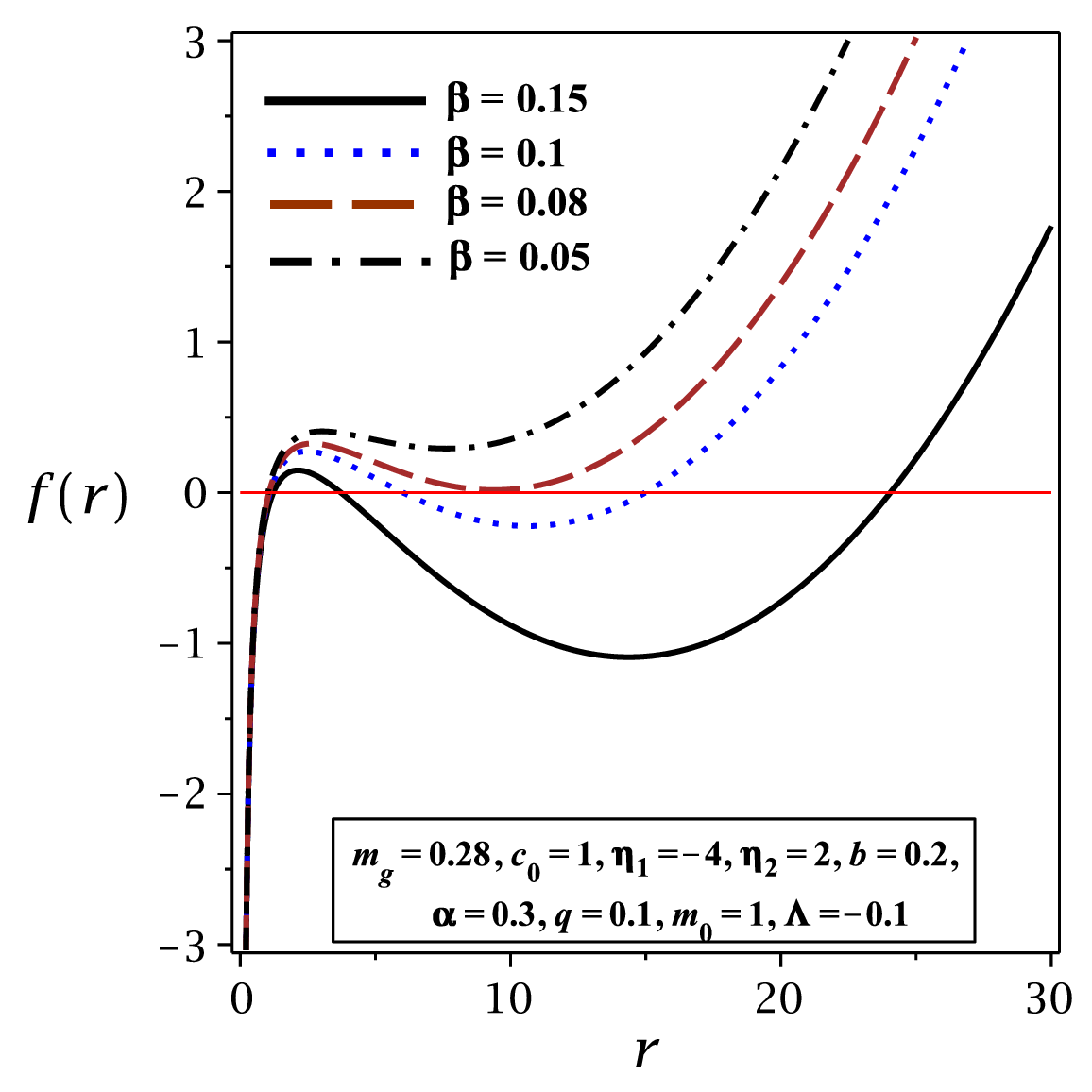} 
		\newline
		\caption{The mertic function $f(r)$ versus $r$ for different values of
			parameters.}
		\label{fig1}
	\end{figure}

	To find the roots of the metric function $f(r)$, we need to solve the
	equation (\ref{f(r)}). Finding an exact solution is complicated, so we have
	plotted $f(r)$ versus $r$ in Fig. \ref{fig1}. Our analysis reveals that
	these black holes can have multiple horizons, depending on the values of
	various parameters. We summarize our findings regarding the effects of these
	parameters in four panels.
	
	1- \textbf{Up-left panel in Fig. \ref{fig1}:} There exists a critical mass
	value ($m_{critical}$) for these black holes, which corresponds to one event
	horizon and external root (indicated by the dashed line in this panel). When 
	$m_{0}<m_{critical}$, there are three real roots: one event horizon and two
	inner roots. However, when the mass exceeds the critical value (i.e., $%
	m_{0}>m_{critical}$), the number of roots decreases from two to one.
	Additionally, our findings highlight that as $m_{0}$ increases, the event
	horizon of the black hole also expands. In other words, the radius of the
	massive black hole increases, as we anticipated.
	
	2- \textbf{Up-right panel in Fig. \ref{fig1}:} Similar to the previous case,
	a critical value for the electrical charge ($q_{\text{critical}}$) exists in
	which the black holes encounter one event horizon and one external root (see
	the dashed line in the up-right panel of Fig. \ref{fig1}). Also, for $q < q_{%
		\text{critical}}$, there are three real roots (one event horizon and two
	inner roots). On the other hand, for $q > q_{\text{critical}}$, the number
	of roots decreases from two to one. In addition, by increasing $q$, the
	radius of the black hole increases.
	
	3- \textbf{Down-left panel in Fig. \ref{fig1}:} There is a critical value
	for $\alpha$, in which the black holes have two roots (one event horizon and
	one external root), see dashed line in the down-left panel in Fig. \ref{fig1}%
	. For $\alpha > \alpha_{\text{critical}}$, there are three real roots (one
	event horizon and two inner roots). For $\alpha<\alpha_{\text{critical}}$,
	the number of roots decreases from two to one root. In other words, black
	holes with large values of the dilaton field encounter multiple horizons.
	Furthermore, a large value of the dilaton field leads to large black holes.
	
	4- \textbf{Down-right panel in Fig. \ref{fig1}:} We adjust the coupling
	constant $\beta$ to examine its influence on the number of roots of the
	metric function. Our findings reveal that for $\beta_{\text{critical}}$, the
	black holes encounter one event horizon and one external root (see the
	dashed line in the down-right panel of Fig. \ref{fig1}). Also, for $%
	\beta>\beta_{\text{criticality}}$, there are three roots. In contrast, for $%
	\beta < \beta_{\text{criticality}}$, the black holes exhibit only one real
	root, corresponding to the event horizon. Additionally, a higher value of
	the coupling constant results in larger black holes.
	
	\section{Thermodynamics and Geometrothermodynamics}
	
	\label{sec2}
	
	\subsection{Thermodynamic Quantities}
	
	In this section, we will examine the conserved and thermodynamic quantities
	of charged black holes in dilaton-dRGT-like massive gravity. We will then
	assess how various parameters of these black holes influence thermal
	stability regions and phase transition points. This will be done using heat
	capacity and geometrothermodynamics approaches.
	
	Applying the surface gravity ($\kappa $), the Hawking temperature of such
	black holes are given by \cite{Yue2024} 
	\begin{eqnarray}
		T &=&\frac{\kappa }{2\pi }=\frac{f^{^{\prime }}\left( r\right) \left\vert
			_{r=r_{+}}\right. }{4\pi }=\frac{-\gamma _{1,1}}{4\pi \gamma _{1,-1}r_{+}}%
		\left( \frac{b}{r_{+}}\right) ^{\frac{-2\alpha ^{2}}{\gamma _{1,1}}}-\frac{%
			\gamma _{1,1}\Lambda r_{+}}{4\pi }\left( \frac{b}{r_{+}}\right) ^{\frac{%
				2\alpha ^{2}}{\gamma _{1,1}}}  \notag \\
		&&  \notag \\
		&&-\frac{\gamma _{1,1}c_{0}\eta _{1}m_{g}^{2}\left( \frac{b}{r_{+}}\right) ^{%
				\frac{-\xi _{1,2,0}}{\gamma _{1,1}}}}{4\pi \xi _{2,2,-1}}-\frac{\gamma
			_{1,1}c_{0}^{2}\eta _{2}m_{g}^{2}\left( \frac{b}{r_{+}}\right) ^{\frac{-\xi
					_{2,2,0}}{\gamma _{1,1}}}}{4\pi \xi _{1,1,-1}r_{+}}+\frac{q^{2}\gamma _{1,1}%
		}{2\pi \gamma _{1,-2}r_{+}^{3}}\left( \frac{b}{r_{+}}\right) ^{\frac{%
				-4\alpha ^{2}}{\gamma _{1,1}}}.  \label{Temp}
	\end{eqnarray}
	
	The entropy of such black holes is obtained in the following form by using
	Wald's approach \cite{Yue2024} 
	\begin{equation}
		S=\pi r_{+}^{2}\left( \frac{b}{r_{+}}\right) ^{\frac{2\alpha ^{2}}{\gamma
				_{1,1}}},  \label{S}
	\end{equation}%
	where the entropy is altered by the presence of a dilaton field. Importantly, the modified entropy, as expressed in equation (\ref{S}), simplifies to the standard form, $S = \pi r_{+}^{2}$, when the dilaton field is not present (i.e., when $\alpha = 0$). In contrast, as $\alpha$ approaches infinity, the modified entropy (Eq. (\ref{S})) converges to a constant value that depends on the coupling constant ($b$). Specifically, we have $\lim_{\alpha \rightarrow \infty}\left( \pi r_{+}^{2}\left( \frac{b}{r_{+}}\right)^{\frac{2\alpha^{2}}{\alpha^{2}+1}}\right) = \pi b^{2}$. This demonstrates one of the effects of the dilaton field on entropy.

	Using the Ashtekar-Magnon-Das (AMD) approach, we can extract the total mass
	of these black holes as follows \cite{Yue2024} 
	\begin{eqnarray}
		M &=&\frac{-r_{+}}{2\gamma _{1,-1}}+\frac{\gamma _{1,1}\Lambda r_{+}^{3}}{%
			2\gamma _{1,-3}}\left( \frac{b}{r_{+}}\right) ^{\frac{4\alpha ^{2}}{\gamma
				_{1,1}}}+\frac{q^{2}\gamma _{1,1}}{\gamma _{1,-1}\gamma _{1,-2}r_{+}}\left( 
		\frac{b}{r_{+}}\right) ^{\frac{-2\alpha ^{2}}{\gamma _{1,1}}}  \notag \\
		&&  \notag \\
		&&-\frac{\gamma _{1,1}^{2}c_{0}\eta _{1}m_{g}^{2}r_{+}^{2}\left( \frac{b}{%
				r_{+}}\right) ^{\frac{\xi _{1,-2,0}}{\gamma _{1,1}}}}{2\xi _{1,2,2}\xi
			_{2,2,-1}}-\frac{\gamma _{1,1}c_{0}^{2}\eta _{2}m_{g}^{2}r_{+}\left( \frac{b%
			}{r_{+}}\right) ^{\frac{\xi _{0,-2,0}}{\gamma _{1,1}}}}{2\xi _{1,1,-1}\xi
			_{1,2,1}},  \label{AMDMass}
	\end{eqnarray}%
	where it reduces to the total mass of black holes in dRGT-like massive gravity in the absence of the dilaton field ($\alpha =0$) as
	\begin{equation}
		\underset{\alpha \rightarrow 0}{\lim }M=\frac{r_{+}}{2}+\frac{c_{0}^{2}\eta
			_{2}m_{g}^{2}r_{+}}{2}+\frac{c_{0}\eta _{1}m_{g}^{2}r_{+}^{2}}{4}-\frac{%
			\Lambda r_{+}^{3}}{6}+\frac{q^{2}}{2r_{+}},  \label{Mmass}
	\end{equation}%
	that it diverges at $r_{+} = 0$ due to the presence of electric charge. However, the total mass (Eq. (\ref{AMDMass})) of charged black holes in dilaton-dRGT-like massive gravity, in the absence of the cosmological constant and with a very large value for $\alpha$, takes the following form
	\begin{equation}
		\underset{\alpha \rightarrow \infty }{\lim }M=-\frac{c_{0}^{2}\eta
			_{2}m_{g}^{2}br_{+}}{2},  \label{Mdilaton}
	\end{equation}
	where the results indicate that the electric charge disappears due to the influence of the strong dilaton field. However, the coupling between $b$ (the coupling constant) and the parameters of massive gravity remains intact in an environment with a strong dilaton field. Additionally, equation (\ref{Mdilaton}) presents two noteworthy points
	
	i) The total mass is zero at $r_{+}=0$ when the dilaton field is strong.
	
	ii) The signature of $\eta_{2}$ must be negative to prevent the total mass from taking on a negative value.
	
	The electric potential $\Phi $ of the charge black holes in dilaton-dRGT-like massive gravity is determined as follows \cite{Yue2024} 
	\begin{equation}
		\Phi =\frac{2\gamma _{1,1}q}{\gamma _{1,-1}\gamma _{1,-2}r_{+}}\left( \frac{b%
		}{r_{+}}\right) ^{\frac{-2\alpha ^{2}}{\gamma _{1,1}}},
	\end{equation}%
	where it reduces to $\Phi =\frac{q}{r_{+}}$ when $\alpha =0$. Also,
		for $\alpha \rightarrow \infty $, $\Phi $ is zero (i.e, $\underset{\alpha \rightarrow \infty }{\lim \Phi =0}$). This indicate that the electric potential disappears when the effect of the dilaton field is
		very strong. 
	
	It is easy to demonstrate that the conserved quantities and thermodynamic
	variables comply with the first law of thermodynamics in the following form 
	\begin{equation}
		dM=TdS+\Phi dq.
	\end{equation}
	
	\subsection{Heat Capacity and Thermal Stability}
	
	Here, we analyze heat capacity and geometrothermodynamics to identify
	thermal stability and phase transition points simultaneously. Specifically, we will examine how various parameters of charged black holes in
	dilaton-massive gravity affects areas of thermal stability and points of
	phase transition by applying both heat capacity and geometrothermodynamics
	approaches.
	
	The study of heat capacity provides crucial information about areas of
	thermal stability and points of phase transition within the context of the
	canonical ensemble. Specifically, a positive heat capacity indicates thermal
	stability, while a negative heat capacity suggests instability. Divergences
	in heat capacity correspond to phase transition points. Additionally, the
	roots of heat capacity determine the boundary points of thermal systems.
	
	The heat capacity is defined as 
	\begin{equation}
		C_{Q}=T\left( \frac{\partial S}{\partial T}\right) _{Q}=\frac{\left( \frac{%
				\partial M(S,Q)}{\partial S}\right) _{Q}}{\left( \frac{\partial ^{2}M(S,Q)}{%
				\partial S^{2}}\right) _{Q}}=\frac{M_{S}}{M_{SS}},  \label{C}
	\end{equation}%
	where $M_{S}=\left( \frac{\partial M(S,Q)}{\partial S}\right) _{Q}$, and $%
	M_{SS}=\left( \frac{\partial ^{2}M(S,Q)}{\partial S^{2}}\right) _{Q}$. So, we have to rewrite the mass as a function of the extensive quantities $S$
	and $Q$ for obtaining the heat capacity (\ref{C}). Using Eqs. (\ref{Temp}), (%
	\ref{S}) and by replacing them in Eq. (\ref{AMDMass}), the mass $M(S,Q)$ is
	found as 
	\begin{eqnarray}
		M(S,Q) &=&\frac{-b}{2\gamma _{1,-1}}\left( \frac{S}{\pi b^{2}}\right) ^{%
			\frac{\gamma _{1,1}}{2}}+\frac{\gamma _{1,1}\Lambda b^{3}}{2\gamma _{1,-3}}%
		\left( \frac{S}{\pi b^{2}}\right) ^{\frac{\gamma _{-1,3}}{2}}+\frac{%
			Q^{2}\gamma _{1,1}}{\gamma _{1,-1}\gamma _{1,-2}b}\left( \frac{S}{\pi b^{2}}%
		\right) ^{\frac{\gamma _{1,-1}}{2}}  \notag \\
		&&  \notag \\
		&&-\frac{\gamma _{1,1}^{2}c_{0}\eta _{1}m_{g}^{2}b^{2}\left( \frac{S}{\pi
				b^{2}}\right) ^{\frac{_{\xi _{1,2,2}}}{2}}}{2\xi _{1,2,2}\xi _{2,2,-1}}-%
		\frac{\gamma _{1,1}c_{0}^{2}\eta _{2}m_{g}^{2}b\left( \frac{S}{\pi b^{2}}%
			\right) ^{\frac{_{\xi _{1,2,1}}}{2}}}{2\xi _{1,1,-1}\xi _{1,2,1}}.
		\label{MSQ}
	\end{eqnarray}%
	where $q=Q$.
	
	By applying Eqs. (\ref{C}) and (\ref{MSQ}), we can express the heat capacity
	in the following manner 
	\begin{equation}
		C_{Q}=\frac{\left( \mathcal{A}_{1}+\mathcal{A}_{2}+\mathcal{A}_{3}+\mathcal{A%
			}_{4}\right) S}{\frac{-\gamma _{1,-1}\mathcal{A}_{1}}{2}+\frac{\mathcal{A}%
				_{2}\xi _{1,2,0}}{2}-\frac{Q^{2}\gamma _{1,-3}\xi _{1,1,-1}\xi
				_{2,2,-1}\left( \frac{S}{\pi b^{2}}\right) ^{_{\xi _{1,0,0}}}}{\gamma _{1,-2}%
			}+\mathcal{A}_{5}},
	\end{equation}
	
	where $\mathcal{A}_{1}$, $\mathcal{A}_{2}$, $\mathcal{A}_{3}$, $\mathcal{A}%
	_{4}$, and $\mathcal{A}_{5}$ are 
	\begin{eqnarray}
		\mathcal{A}_{1} &=&\frac{\xi _{1,1,-1}\xi _{2,2,-1}\Lambda S^{2}}{\pi ^{2}},
		\\
		\mathcal{A}_{2} &=&c_{0}\eta _{1}m_{g}^{2}b^{3}\gamma _{1,1}\xi
		_{1,1,-1}\left( \frac{S}{\pi b^{2}}\right) ^{\frac{_{\xi _{2,2,3}}}{2}}, \\
		\mathcal{A}_{3} &=&c_{0}^{2}\eta _{2}m_{g}^{2}b^{2}\xi _{2,2,-1}\left( \frac{%
			S}{\pi b^{2}}\right) ^{\xi _{1,1,1}}, \\
		\mathcal{A}_{4} &=&\left( \frac{-2Q^{2}}{\gamma _{1,-2}\left( \frac{S}{\pi
				b^{2}}\right) ^{\alpha ^{2}}}-\frac{b^{2}\left( \frac{S}{\pi b^{2}}\right)
			^{\gamma _{1,1}}}{\gamma _{1,-1}}\right) \xi _{1,1,-1}\xi _{2,2,-1}, \\
		\mathcal{A}_{5} &=&\left[ c_{0}^{2}\eta _{2}m_{g}^{2}\xi _{1,2,-1}\left( 
		\frac{S}{\pi b^{2}}\right) ^{\xi _{0,1,0}}+\xi _{1,1,-1}\right] b^{2}\left( 
		\frac{S}{\pi b^{2}}\right) ^{\gamma _{1,1}}.
	\end{eqnarray}
	
	We focus on studying heat capacity and temperature to determine four key
	properties of black holes:
	
	i) Thermal stability is characterized by the condition $C_{Q}>0$.
	
	ii) The point where the heat capacity $C_{Q}=T=0$ is recognized as a
	physical limitation point.
	
	iii) Positive and negative temperature values indicate physical and
	non-physical black holes, respectively.
	
	iv) The divergences in heat capacity are associated with the critical points
	of phase transitions in black holes.
	
\begin{figure}[h]
		\centering
		\includegraphics[width=80mm]{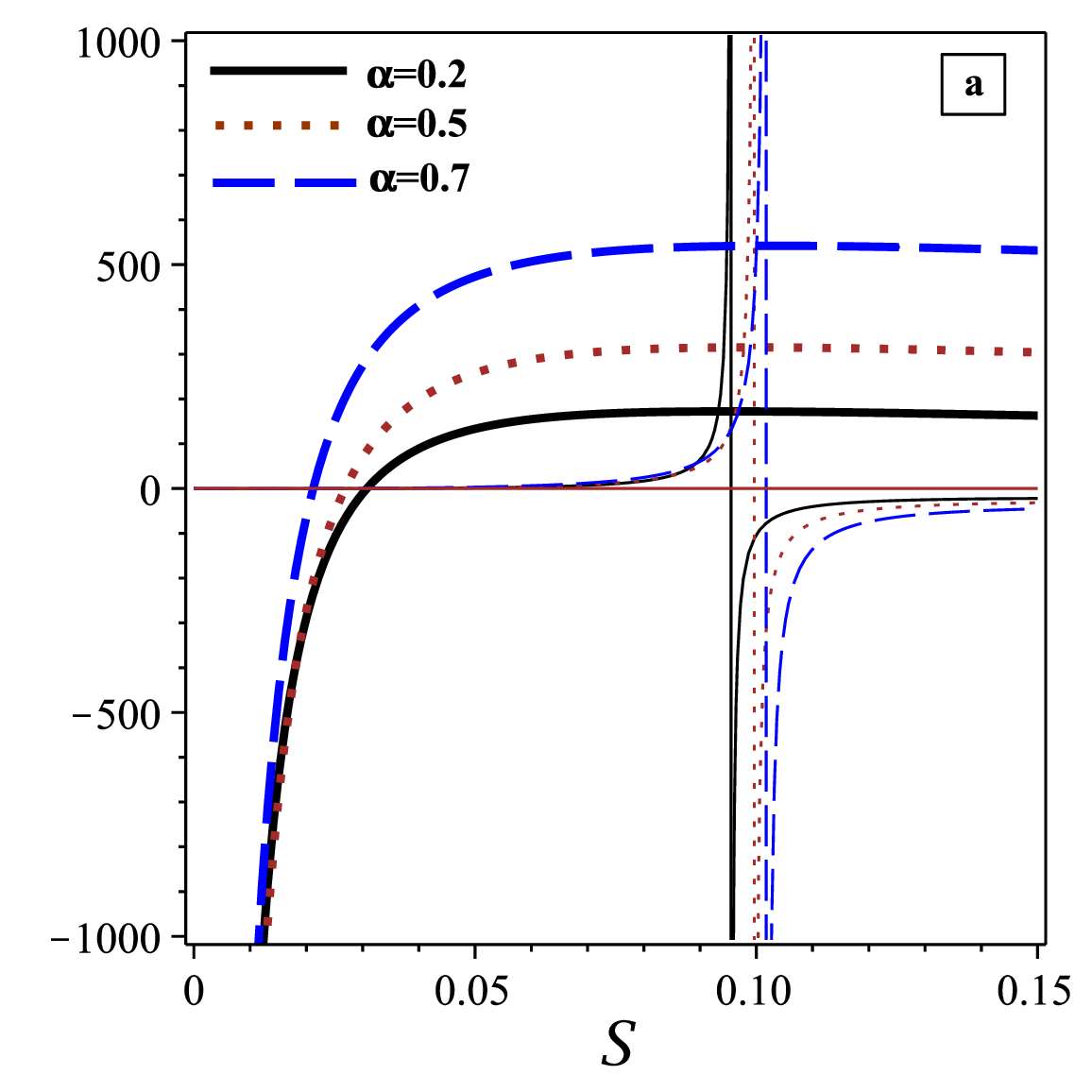} \includegraphics[width=80mm]{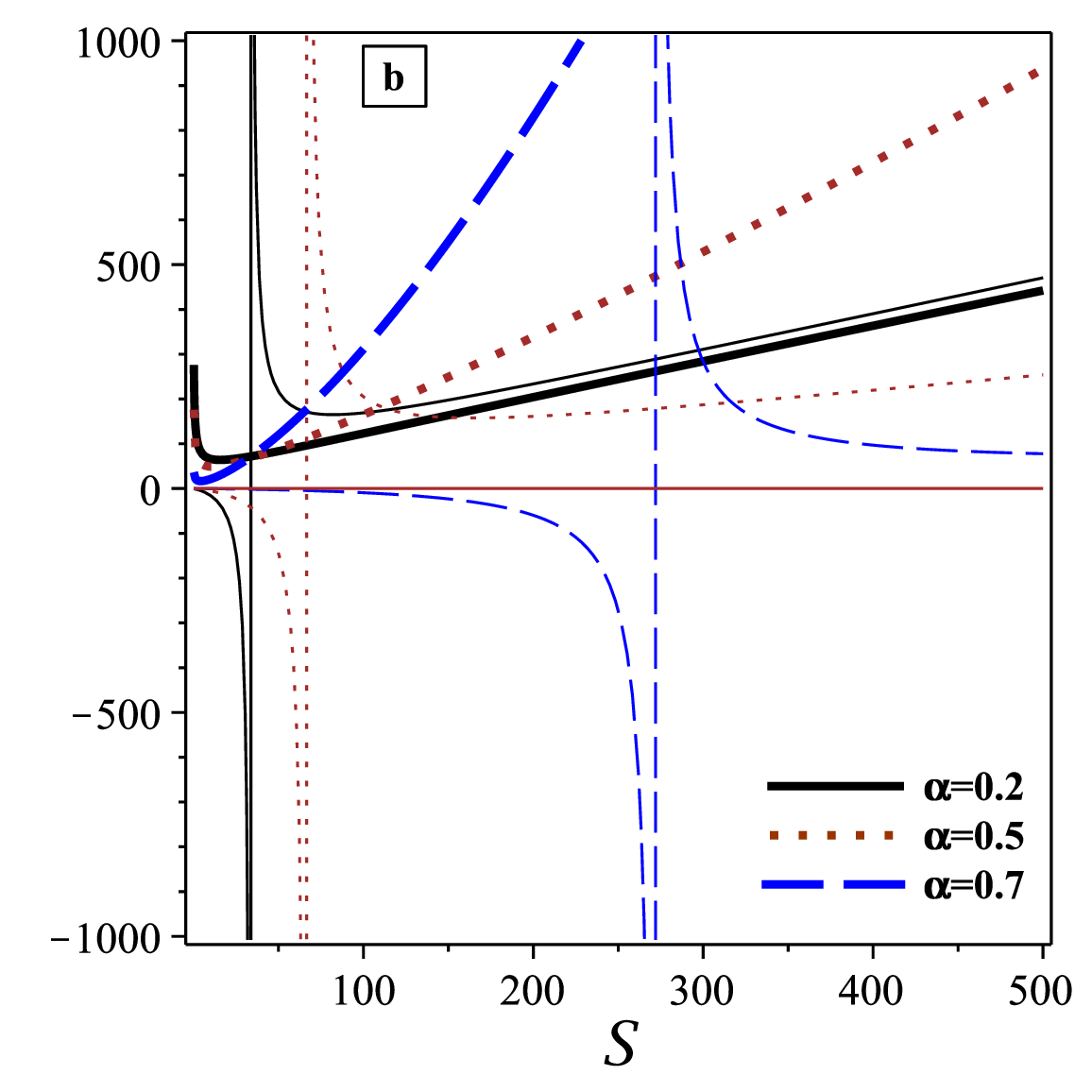} 
		\newline
		\caption{The heat capacity ($C_{Q}$) and the Hawking temperature ($T$)
			versus entropy ($S$) for different values of $\protect\alpha $ by
			considering $Q=b=\protect\beta =-\protect\eta _{2}=0.1$, $\protect\eta %
			_{1}=0.5$, $\Lambda =-0.1$, $c_{0}=1$, and $m_{g}=0.2$. Bold lines represent the Hawking temperature ($T$), while thin lines indicate the heat capacity ($C_{Q}$)}.
		\label{fig2}
\end{figure}

\begin{figure}[h]
		\centering
		\includegraphics[width=80mm]{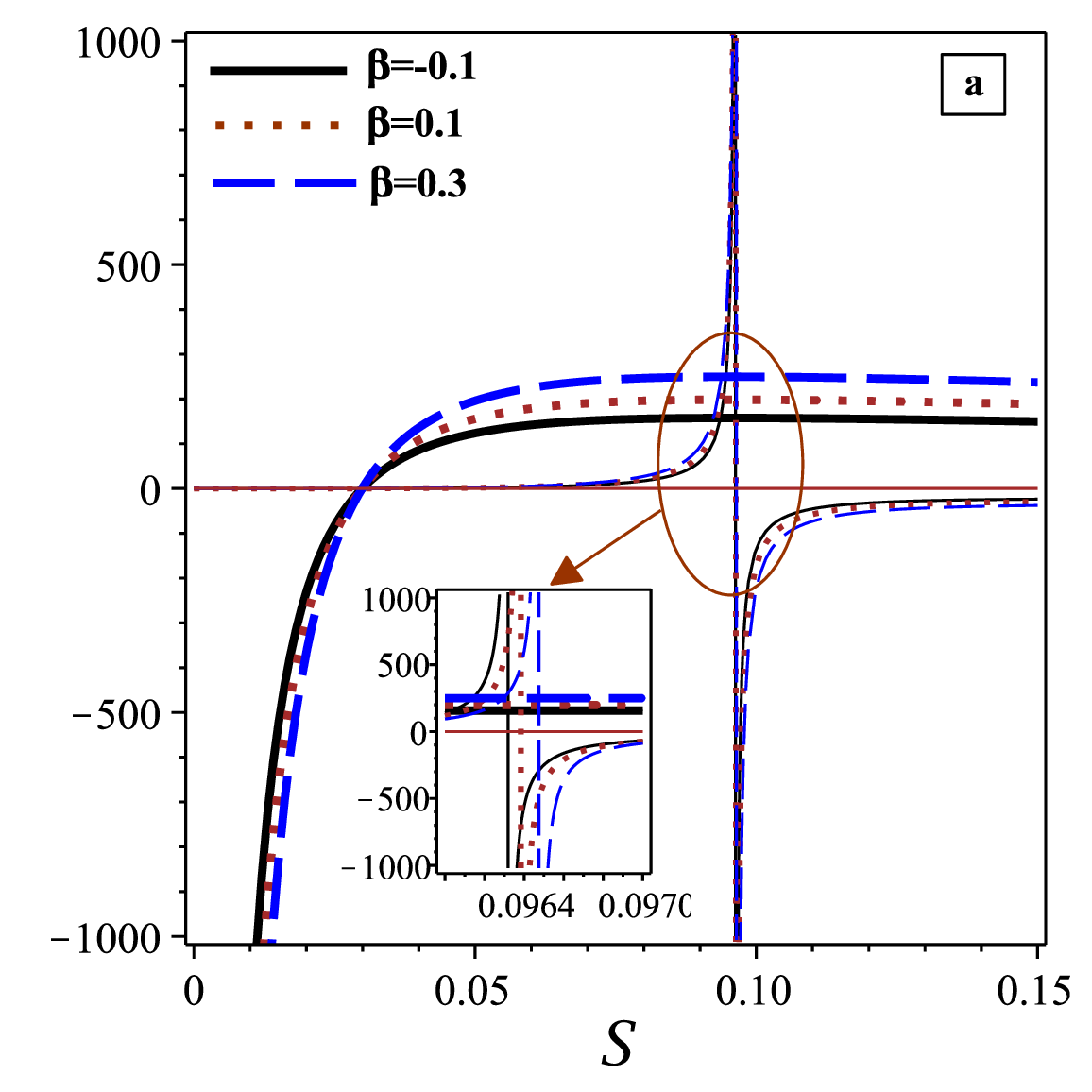} \includegraphics[width=80mm]{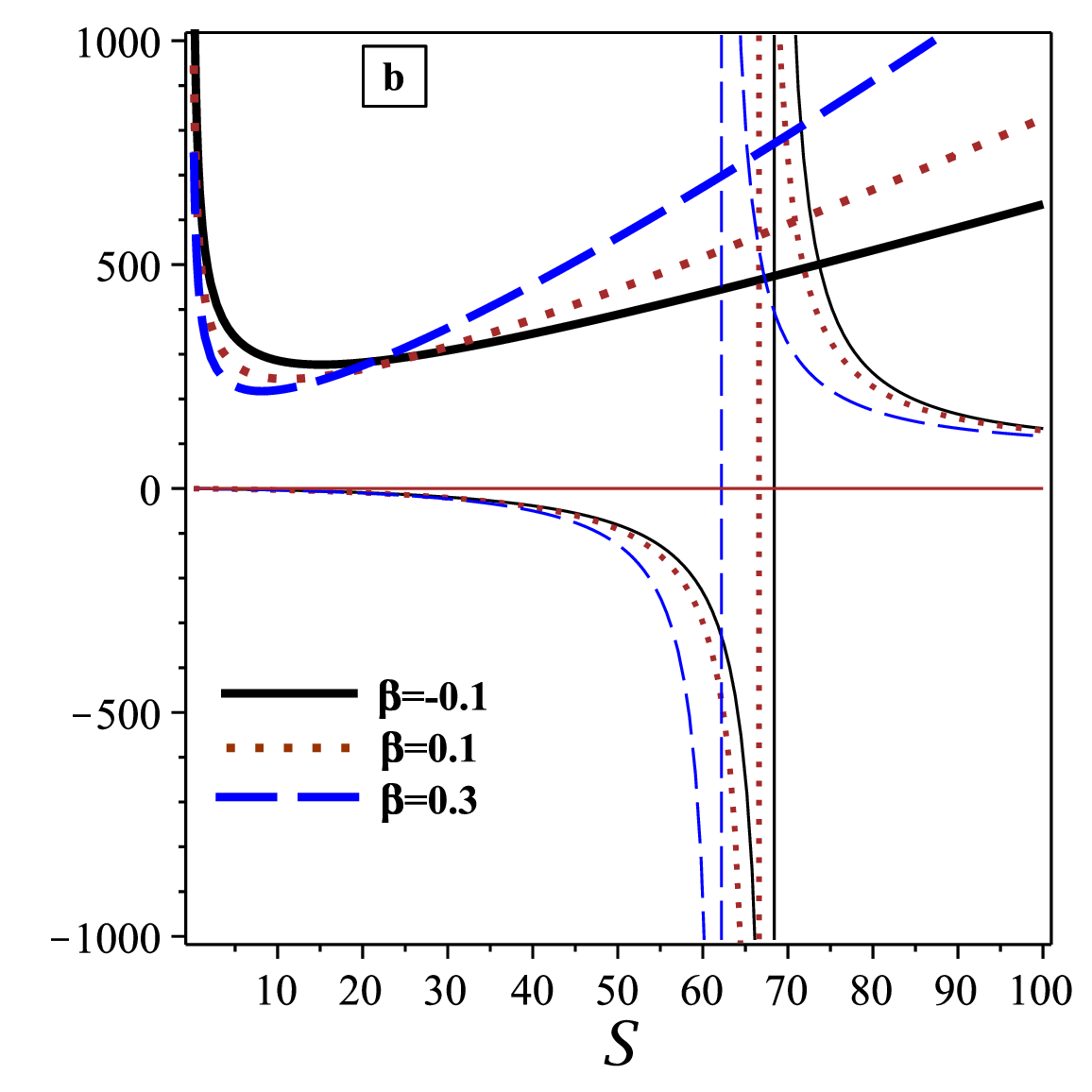} 
		\newline
		\caption{The heat capacity ($C_{Q}$) and the Hawking temperature ($T$)
			versus entropy ($S$) for different values of $\protect\beta $ by considering 
			$Q=b=-\protect\eta _{2}=0.1$, $\protect\alpha =0.3$, $\protect\eta _{1}=0.5$%
			, $\Lambda =-0.1$, $c_{0}=1$, and $m_{g}=0.2$. Bold lines represent the Hawking temperature ($T$), while thin lines indicate the heat capacity ($C_{Q}$)}.
		\label{fig3}
	\end{figure}

\begin{figure}[h]
		\centering
		\includegraphics[width=80mm]{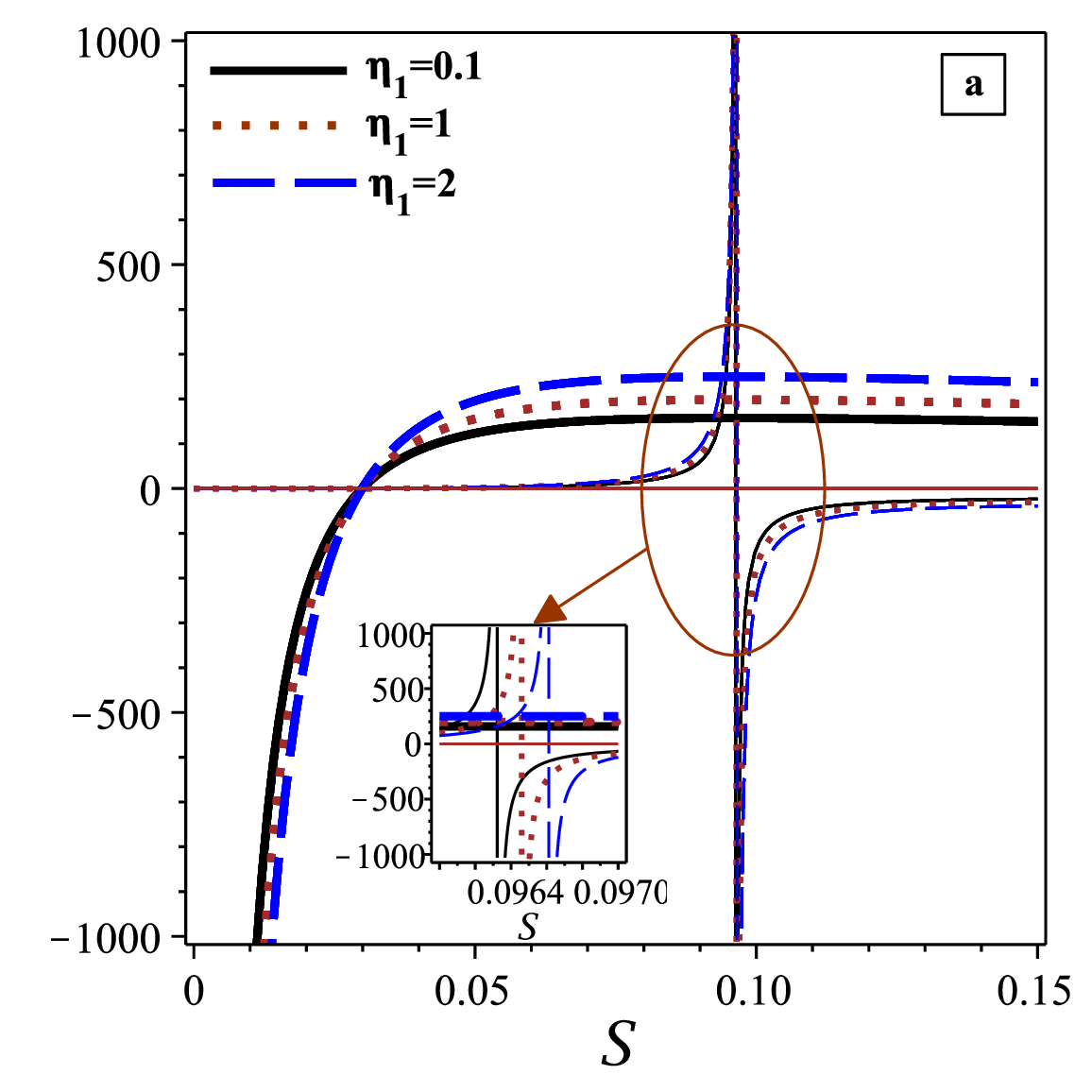} \includegraphics[width=80mm]{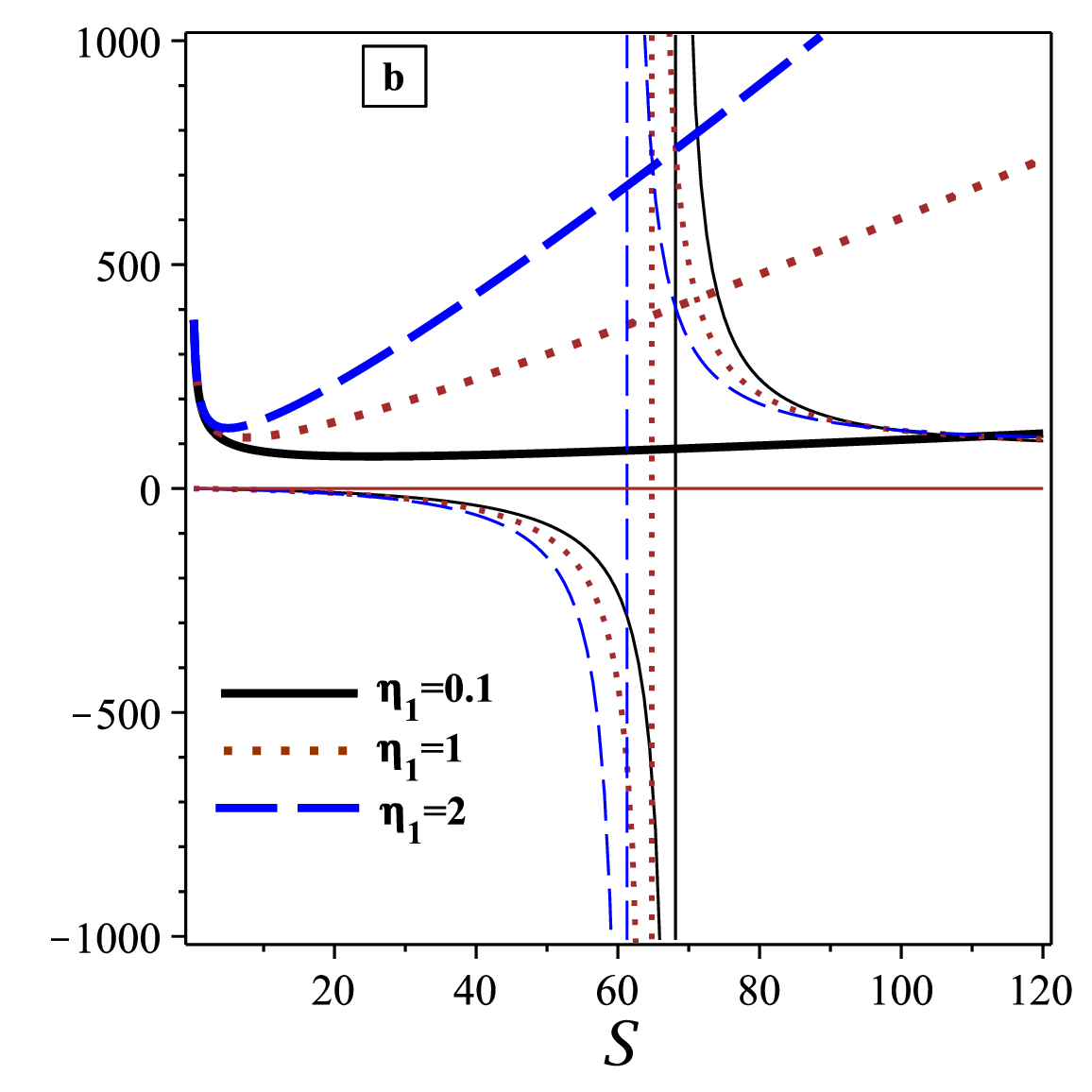} 
		\newline
		\caption{The heat capacity ($C_{Q}$) and the Hawking temperature ($T$)
			versus entropy ($S$) for different values of $\protect\eta _{1}$ by
			considering $Q=b=\protect\beta =-\protect\eta _{2}=0.1$, $\protect\alpha %
			=0.3 $, $\Lambda =-0.1$, $c_{0}=1$, and $m_{g}=0.2$. Bold lines represent the Hawking temperature ($T$), while thin lines indicate the heat capacity ($C_{Q}$)}.
		\label{fig4}
	\end{figure}

\begin{figure}[h]
		\centering
		\includegraphics[width=80mm]{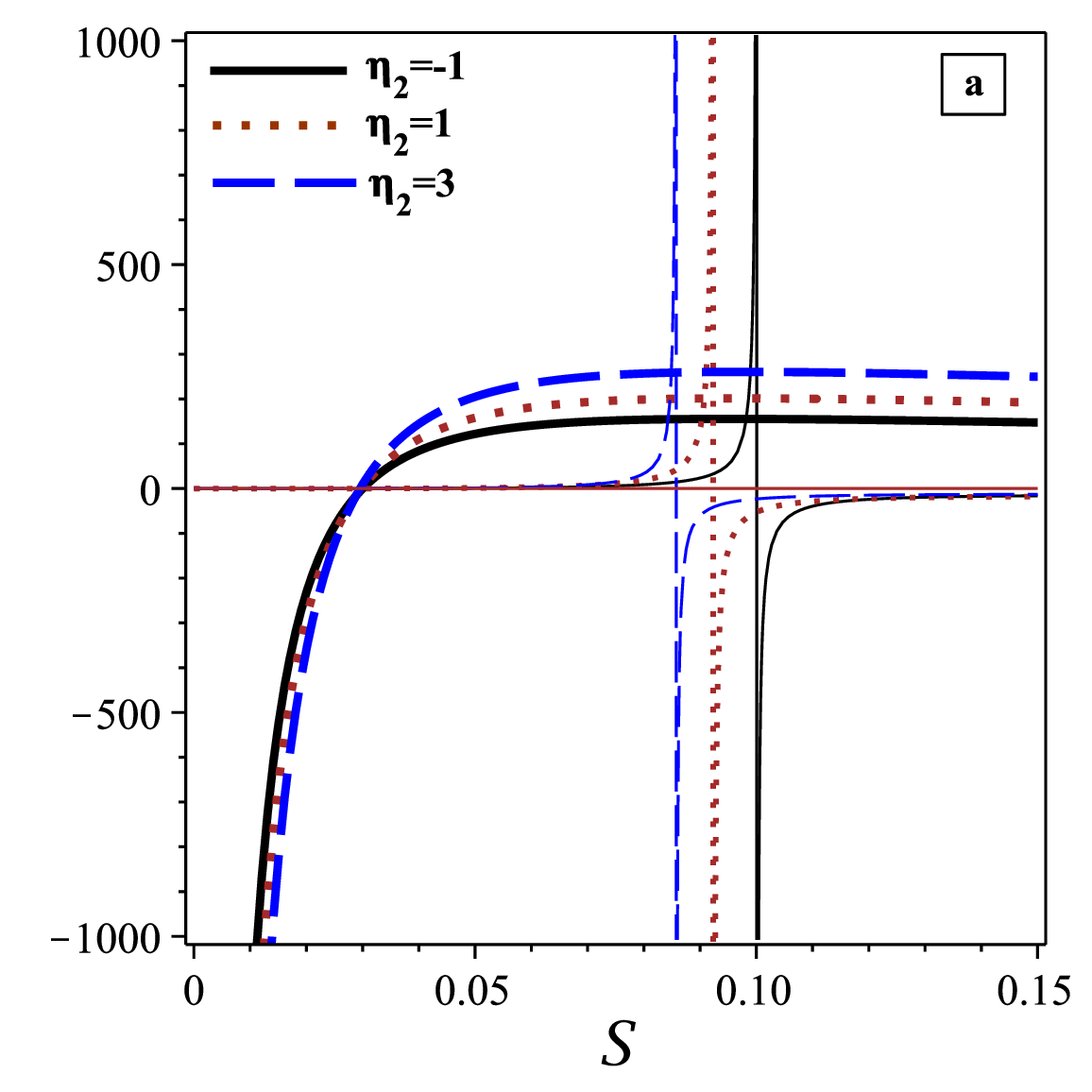} \includegraphics[width=80mm]{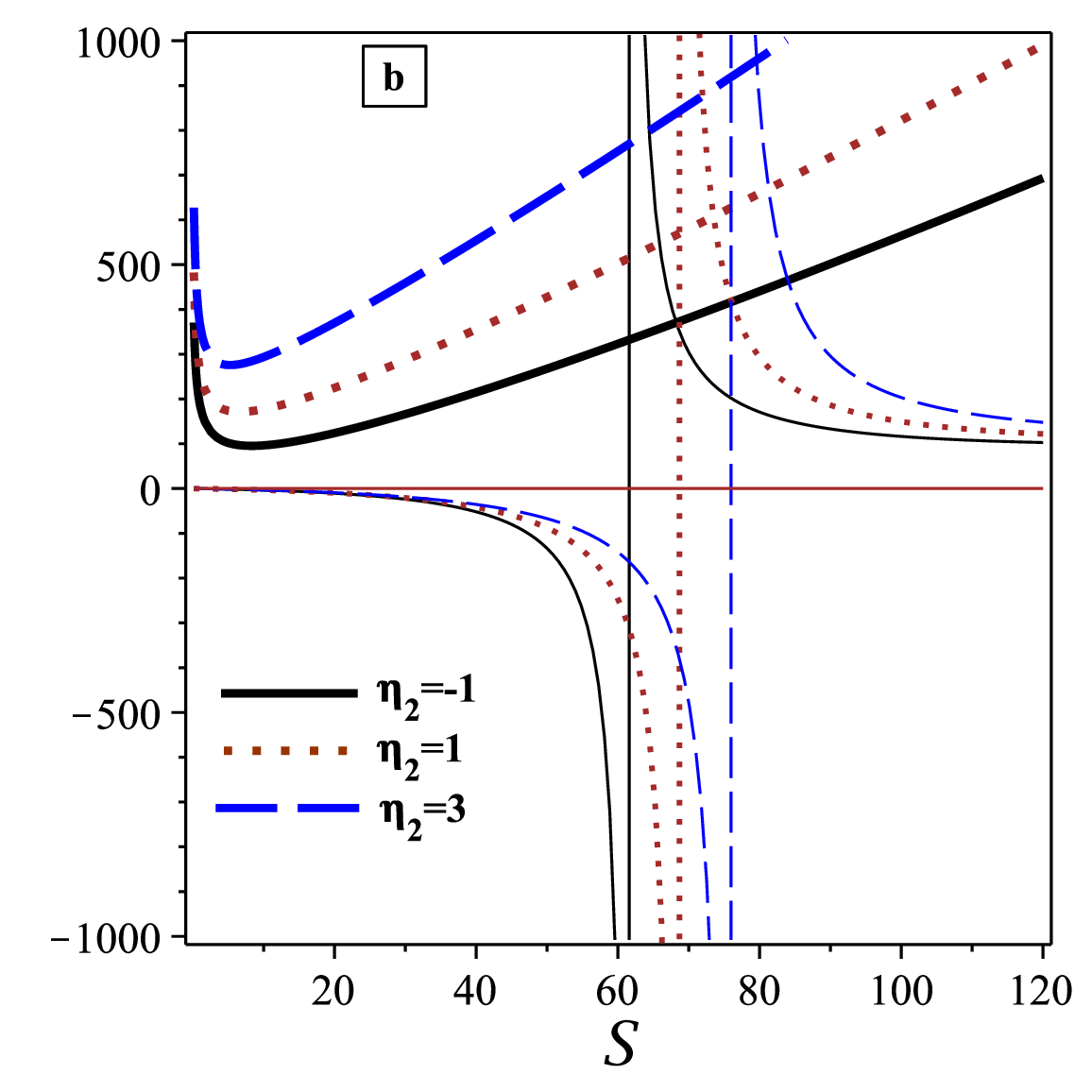} 
		\newline
		\caption{The heat capacity ($C_{Q}$) and the Hawking temperature ($T$)
			versus entropy ($S$) for different values of $\protect\eta_{2}$ by
			considering $Q=b=\protect\beta=0.1$, $\protect\eta_{1}=1$, $\protect\alpha%
			=0.3$, $\Lambda=-0.1$, $c_{0}=1$, and $m_{g}=0.2$. Bold lines represent the Hawking temperature ($T$), while thin lines indicate the heat capacity ($C_{Q}$)}.
		\label{fig5}
	\end{figure}

	Based on the reasons provided, we plot the Hawking temperature ($T$) and
	heat capacity ($C_{Q}$) against entropy ($S$) in Figs. \ref{fig2}-\ref{fig5}%
	. This allows us to examine the effects of the parameters of charged black
	holes in dilaton-dRGT-like massive gravity on thermal stability and phase
	transition points.
	
	Our analysis reveals two divergence points and one zero point for the heat
	capacity, which we denote as $S_{{div}_{1}}$ and $S_{{div}_{2}}$ for the
	first and second divergence points, respectively. Additionally, we identify
	the root of the heat capacity as $S{0}$. We categorize the behavior of heat
	capacity into four distinct areas (see Figs. \ref{fig2}-\ref{fig5}):
	
	\textbf{First area (}$S_{1}$\textbf{)}: This area is defined as being
	between $0$ and $S_{0}$ (i.e., $0 \leq S < S_{0}$). We refer to these as
	very small black holes. These black holes are considered non-physical
	objects because they have a negative temperature, and their heat capacity is
	also negative.
	
	\textbf{Second area (}$S_{2}$\textbf{)}: This area lies between $S_{0}$ and
	the first divergence point of the heat capacity (i.e., $S_{0}<S<S_{{div}%
		_{1}} $). We refer to this area as small black holes. In this area, both the
	temperature and heat capacity of the small black holes are positive.
	Therefore, the small charged black holes in dilaton-dRGT-like massive
	gravity meets the criteria for thermal stability and physical conditions.
	
	\textbf{Third area (}$S_{3}$\textbf{)}: This area lies between $S_{{div}%
		_{1}} $ and $S_{{div}_{2}}$ (i.e., $S_{{div}_{1}}<S<S_{{div}_{2}}$). We
	refer to this area as medium black holes. While the temperature of a medium
	black hole is positive, its heat capacity is negative. As a result, these
	black holes are unstable because they do not meet the conditions for thermal
	stability.
	
	\textbf{Fourth area (}$S_{4}$\textbf{)}: This area pertains to $S > S_{{%
			\text{div}}_{2}}$, which we refer to as large black holes. In this region,
	both the temperature and heat capacity of the large black holes are positive, indicating that they can simultaneously satisfy thermal stability
	and physical conditions.
	
	We study the effects of $\alpha $, $\beta $, $\eta _{1}$, and $\eta_{2}$ on
	the temperature and heat capacity of charged black holes in
	dilaton-dRGT-like massive gravity in figures \ref{fig2}-\ref{fig5}.\\

\textbf{I. The impact of $\alpha$:} We investigate the effect of $\alpha$ on the thermal stability area shown in
Fig. \ref{fig2}. Our findings indicate that increasing the value of $\alpha$ causes the first divergence point ($S_{{div}_{1}}$) and the zero point of the heat capacity ($S_{0}$) to shift toward larger and smaller entropies, respectively. This shift results in an expansion of the stable area $S_{2}$ ($S_{0} < S < S_{{div}{1}}$). However, the stable area for large black holes (denoted as the fourth area or $S_{4}$) decreases as $\alpha$ increases
because $S_{{div}_{2}}$ moves to larger entropy values. Since the change in $S_{4}$ is greater than that in $S_{2}$, the overall thermal stability areadecreases with an increase in $\alpha$.
	
		\textbf{II. The impact of $\beta$}:	We assess the impact of $\beta$ on the thermal stability areas illustrated
	in Fig. \ref{fig3}. Our findings indicate that the thermal stability areas ($%
	S_{2}$ and $S_{4}$) of charged black holes in dilaton-dRGT-like massive
	gravity expand as the value of $\beta$ increases. Specifically, as $\beta$
	rises, $S_{{div}_{1}}$ shifts to higher entropy, while $S_{0}$ remains
	unchanged. This results in an increased stable area for $S_{2}$ (where $%
	S_{0}<S< S_{{div}{1}}$). Additionally, $S_{{div}_{2}}$ moves towards lower
	entropy with increasing $\beta$, further enhancing the thermal stability
	area for larger black holes.
	
		\textbf{III. The impact of $\eta_1$}: The impact of $\eta_{1}$ on the thermal stability areas of charged black
	holes in dilaton-dRGT-like massive gravity is illustrated in Fig. \ref{fig4}%
	. As $\eta_{1}$ increases, $S_{{div}_{1}}$ shifts to higher entropy values,
	while $S_{0}$ remains unchanged. This shift expands the thermal stability
	area within the range $S_{0} < S < S_{{div}_{1}}$. Furthermore, the second
	divergence point decreases as $\eta_{1}$ increases, which further enhances
	the thermal stability area of $S_{2}$. Consequently, the thermal stability
	areas of charged black holes in dilaton-dRGT-like massive gravity expand
	when the value of $\eta_{1}$ is large.\\
	
		\textbf{IV. The impact of $\eta_2$}: We examine the impact of $\eta_{2}$ on the thermal stability areas of
	charged black holes in dilaton-dRGT-like massive gravity, as illustrated in
	Fig. \ref{fig5}. Increasing $\eta_{2}$ causes $S_{{div}_{1}}$ and $S_{{div}%
		_{2}}$ to shift towards smaller and larger entropy values, respectively,
	while $S_{0}$ remains unchanged. This results in a reduction of thermal
	stability regions in both the ranges $S_{0}<S<S_{{div}_{1}}$ and $S>S_{{div}%
		_{2}}$. Consequently, the thermal stability areas of these black holes
	diminish as $\eta_{2}$ increases.\\
	
	By examining the heat capacity in Figs. \ref{fig2}-\ref{fig5}, we identify
	two divergences in the heat capacity, denoted as $S_{{div}_{1}}$ and $S_{{div%
		}_{2}}$. Between these two divergence points, the black holes are physical
	and unstable, characterized by positive temperature and negative heat
	capacity. This leads to a phase transition from small entropy black holes to
	large entropy black holes. This phase transition is analogous to the van der
	Waals phase transition is observed in fluids. Research has shown that the
	thermodynamics of an asymptotically AdS metric in four-dimensional spacetime
	closely resembles that of a van der Waals fluid. Specifically, when the
	charge is below a critical value, the isocharge in the temperature-entropy plane exhibits one unstable branch and two stable branches, along with a
	second-order critical point at a critical charge. Consequently, we observe a
	van der Waals-like phase transition in charged black holes within
	dilaton-dRGT-like massive gravity. This behavior has been documented in Ref. 
	\cite{Yue2024}, which explores the extended phase space by treating the
	cosmological constant as thermodynamic pressure ($P=-\frac{\Lambda}{8\pi}$) 
	\cite{VdWI,VDWI2,VdWII,VdWIII,VdWIV,VdWV}, with its conjugate quantity
	considered as thermodynamic volume. Under this framework, a distinct
	pressure-volume oscillatory behavior emerges, linking the small-large black
	hole phase transition to the liquid-gas phase transition of the van der
	Waals fluid \cite{KM}. This phenomenon has been further investigated in
	detail for black holes in Refs. \cite{PTI,PTII,PTIII,PTIV,PTV,PTVI,PTVII}.
	
As we mentioned in the previous paragraph, the phase transition of Maxwell-dilaton-dRGT-like black holes have been examined in greater detail in Ref. \cite{Yue2024}. The authors evaluated the effects of the parameters $\alpha$ and $\gamma$ on the phase transition within both canonical and grand canonical ensembles. Specifically, they found that, when considering black holes in the context of dilaton-dRGT-like massive gravity and under grand canonical ensembles, van der Waals-like critical behavior necessitates a coupling between the graviton and dilaton fields ($\gamma \neq 0$). Furthermore, when $\alpha \neq 0$, additional phenomena emerge, potentially including first-order phase transitions between small black holes (SBHs) and large black holes (LBHs). An analysis of the Gibbs free energy indicates the presence of a triple critical point and a zero-order reverse reentrant phase transition (SBHs $\leftrightarrow$ LBHs $\leftrightarrow$ SBHs), which is thermodynamically the opposite of the standard reentrant transition. In contrast, in canonical ensembles, van der Waals-like transitions can occur without coupling; however, an examination of the Helmholtz free energy also reveals that reverse reentrant transitions exist when $\alpha \neq 0$. In the context of the phase transition in the canonical ensemble, we examined the heat capacity in greater detail. We found that black holes in the Maxwell-dilaton-dRGT-like massive theory can undergo a phase transition between small black holes (SBHs) and large black holes (LBHs). Our findings are consistent with the results obtained in Ref. \cite{TSHendi} within the canonical ensemble.
	
	\subsection{Geometrothermodynamics}
	
	Geometrothermodynamics offers a valuable approach for exploring the critical
	points of phase transitions, specifically the divergence points of heat
	capacity in black holes. Historically, several thermodynamic metrics have
	been developed to identify these critical points, including those proposed
	by Weinhold \cite{WeinholdI,WeinholdII}, Ruppeiner \cite%
	{RuppeinerI,RuppeinerII}, Quevedo \cite{QuevedoI,QuevedoII}, and HPEM \cite%
	{HPEM}. However, references indicate that the Ruppeiner, Weinhold, and
	Quevedo metrics may not fully align with all divergence and zero points of
	heat capacity for certain black holes \cite{HPEMI,HPEMIV}. In contrast, there have been no reported inconsistencies with the HPEM metric regarding the identification of these divergence points and zero heat capacity. In this study, we will examine four thermodynamic metrics to determine which one accurately corresponds to the divergence and zero points of heat
	capacity in charged black holes within the framework of dilaton-dRGT-like
	massive gravity.
	
	The Weinhold metric is introduced as follows \cite{WeinholdI,WeinholdII} 
	\begin{equation}
		ds_{W}^{2}=g_{ab}^{W}dX^{a}dX^{b},  \label{Wein}
	\end{equation}%
	where $g_{ab}^{W} = \frac{\partial^2 M(X^{c})}{\partial X^{a} \partial X^{b}}
	$ and $X^{a} \equiv X^{a}(S, N^{i})$, with $N^{i}$ representing other
	extensive variables of the system.
	
	Using the metric (\ref{Wein}), we can express the denominator of Weinhold's
	Ricci scalar for charged black holes in dilaton-dRGT-like massive gravity as
	follows 
	\begin{equation}
		denom(R_{W})=\left( M_{SS}M_{QQ}-M_{SQ}^{2}\right) ^{2}M^{2},  \label{RW}
	\end{equation}%
	where $M=M(S,Q)$. Comparing Eq. (\ref{C}) and Eq. (\ref{RW}), we observe
	that the denominator of the Weinhold Ricci scalar contains two additional
	terms, $M_{SQ}^{2}$ and $M_{QQ}$. These extra terms create a mismatch
	between the divergence and zero points of the heat capacity and the
	divergences of the Weinhold Ricci scalar. Consequently, the Weinhold metric
	exhibits additional divergences that are not associated with any bounds or
	phase transition points of the heat capacity.
	
\begin{figure}[h]
		\centering
		\includegraphics[width=50mm]{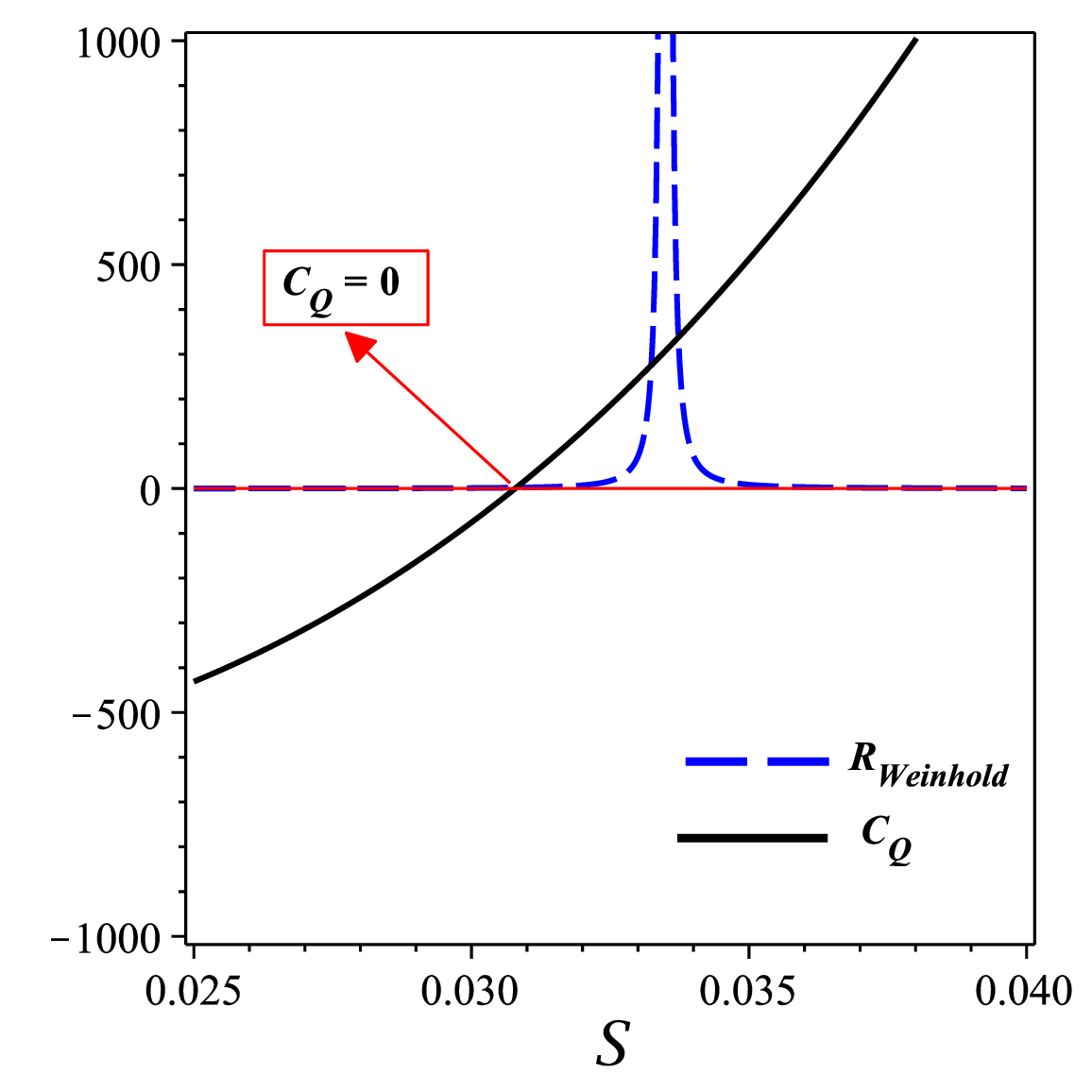} \includegraphics[width=50mm]{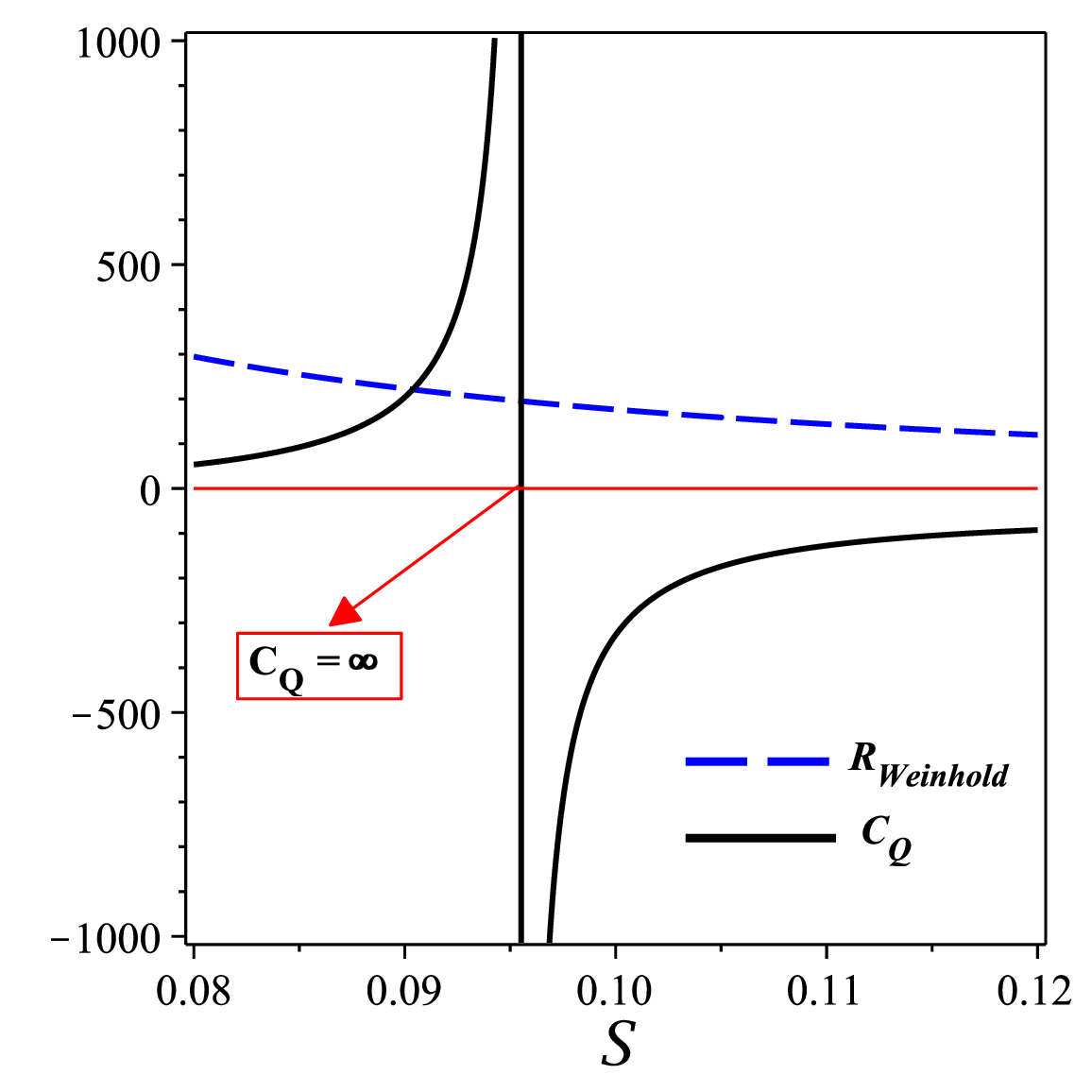} %
		\includegraphics[width=50mm]{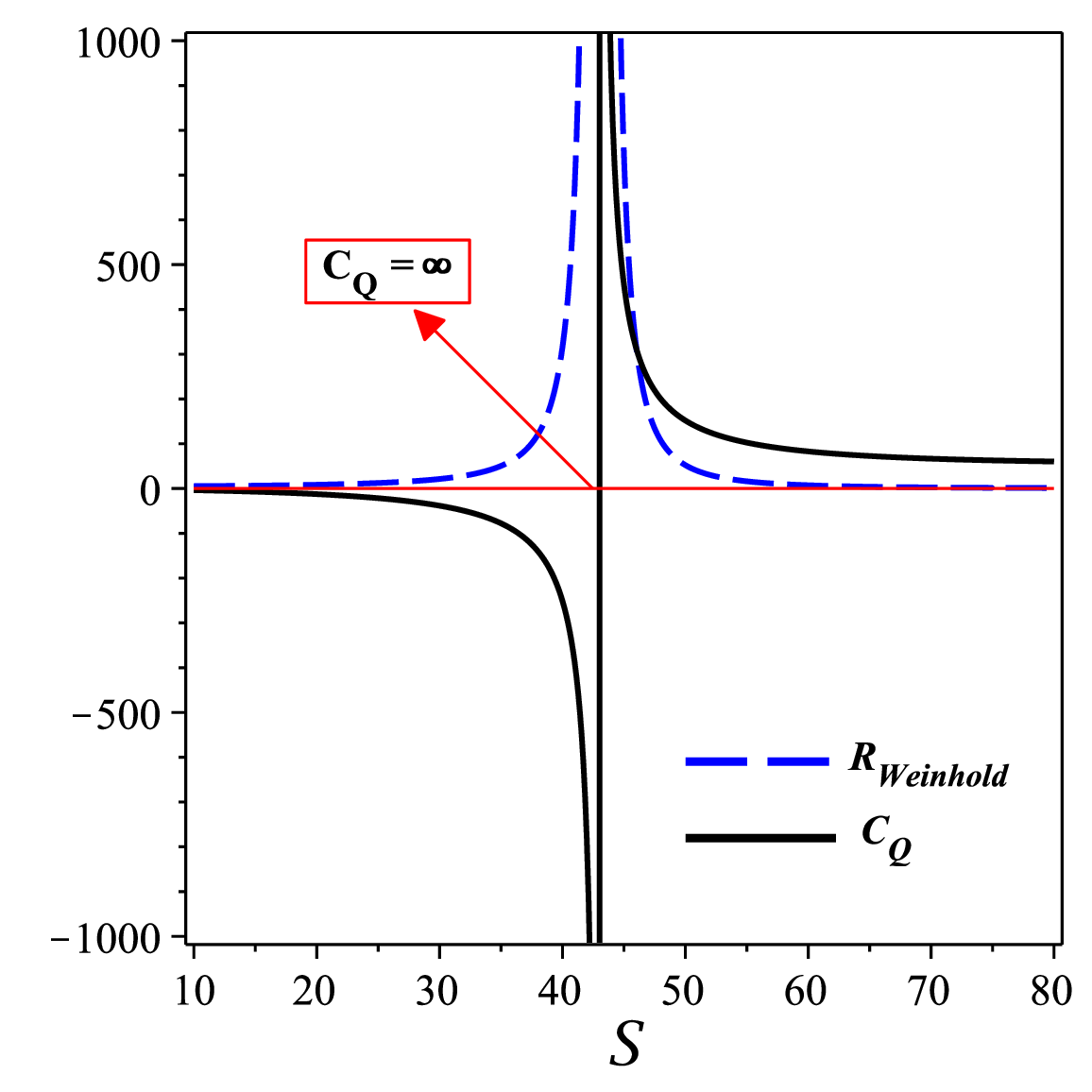} \newline
		\caption{The heat capacity ($C_{Q}$) and the Ricci scalar of the Weinhold
			metric ($R_{Weinhold}$) versus entropy ($S$) for $Q=b=\protect\beta =-%
			\protect\eta _{2}=0.1$, $\protect\eta _{1}=0.5$, $\Lambda =-0.1$, $c_{0}=1$,
			and $m_{g}=\protect\alpha =0.2$. Continuous line corresponds to the heat capacity ($C_{Q}$), while the dashed line represents the Ricci scalar of the Weinhold metric ($R_{Weinhold}$)}.
		\label{fig6}
	\end{figure}

	For further details, we present a plot of the heat capacity and the Ricci
	scalar of the Weinhold metric versus entropy in Fig. \ref{fig6}. Our findings indicate that the physical limitations (roots of the heat capacity)
	and the critical points of phase transitions (divergence points of heat
	capacity) do not coincide with the divergences of the Ricci scalar of the
	Weinhold metric.
	
	The Ruppeiner metric is defined as follows \cite{RuppeinerI,RuppeinerII}
	
	\begin{equation}
		ds_{R}^{2}=\frac{1}{T}ds_{W}^{2},  \label{Rupp2}
	\end{equation}
	where the metrics of Ruppeiner and Weinhold are conformally equivalent, with the inverse of the temperature serving as the conformal factor.
	
	Considering equation (\ref{Rupp2}), we derive the denominator of Ruppeiner's
	Ricci scalar for charged black holes in dilaton-dRGT-like massive gravity as
	follows 
	\begin{equation}
		denom(R_{R})=\left( M_{SS}M_{QQ}-M_{SQ}^{2}\right) ^{2}M^{2}T,  \label{RR}
	\end{equation}%
	where $T=T(S,Q)$. Like the Weinhold metric, the Ruppeiner Ricci scalar
	includes two additional terms in its denominator ($M_{SQ}^{2}$ and $M_{QQ}$)
	that do not correspond to any bounds or phase transition points of the heat
	capacity. As a result, this metric cannot fully capture the phase
	transitions and bound points of certain thermodynamic systems due to these
	extra terms in the denominator.
	
\begin{figure}[h]
		\centering
		\includegraphics[width=50mm]{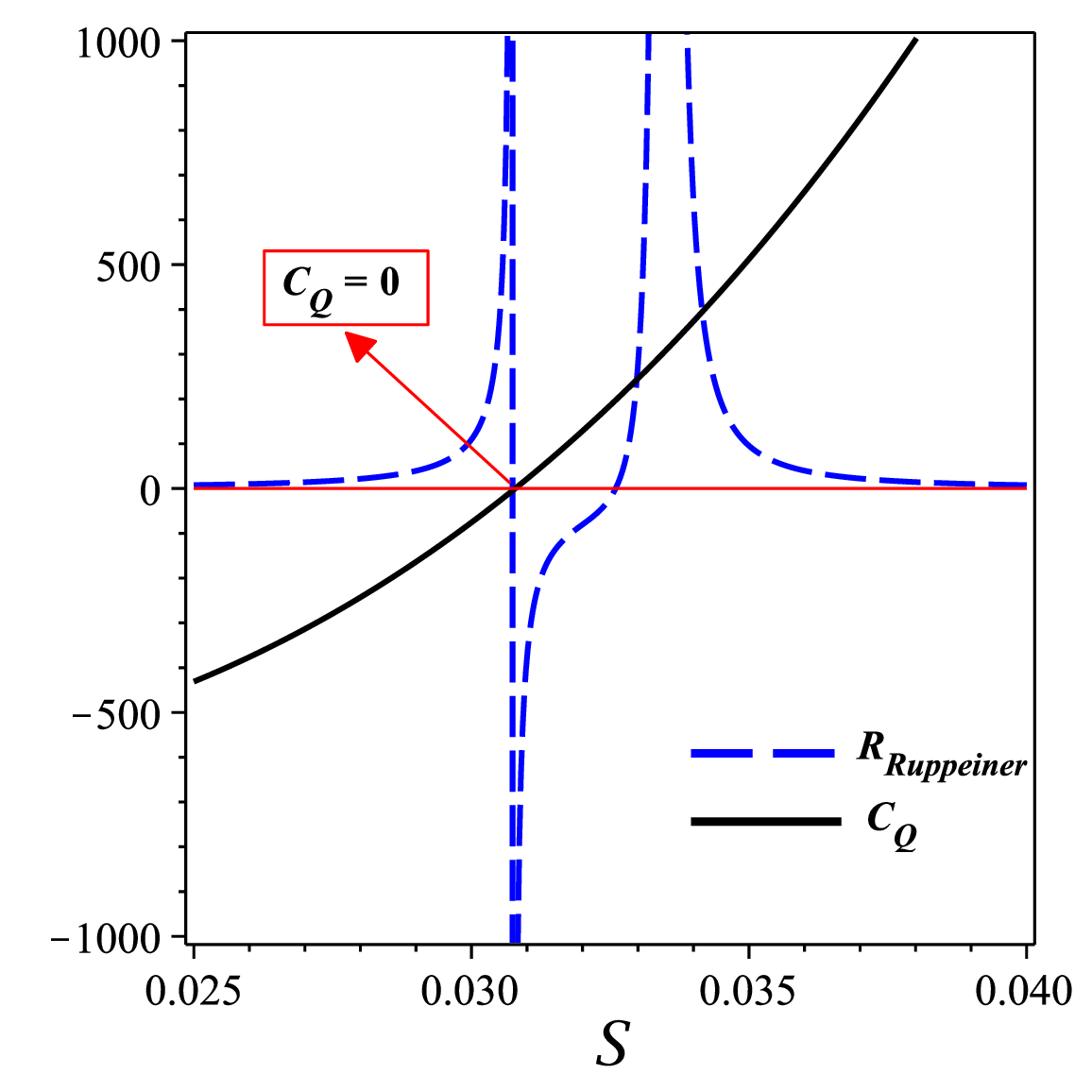} \includegraphics[width=50mm]{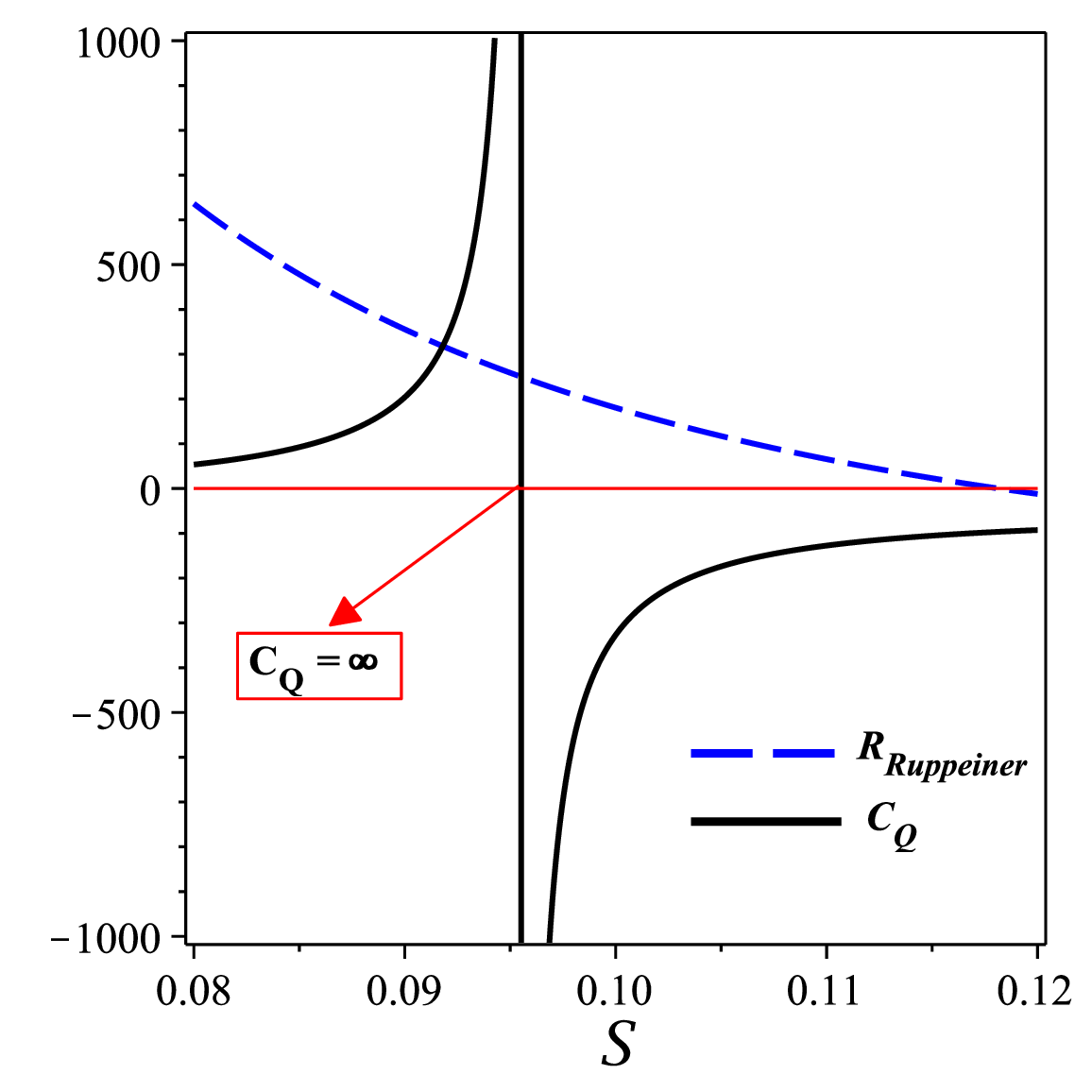} %
		\includegraphics[width=50mm]{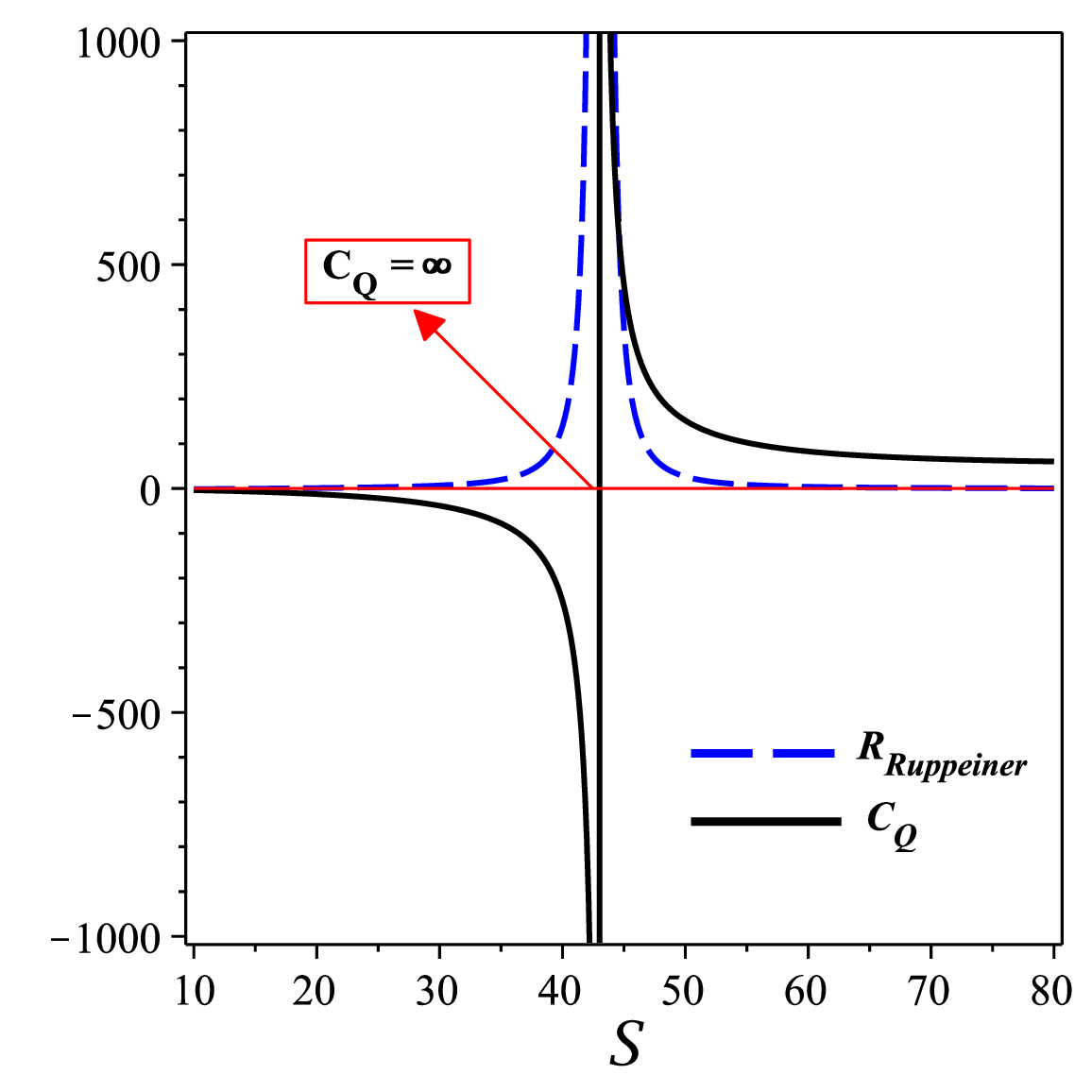} \newline
		\caption{The heat capacity ($C_{Q}$) and the Ricci scalar of the Ruppeiner
			metric ($R_{Ruppeiner}$) versus entropy ($S$) for $Q=b=\protect\beta =-%
			\protect\eta _{2}=0.1$, $\protect\eta _{1}=0.5$, $\Lambda =-0.1$, $c_{0}=1$,
			and $m_{g}=\protect\alpha =0.2$. Continuous line corresponds to the heat capacity ($C_{Q}$), while the dashed line represents the Ricci scalar of the Ruppeiner metric ($R_{Ruppeiner}$)}.
		\label{fig7}
	\end{figure}

	To investigate the behavior of the Ruppeiner metric, we plot the heat
	capacity and the Ricci scalar of the Ruppeiner metric versus entropy in Fig. %
	\ref{fig7}. Our findings show that the physical limitations and the critical
	points of phase transitions do not coincide with the divergences of the
	Ricci scalar of the Ruppeiner metric.
	
	The Quevedo metric is given by \cite{QuevedoI,QuevedoII}
	
	\begin{equation}
		ds_{Q}^{2}=\Omega \left( -M_{SS}dS^{2}+M_{QQ}dQ^{2}\right) ,  \label{Quevedo}
	\end{equation}%
	where $M_{QQ}=\left( \frac{\partial ^{2}M(S,Q)}{\partial Q^{2}}\right) _{S}$%
	, and $\Omega $ is
	
	\begin{equation}
		\Omega =\left\{ 
		\begin{array}{ccc}
			SM_{S}+QM_{Q} &  & caseI \\ 
			&  &  \\ 
			SM_{S} &  & caseII%
		\end{array}%
		\right. .
	\end{equation}
	
	Applying the Quevedo metrics (\ref{Quevedo}), we can extract the denominator
	of Quevedo's Ricci scalar for charged black holes in dilaton-dRGT-like
	massive gravity, which leads to 
	\begin{equation}
		denom(R_{Q})=\left\{ 
		\begin{array}{ccc}
			\left( SM_{S}+QM_{Q}\right) ^{3}M_{SS}^{2}M_{QQ}^{2} &  & caseI \\ 
			&  &  \\ 
			S^{3}M_{S}^{3}M_{SS}^{2}M_{QQ}^{2} &  & caseII%
		\end{array}%
		\right. .
	\end{equation}
	
\begin{figure}[h]
		\centering
		\includegraphics[width=50mm]{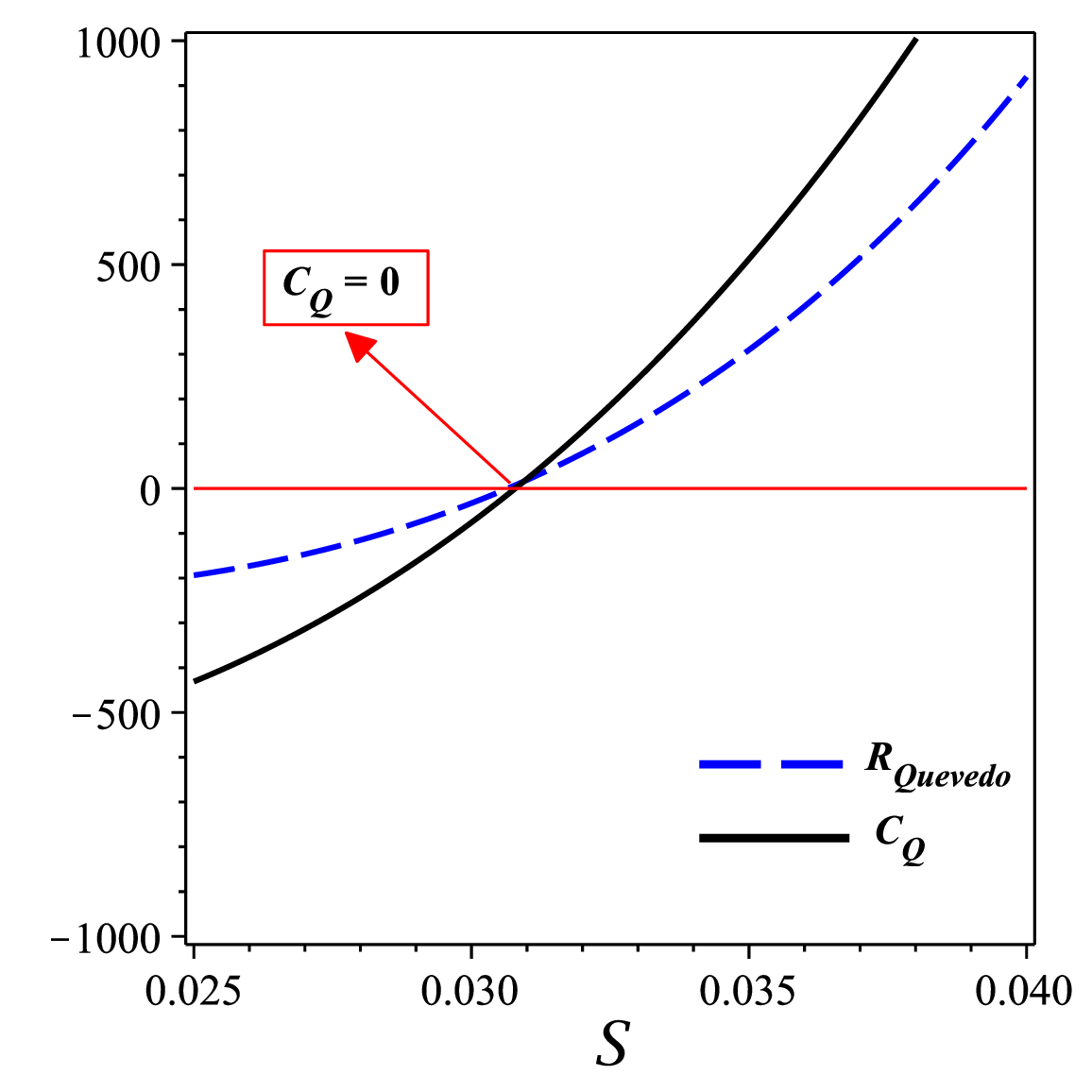} \includegraphics[width=50mm]{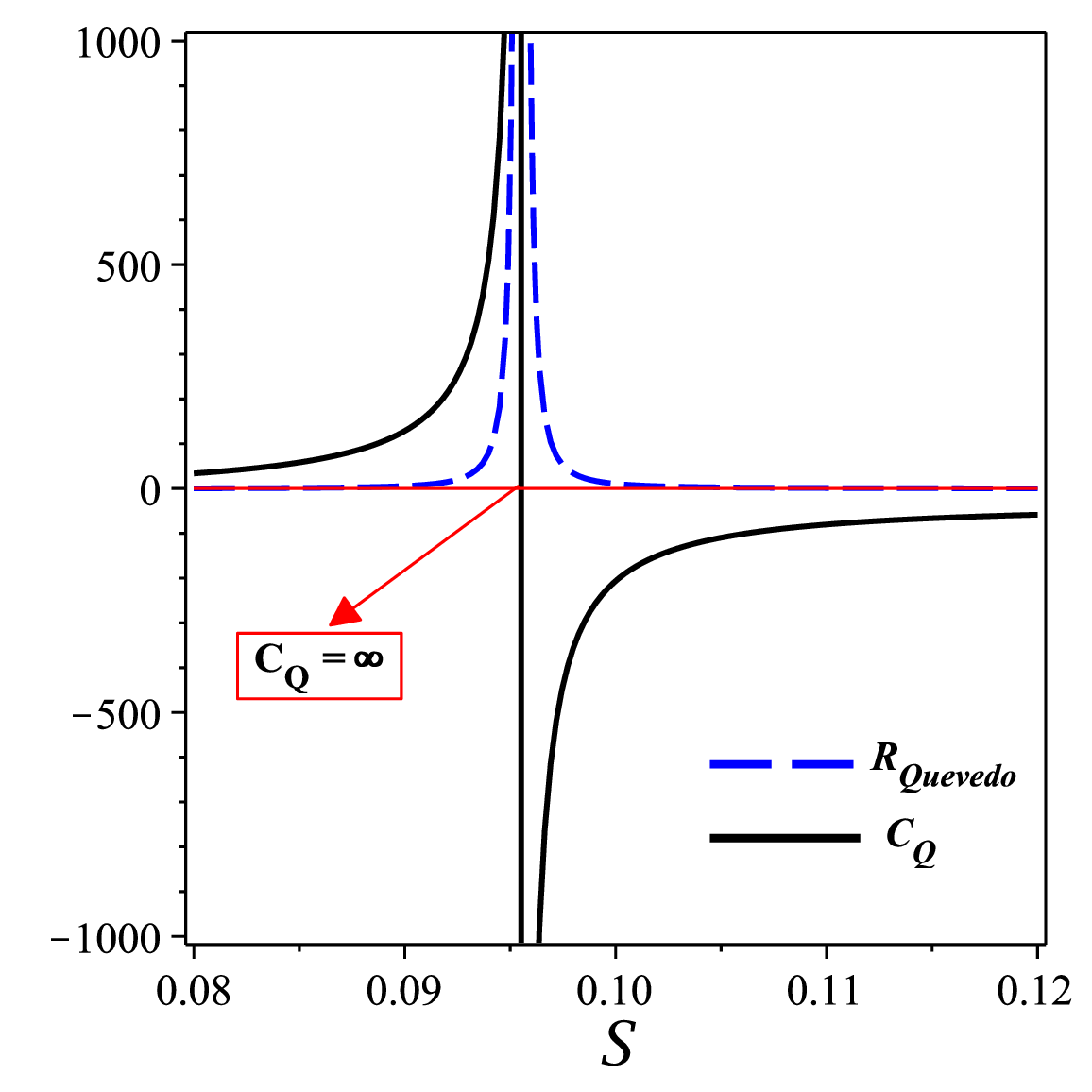} %
		\includegraphics[width=50mm]{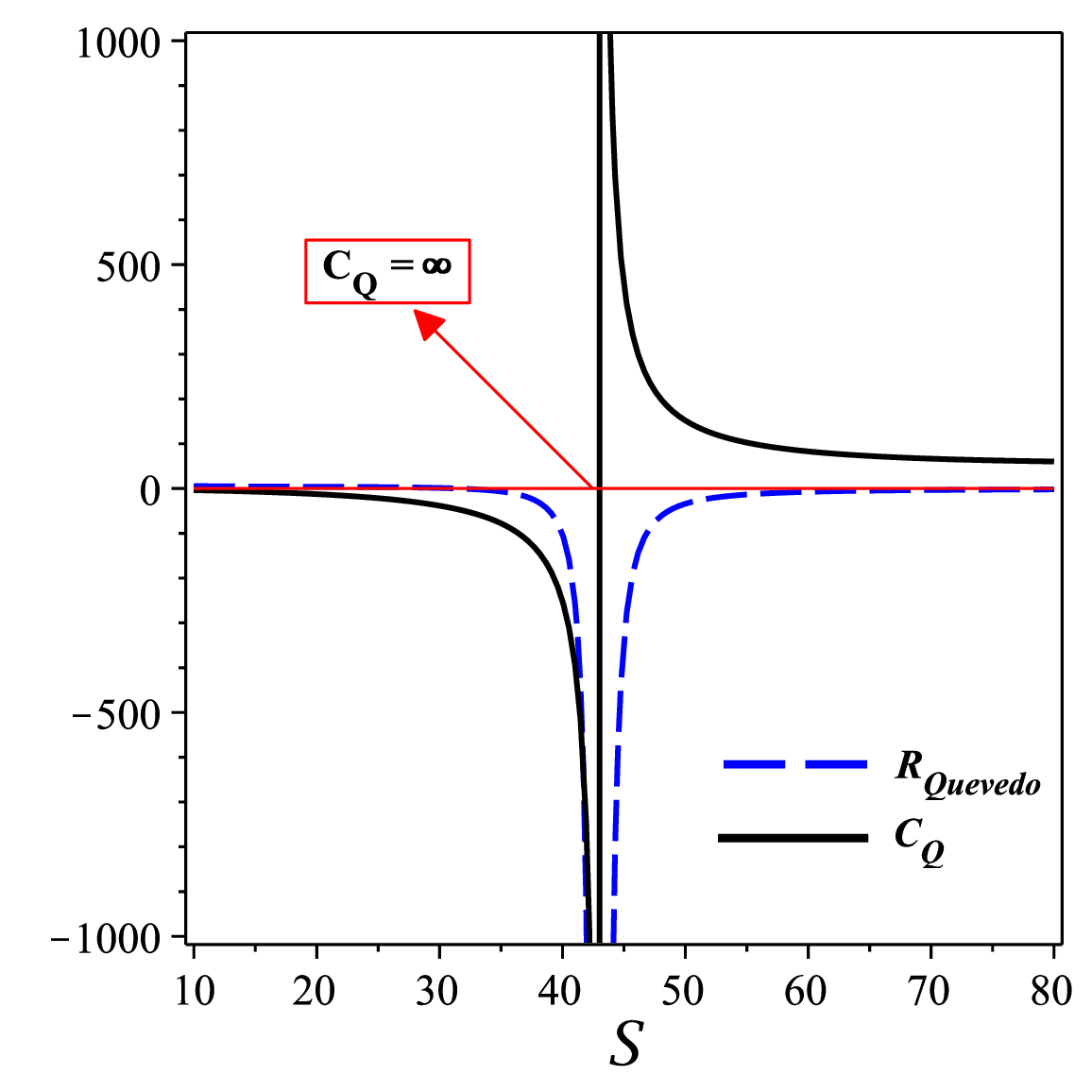} \newline
		\caption{The heat capacity ($C_{Q}$) and the Ricci scalar of the Quevedo
			metric ($R_{Quevedo}$) versus entropy ($S$) for $Q=b=\protect\beta =-\protect%
			\eta _{2}=0.1$, $\protect\eta _{1}=0.5$, $\Lambda =-0.1$, $c_{0}=1$, and $%
			m_{g}=\protect\alpha =0.2$. Continuous line corresponds to the heat capacity ($C_{Q}$), while the dashed line represents the Ricci scalar of the Quevedo metric ($R_{Quevedo}$)}.
		\label{fig8}
	\end{figure}

	Due to the presence of $M_{Q}=\left( \frac{\partial M(S,Q)}{\partial Q}
	\right) {S}$ and $M{QQ}^{2}$ in Quevedo's Ricci scalars, we may encounter
	additional divergence points that do not correspond to the phase transition
	or the critical points of the heat capacity. For further details, we plot
	this metric versus entropy in Fig. \ref{fig8}. As shown in the left panel of
	Fig. \ref{fig8}, there are no divergence points of Quevedo's Ricci scalar
	that coincide with the zero point of the heat capacity.
	
	The HPEM metric is introduced in reference \cite{HPEM} as 
	\begin{equation}
		dS_{HPEM}^{2}=\frac{SM_{S}}{M_{QQ}^{3}}\left(
		-M_{SS}dS^{2}+M_{QQ}dQ^{2}\right) .  \label{HPEM}
	\end{equation}
	
	We derive the denominator of HPEM's Ricci scalar for charged black holes in
	dilaton-dRGT-like massive gravity as follows 
	\begin{equation}
		denom(R_{HPEM})=S^{3}M_{S}^{3}M_{SS}^{2},  \label{RHPEM}
	\end{equation}%
	this ensures that all phase transition and bound points align with the
	divergences of the HPEM's Ricci scalar, without introducing any additional
	terms that could create extra divergences.
	
\begin{figure}[h]
		\centering
		\includegraphics[width=50mm]{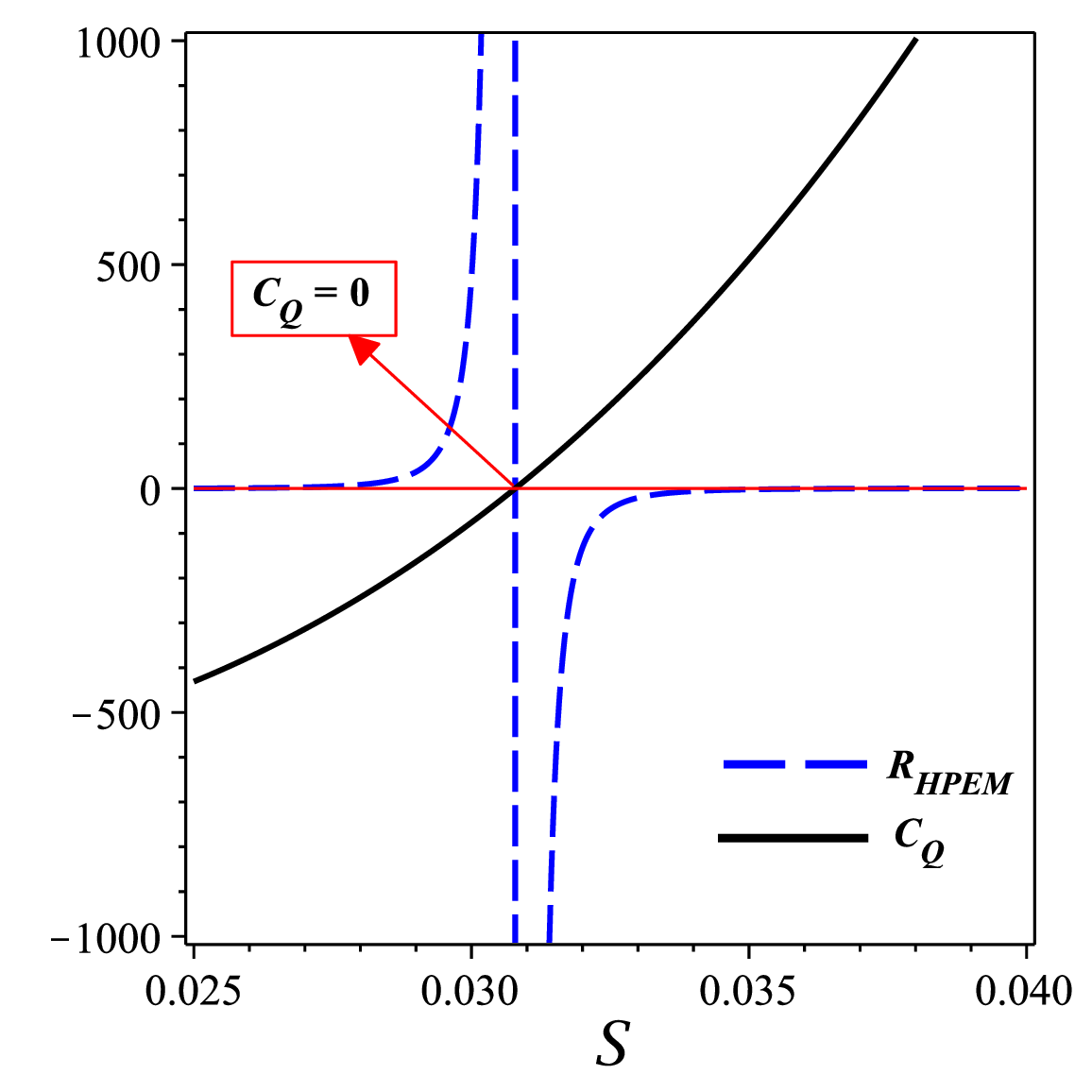} \includegraphics[width=50mm]{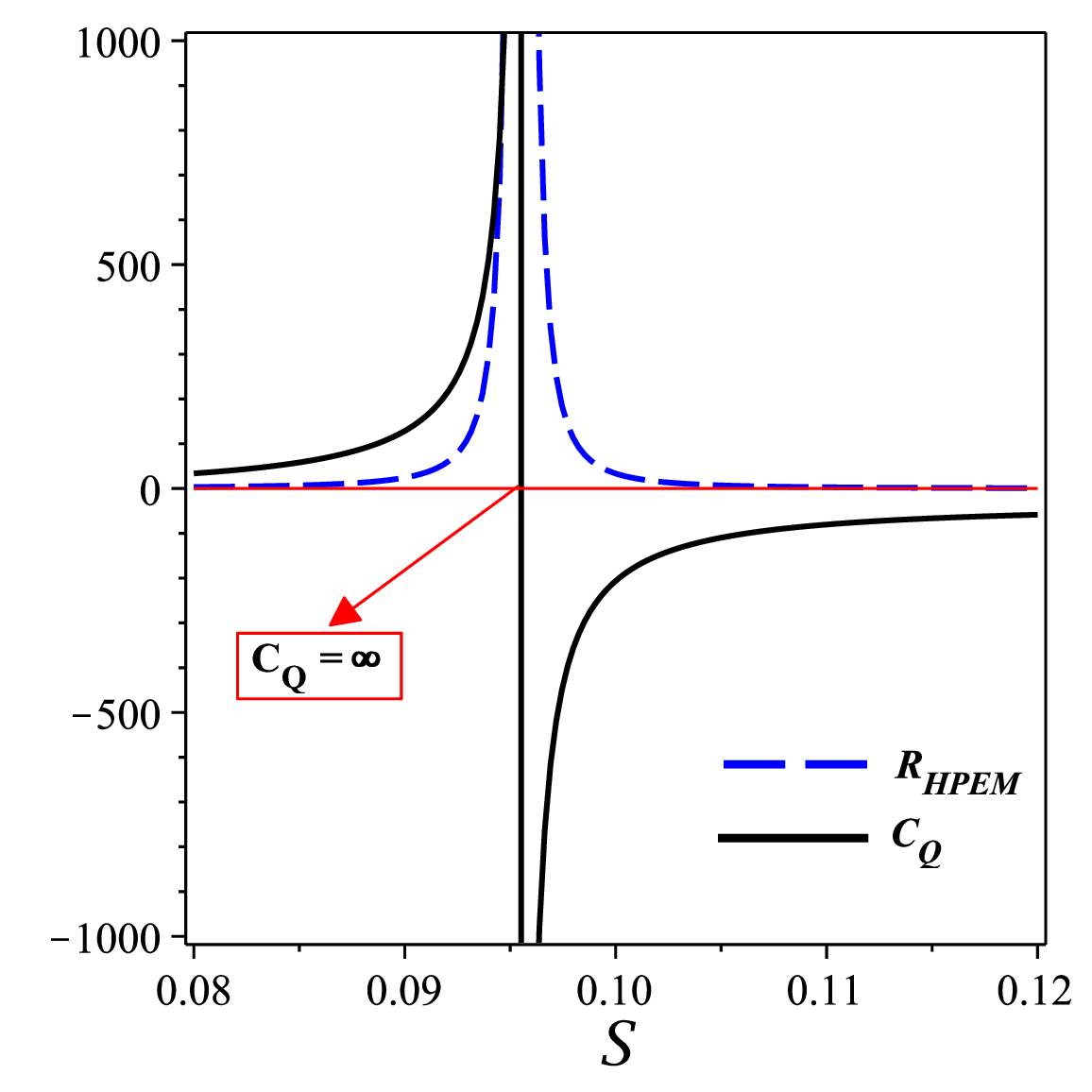} %
		\includegraphics[width=50mm]{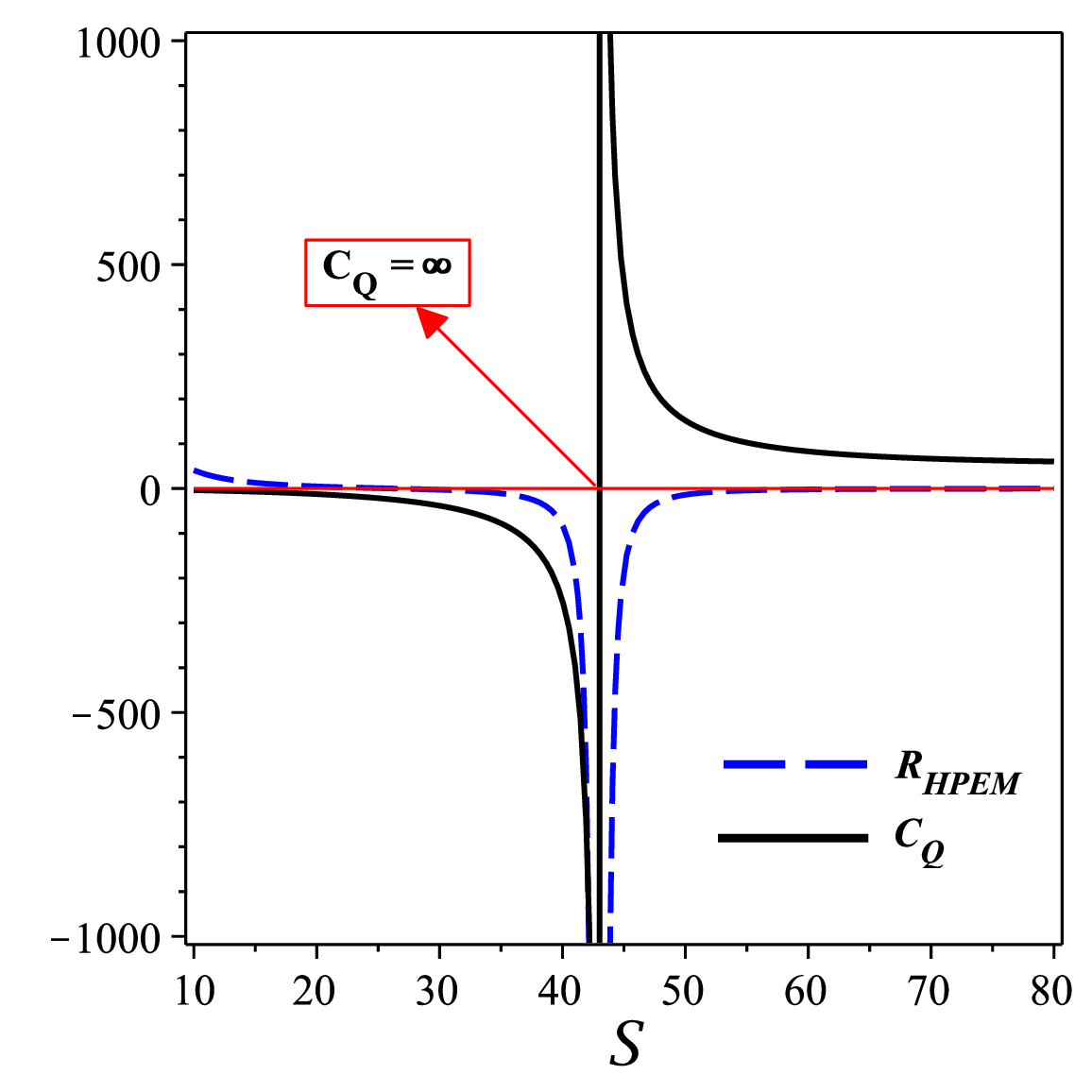} \newline
		\caption{The heat capacity ($C_{Q}$) and the Ricci scalar of the HPEM metric
			($R_{HPEM}$) versus entropy ($S$) for $Q=b=\protect\beta =-\protect\eta %
			_{2}=0.1$, $\protect\eta _{1}=0.5$, $\Lambda =-0.1$, $c_{0}=1$, and $m_{g}=%
			\protect\alpha=0.2$. Continuous line corresponds to the heat capacity ($C_{Q}$), while the dashed line represents the Ricci scalar of the HPEM metric ($R_{HPEM}$)}.
		\label{fig9}
	\end{figure}

	Our findings indicate that the divergence points of the Ricci scalar of the
	HPEM metric coincides with both the roots and divergence points of the heat
	capacity. For further details, please refer to Fig. \ref{fig9}. This means
	that all physical limitations (roots of the heat capacity) and the critical
	points of phase transitions (divergence points of heat capacity) are
	reflected in the divergences of the Ricci scalar of the HPEM metric (see
	Fig. \ref{fig9}). Another important result concerning the HPEM metric is the
	differing behavior of the Ricci scalar before and after its divergence
	points. Specifically, the Ricci scalar exhibits distinct behavior at
	divergence points associated with physical limitations and phase transition
	critical points. In other words, the sign of the Ricci scalar changes before
	and after divergences when the heat capacity is zero (see Fig. \ref{fig9}).
	However, the signs of the Ricci scalar remain the same when the heat
	capacity encounters divergences (see Fig. \ref{fig9}). These divergences are
	referred to as $\Lambda$ divergences. Therefore, this approach allows us to
	differentiate between physical limitations and phase transition critical
	points.
	
	\section{Optical features}
	
	\label{sec:photon}
	
	The groundbreaking image of a black hole captured by the Event Horizon Telescope \cite{akiyama2022firstSgrA,akiyama2019firstM87} has highlighted the significance of studying shadow phenomena and geodesic structures for a deeper understanding of black hole characteristics \cite{volker,perlick2015influence,vagnozzi2022horizon, okyay2022,Khan20241,Khan2022,Rahmatov2025,ref01,ref02}. This study focuses on deriving null geodesic equations and investigating how Maxwell-dilaton-dRGT-like massive gravity influences photon trajectories through Lagrangian formalism in the following sections.
	
	\subsection{Photonic radius}
	
	In this part, we examine the photonic radius in the Maxwell-dilaton-dRGT-like massive framework. The Lagrangian is defined as $2\mathcal{L}=g_{\mu \nu }\dot{x}^{\mu }\dot{x}^{\nu }$, and the dot denotes differentiation concerning an arbitrary affine parameter. Assuming motion is confined to the equatorial plane, i.e., $\theta =\frac{\pi }{2}$, the Lagrangian for a photon as a massless particle simplifies to 
	\begin{equation}
		2\mathcal{L}=-f(r)\dot{t}^{2}+\frac{\dot{r}^{2}}{f(r)}+r^{2}R^{2}(r)\dot{\phi%
		}^{2}=0.  \label{lagrangian}
	\end{equation}
	
	Moreover, the presence of two Killing vectors within the system results in two conserved quantities: energy, denoted as $E=f(r)\dot{t}$, and angular momentum, represented as $L=r^{2}R^{2}(r)\dot{\phi}$. Considering these assumptions, Eq. \eqref{lagrangian} leads to the radial equation 
	\begin{equation}
		\dot{r}^{2}+E^2=V_{\text{eff}},
	\end{equation}%
	where the effective potential is expressed as 
	\begin{equation}
		V_{\text{eff}}=\frac{L^{2}f(r)}{r^{2}R^{2}(r)}.
	\end{equation}
	
\begin{figure}[h]
			\centering
			\includegraphics[width=0.4\linewidth]{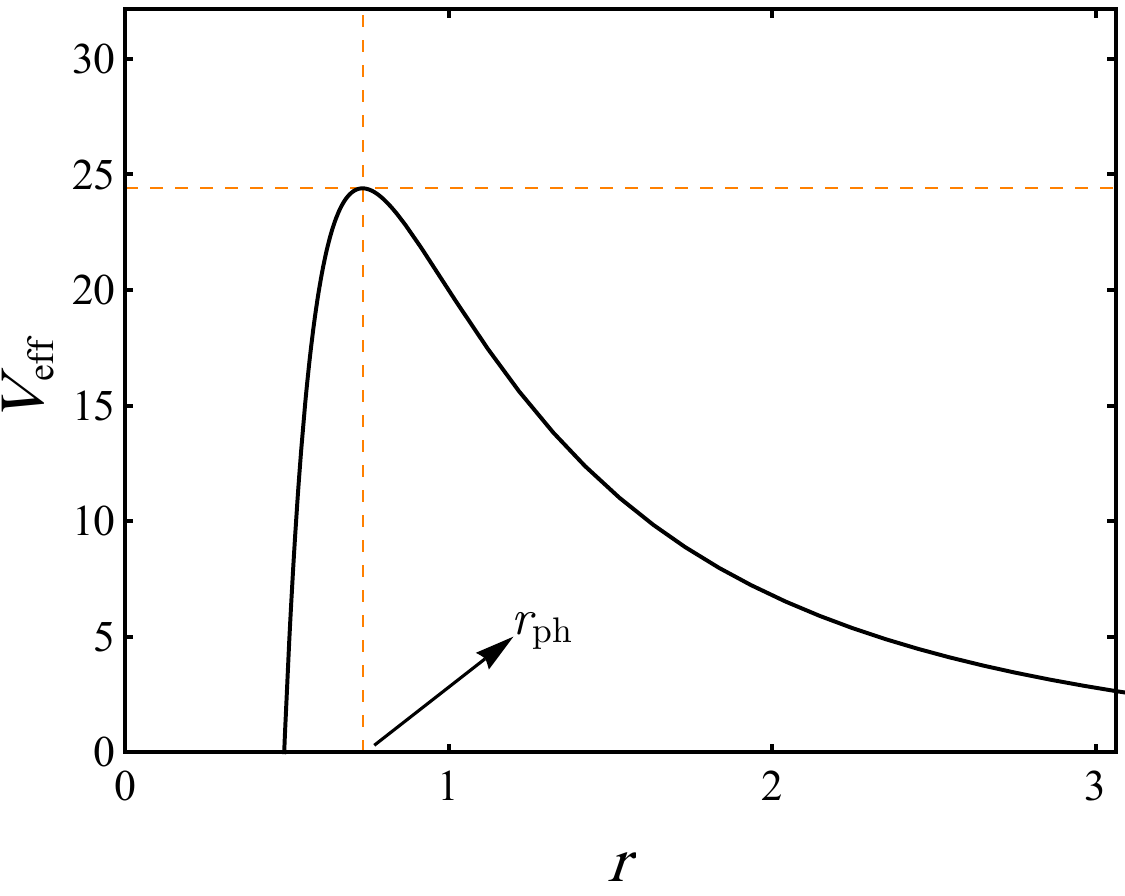}
			\caption{Typical shape of the effective potential $V_{\text{eff}}(r)$ for null geodesics, showing a maximum at the photon sphere radius $r = r_{\text{ph}}$. The condition $\frac{\partial^2 V_{\text{eff}}}{\partial r^2} \bigg|_{r = r_{\text{ph}}} < 0$ confirms the instability of photon orbits at this radius.}
			\label{fig:Veff}
	\end{figure}

	The photonic radius $r_{\text{ph}}$ is calculated by finding the critical radius in the effective potential
	\begin{equation}
		V_{\text{eff}}=\frac{\partial V_{\text{eff}}}{\partial r}=0,  \label{eq:Veff}
	\end{equation}%
The unstable critical orbits determine the photonic radius $r = r_{\text{ph}}$, which can be found by the sign of the second derivative of the effective potential. For an unstable critical orbit (which is the physically relevant case for black hole photon spheres), the following condition must be satisfied. 
		\begin{equation}
			\frac{\partial^2 V_{\text{eff}}}{\partial r^2} \bigg|_{r = r_{\text{ph}}} < 0
		\end{equation}
		Fig.~\ref{fig:Veff} shows the typical shape of $V_{\text{eff}}$ for fixed parameters at $m_0 = 0.5; \alpha = b_0 = \beta = q = -\eta_2= - \Lambda = m_g =  0.1$; $c_0 = \eta_1 = 1$ and $L = 6$. The plot demonstrates clearly that $r_{\text{ph}}$ corresponds to a local maximum, confirming the unstable nature of the photon orbits.
		
	Now, we calculate the photonic radius by utilizing Eq. \eqref{eq:Veff}, which leads to the following equation 
	\begin{align}
		& r_{\text{ph}}\left( \left( \alpha ^{2}+1\right) \left(
		c_{0}m_{g}^{2}\left( \frac{b}{r_{\text{ph}}}\right) {}^{-\frac{2\alpha \beta 
			}{\alpha ^{2}+1}}\left( -\frac{\eta _{1}r_{\text{ph}}\left( \frac{b}{r_{%
					\text{ph}}}\right) {}^{\frac{\alpha ^{2}}{\alpha ^{2}+1}}}{\alpha
			^{2}+2\alpha \beta +2}-\frac{2c_{0}\text{$\eta _{2}$}}{\alpha ^{2}+2\alpha
			\beta -1}\right) -\frac{4q^{2}\left( \frac{b}{r_{\text{ph}}}\right) {}^{%
				\frac{2}{\alpha ^{2}+1}}}{\left( \alpha ^{2}-1\right) {b}^{2}}\right)
		-2\right)  \notag  \label{eq:rph} \\
		& -\frac{\left( \alpha ^{2}-3\right) m_{0}{b}^{\frac{2\alpha ^{2}}{\alpha
					^{2}+1}}}{\alpha ^{2}+1}=0.
	\end{align}
	

	Table. \ref{Tab:rph} presents the dependence of the photonic radius $r_{\text{ph}}$ on the key parameters: the parameters of the dilaton field ($\alpha $), massive gravity parameter ($\eta_{1}$), charge ($q$) and also the graviton mass ($m{g}$) in the framework of Maxwell-dilaton-dRGT-like massive gravity, with all other variables held constant on $m_{0}=0.5$, $b_{0}=0.1$, $c_{0}=1$, $\eta _{2}=-0.1$, $\beta =0.1$, and $q=0.1$. It is worth mentioning that the photonic radius is independent of the cosmological constant based on Eq. (\eqref{eq:rph}). The results offer valuable insights into how these physical quantities influence the behavior of photon orbits around black holes.

		Based on Eq.~\eqref{metric} and the discussion in Sec.~\ref{Sec:fr}, the spacetime structure is governed by several key parameters, whose variations can significantly influence its physical characteristics. Accordingly, when deriving a general expression, particular attention must be paid to the behavior of parameters. The results presented here have been obtained for a specific set of parameter values, and while they effectively illustrate the underlying physical features, their applicability should be regarded as limited to this chosen parameter regime.
		
		Table~\ref{Tab:rph} summarizes the behavior of the photon radius $r_{\text{ph}}$ for different choices of the dilaton field parameter $\alpha$, the graviton mass $m_g$, the electric charge $q$, and the parameter $\eta_1$. The corresponding trends are interpreted below.\\

			\begin{itemize}
			 \item \textbf{Dilaton field (\(\alpha\)):} Starting with the variation of the dilaton field parameter $\alpha$, for fixed $\eta_{1} = -2$, $q = 0.5$, and $m_{g}=0.2$, Table~\ref{Tab:rph} shows that $r_{\text{ph}}$ decreases as $\alpha$ increases. This trend reflects the role of the dilaton field in strengthening the effective gravitational attraction near the horizon. A larger $\alpha$ enhances the coupling between the scalar and electromagnetic sectors, effectively deepening the potential well and drawing the photon orbit closer to the black hole. Consequently, the photon sphere becomes more compact for a stronger dilaton field.\\

			\item \textbf{Graviton mass (\(m_g\)):} In contrast to the role of $\alpha$, an opposite behavior is observed when varying the graviton mass $m_g$ while keeping $\alpha = 0.1$, $q = 0.5$, and $\eta_1 = -2$ fixed. The photon radius $r_{\text{ph}}$ increases with increasing $m_g$. Physically, this can be attributed to the modification of spacetime curvature introduced by the massive graviton term. A higher graviton mass effectively weakens the gravitational field at a given radius, reducing the overall curvature strength and allowing the photon sphere to extend outward.\\
		
			\item \textbf{Electric charge (\(q\)):} The influence of the charge parameter $q$ is also evident from Table~\ref{Tab:rph} when $\alpha = 0.1$, $m_g = 0.2$, and $\eta_1 = -2$ are held constant. Here, $r_{\text{ph}}$ grows with increasing $q$. This arises from the repulsive contribution of the electromagnetic field, which counteracts the gravitational attraction. As the electric charge increases, the net potential barrier moves outward, resulting in a larger photon sphere radius.\\
		
				\item \textbf{Massive gravity parameter (\(\eta_1\)):} Finally, the parameter $\eta_1$ exerts an opposite effect. For fixed $\alpha = 0.1$, $m_g = 0.2$, and $q = 0.5$, higher values of $\eta_1$ lead to smaller $r_{\text{ph}}$. This indicates that positive $\eta_1$ strengthens the attractive component of the massive gravity potential, effectively confining photons more tightly near the black hole.
		\end{itemize}
		Overall, these trends highlight the delicate interplay among the dilaton field, electromagnetic charge, and massive gravity effects in shaping the photon sphere structure. The variation of $r_{\text{ph}}$ with different parameters provides valuable insights into how scalar, vector, and tensor degrees of freedom collectively influence the near-horizon geometry in the Maxwell-dilaton-dRGT-like massive gravity framework. The dependence of the photonic radius on these parameters will, in turn, have direct implications for observable features such as the black hole shadow, which will be examined in the following section.
		
		\begin{table}[H]
			\centering
			\caption{Variation of the photon radius $r_\text{ph}$ with respect to the model parameters $\alpha$, $m_g$, $q$, and $\eta_1$ in the Maxwell-dilaton-dRGT-like massive gravity background. The photonic radius is computed for specific parameters set to fixed values; $m_{0} = 1$, $b = 0.2$, $\beta = 0.15$, $\eta_{2} = -0.1$, $\Lambda = -0.1$, and $c_{0} = 1$.}
			\label{Tab:rph}
			\begin{tabular}{|c|c|c|c|c|c|c|c|}
				\hline
				$\alpha$ & 0.00 & 0.05 & 0.10 & 0.15 & 0.20 & 0.25 & 0.30 \\ \hline
				$r_\text{ph}$ & 1.8327 & 1.8220 & 1.7896 & 1.7413 & 1.6858 & 1.6331 & 1.5935 \\ \hline \hline
				$m_g$ & 0.0 & 0.2 & 0.4 & 0.6 & 0.8 & 1.0 & 1.2 \\ \hline
				$r_\text{ph}$ & 1.7398 & 1.7427 & 1.7517 & 1.7670 & 1.7896 & 1.8208 & 1.8627 \\ \hline \hline
				$q$ & 0.0 & 0.1 & 0.2 & 0.3 & 0.4 & 0.5 & 0.6 \\ \hline
				$r_\text{ph}$ & 1.4718 & 1.4870 & 1.5307 & 1.5990 & 1.6869 & 1.7896 & 1.9037 \\ \hline \hline
				$\eta_1$ & -3 & -2 & -1 & 0 & +1 & +2 & +3 \\ \hline
				$r_\text{ph}$ & 1.8211 & 1.7896 & 1.7604 & 1.7332 & 1.7077 & 1.6837 & 1.6611 \\ \hline
			\end{tabular}
		\end{table}

	
	\subsection{Shadow radius}
	
	Once the photon sphere radius $r_{\text{ph}}$ has been determined, we can proceed to evaluate the corresponding shadow radius. For a spacetime expressed as $ds^{2}=-g_{tt}dt^{2}+g_{rr}dr^{2}+g_{\theta \theta }d\theta^{2}+g_{\varphi \varphi }d\varphi ^{2}$, the angular size of shadow $\mathcal{\alpha}_{\text{sh}}$ is assumed with the following expression \cite{volker}	
	\begin{equation}\label{alphash}
	\sin{{\alpha}_{\text{sh}}}=\frac{{h}(r_{\text{ph}})}{h(r_{\text{O}})},
	\end{equation}
	where $h(r)$ is defined as 
	\begin{equation}\label{eq:hr}
		h(r)^{2}=\frac{g_{\varphi \varphi }}{g_{tt}}=\frac{r^{2}R^{2}(r)}{f(r)}.
	\end{equation}
	and based on Eq. \eqref{alphash}, the shadow radius detected by an observer located at $\tilde{r}_{\text{O}}$ is approximately defined by 
	\begin{equation}
		\alpha_{\text{sh}}\approx \frac{\mathcal{R}_{\text{sh}}}{\tilde{r}_{\text{O}}}.
	\end{equation}
	By utilizing the definition of $h(r)$ in Eq. \eqref{eq:hr} and the photonic radius obtained by Eq. \eqref{eq:rph}, we explore the shadow radius with the help of the following equation
	\begin{equation}
		\mathcal{R}_\text{sh}=\frac{r_{\text{ph}}R(r_{\text{ph}})}{\sqrt{f(r_{\text{ph}})}}\sqrt{f(r_{\text{O}})}
	\end{equation}
	
\begin{figure}[h]
			\centering
			\includegraphics[width=75mm]{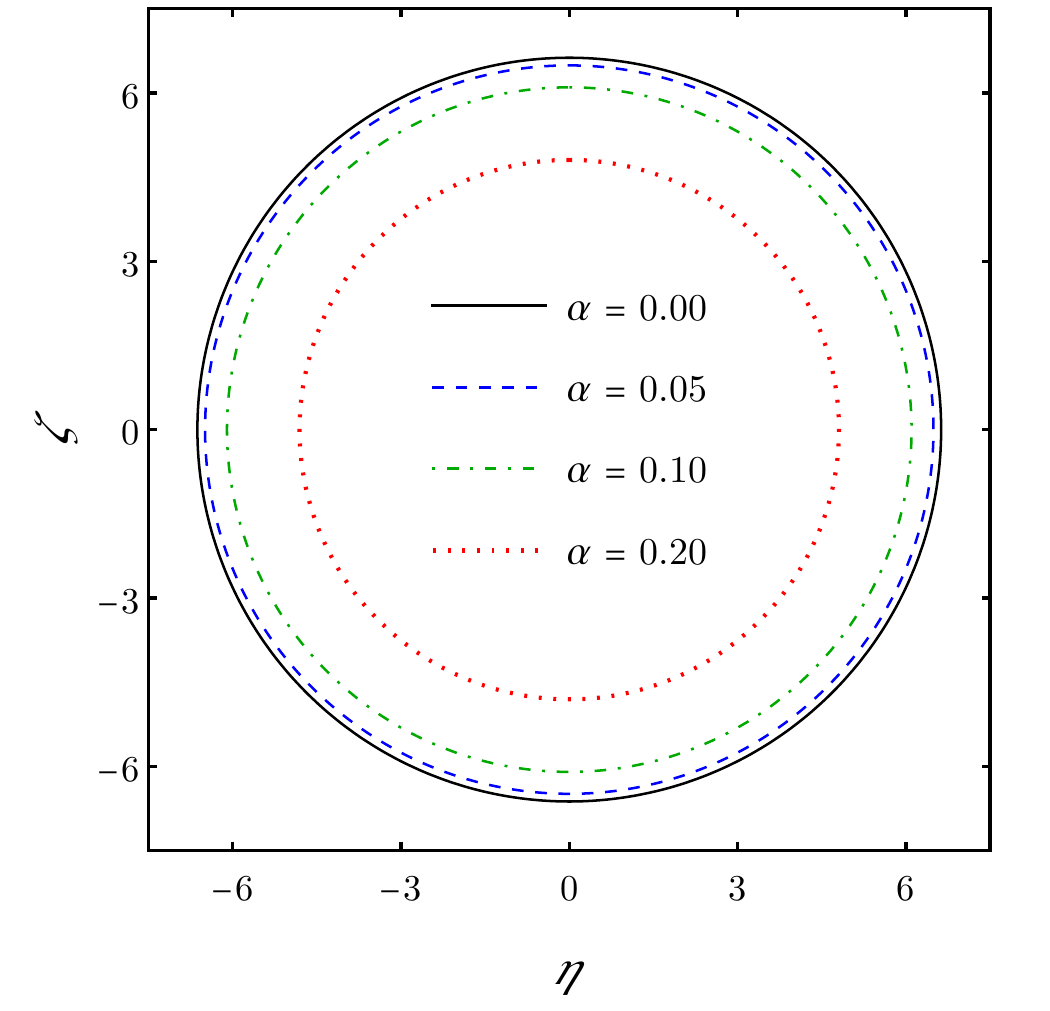}\quad %
			\includegraphics[width=75mm]{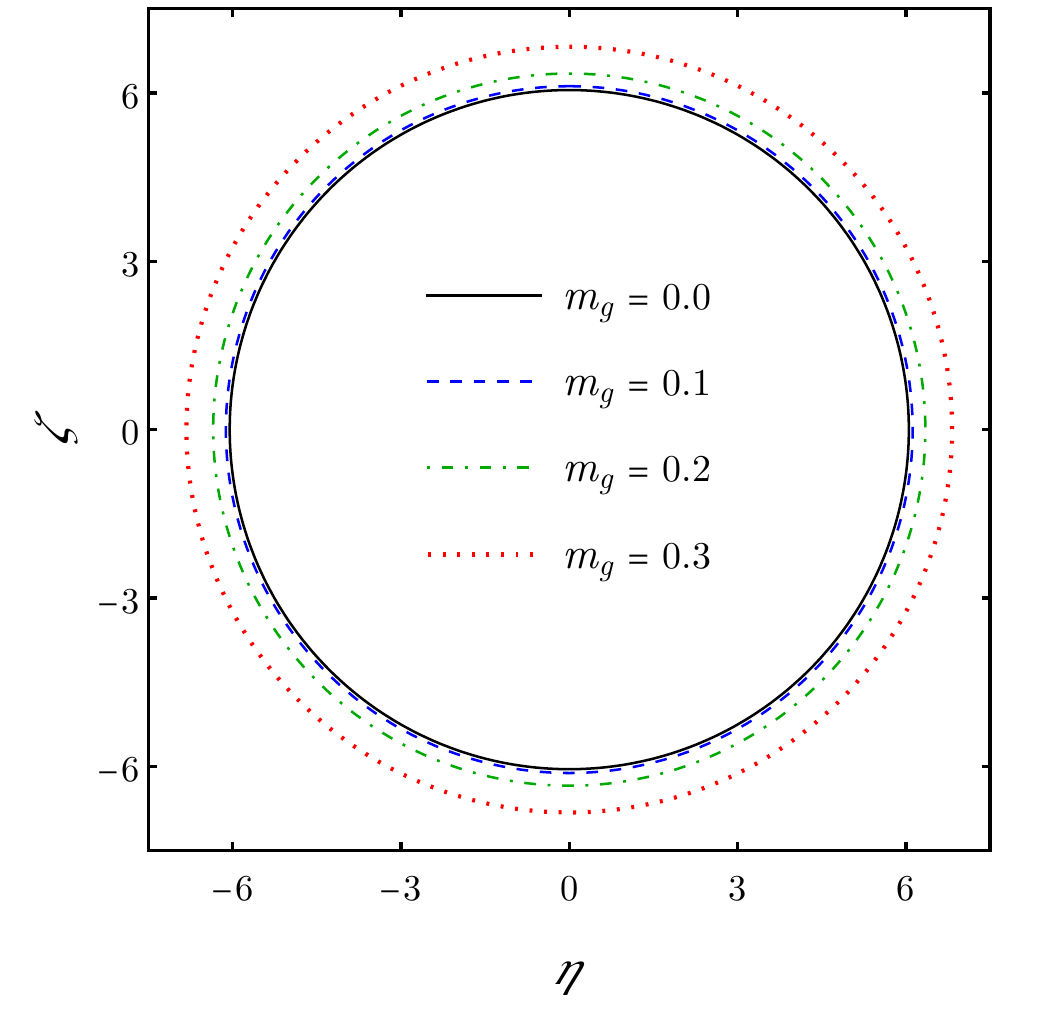}\quad %
			\includegraphics[width=75mm]{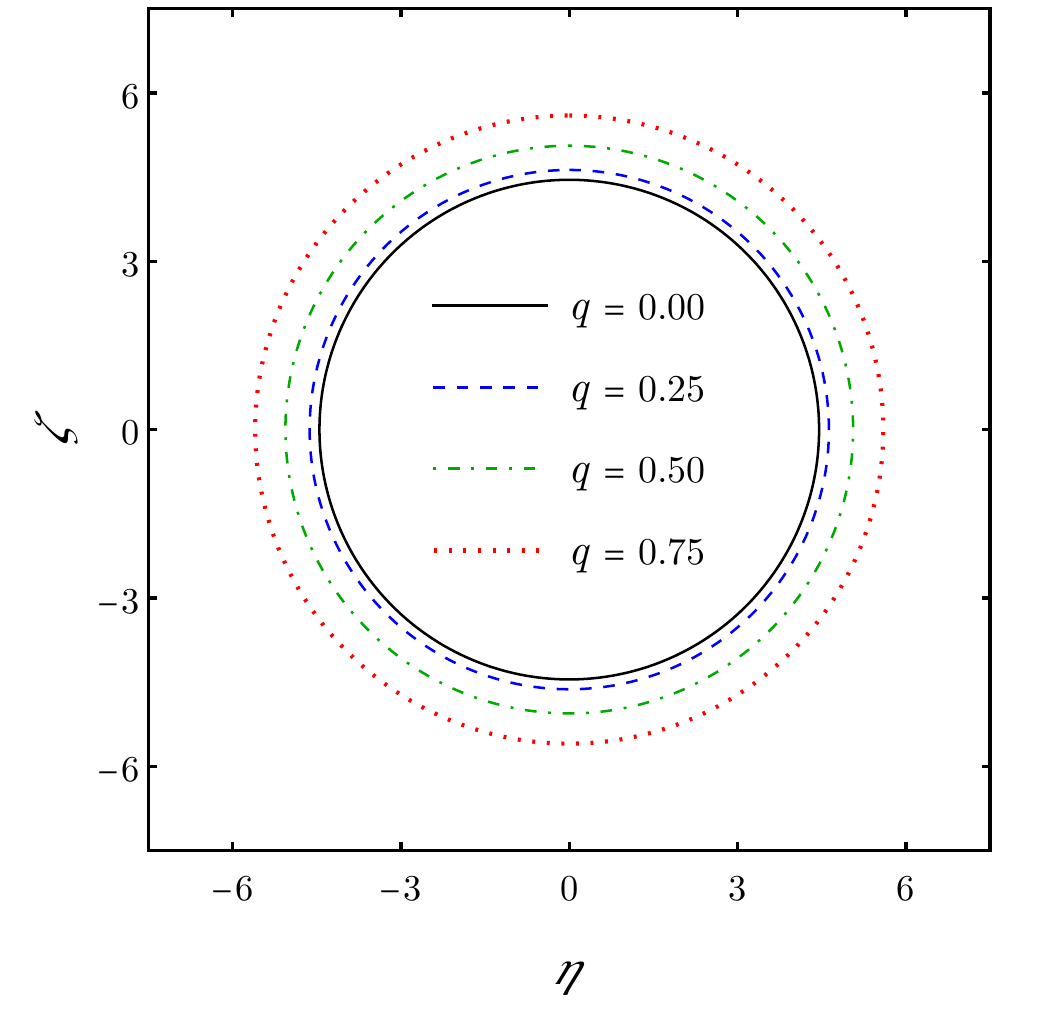}\quad %
			\includegraphics[width=75mm]{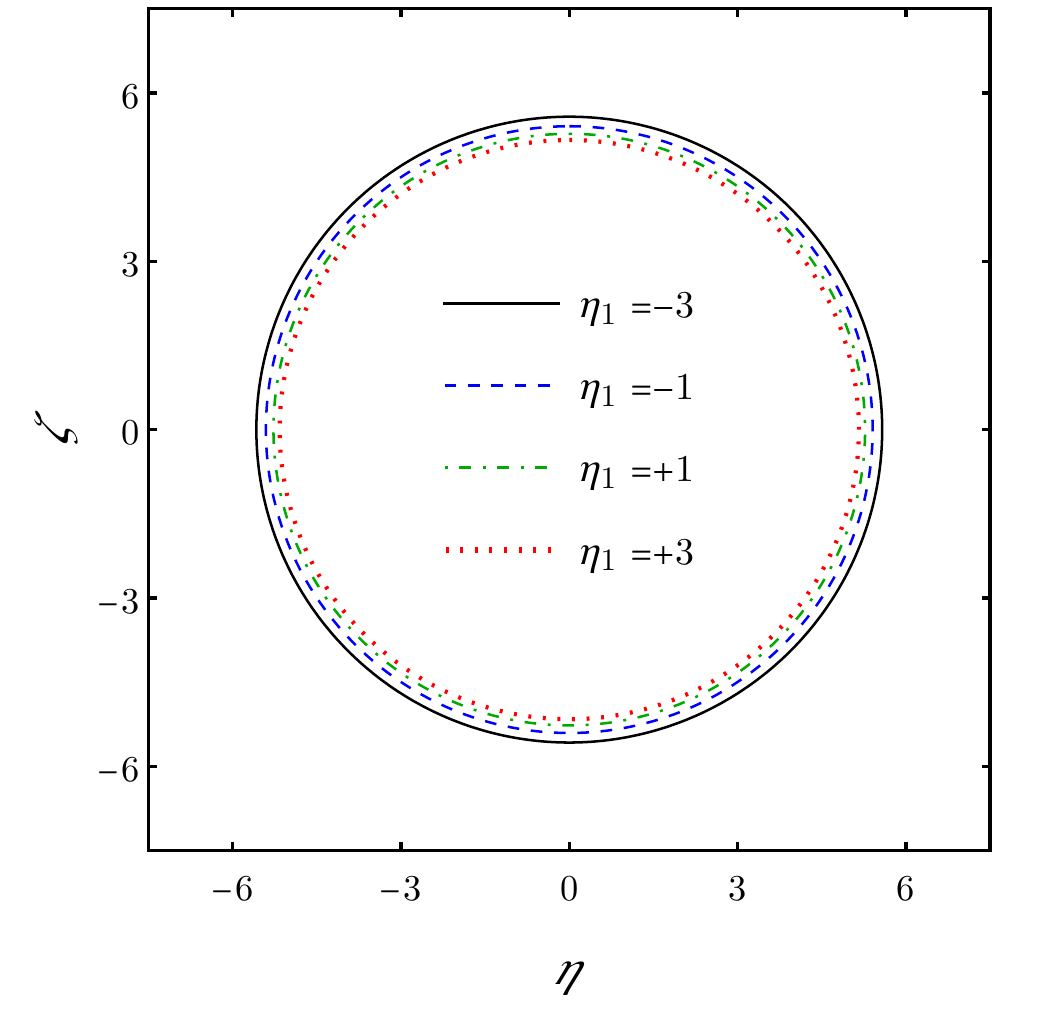}
			\caption{The shadow radius \(\mathcal{R}_{\text{sh}}\) as a function of the model parameters in Maxwell-dilaton-dRGT-like massive gravity. The following fixed parameters are used for all plots: \(m_0 = 1\), \(b = 0.2\), \(c_0 = 1\), \(\eta_2 = -0.1\), \(\Lambda = -0.1\), \(\beta = 0.15\), and the observer radius is set at \(r_\text{O} = 10\).}
			\label{fig:Rsh}
		\end{figure}

	As the shadow radius \(\mathcal{R}_{\text{sh}}\) is a key observable that characterizes the apparent size of a black hole's dark silhouette against the background light. We now analyze its dependence on the parameters of our Maxwell-dilaton-dRGT-like massive gravity model. The behavior of \(\mathcal{R}_{\text{sh}}\), plotted in the celestial coordinates $\xi$ and $\eta$ \cite{volker,Khan2020}, as a function of the model parameters in Figs.~\ref{fig:Rsh}. In these figures, the shadow size is calculated for a distant observer, and its dependence on the parameters \(\alpha\), \(m_g\), \(q\), and \(\eta_1\) reflects the underlying modifications to the spacetime geometry. The analysis for each parameter was conducted by varying it while keeping the others fixed at the following baseline values: $m_{0} = 1$, $b = 0.2$, $\beta = 0.15$, $\eta_{2} = -0.1$, $\Lambda = -0.1$, and $c_{0} = 1$.
		
		\begin{itemize}
			\item \textbf{Dilaton field (\(\alpha\)):} 
			The upper-left panel of Fig.~\ref{fig:Rsh} illustrates the dependence of the shadow radius \(\mathcal{R}_{\text{sh}}\) on the dilaton field parameter \(\alpha\) for fixed values \(m_g = 0.1\), \(\eta_1 = -1\), and \(q = 0.7\). It is evident that \(\mathcal{R}_{\text{sh}}\) \textbf{decreases} as \(\alpha\) increases. This trend parallels the reduction of the photon sphere radius \(r_{\text{ph}}\), as a stronger dilaton field enhances the coupling between the scalar and electromagnetic sectors. Consequently, the effective gravitational potential near the horizon deepens, leading to a more tightly bound photon orbit and a smaller apparent shadow. In this sense, the dilaton field acts to concentrate the spacetime curvature, resulting in a more compact shadow profile.
			
			\item \textbf{Graviton mass (\(m_g\)):} 
			The upper-right panel of Fig.~\ref{fig:Rsh} displays the behavior of \(\mathcal{R}_{\text{sh}}\) with respect to the graviton mass, for \(\alpha = 0.1\), \(\eta_1 = -2\), and \(q = 0.7\). The shadow radius \textbf{increases} as \(m_g\) grows. This is consistent with the outward displacement of \(r_{\text{ph}}\) observed earlier. Physically, the introduction of a finite graviton mass modifies the curvature of spacetime, effectively weakening the central gravitational pull. As a result, photons can orbit at larger radii, and the corresponding shadow appears larger. This behavior highlights the role of massive gravity terms in altering the effective lensing geometry of the black hole.
			
			\item \textbf{Electric charge (\(q\)):} 
			The lower-left panel of Fig.~\ref{fig:Rsh} examines the impact of the electric charge for fixed parameters \(\alpha = 0.1\), \(m_g = 0.2\), and \(\eta_1 = -2\). The shadow radius \textbf{increases} with increasing \(q\). This outcome aligns with the behavior of \(r_{\text{ph}}\), where the repulsive electromagnetic interaction counteracts the gravitational attraction. As \(q\) increases, the balance between these two effects shifts outward, allowing photons to orbit farther from the black hole. Consequently, the shadow expands, reflecting the reduced effective curvature of the surrounding spacetime. 
			
			\item \textbf{Massive gravity parameter (\(\eta_1\)):} 
			Finally, the lower-right panel of Fig.~\ref{fig:Rsh} shows that \(\mathcal{R}_{\text{sh}}\) \textbf{decreases} as \(\eta_1\) increases, for \(\alpha = 0.1\), \(m_g = 0.2\), and \(q = 0.7\). This behavior underscores the role of \(\eta_1\) in strengthening the attractive component of the massive gravity potential. A larger \(\eta_1\) enhances the effective gravitational pull, confining photons to tighter orbits and reducing the shadow size. The dependence of \(\mathcal{R}_{\text{sh}}\) on \(\eta_1\) thus provides a direct probe of how massive gravity corrections influence the strong-field region around the black hole.
		\end{itemize}
		
		The results align with and visually confirm the trends previously established for the photon orbit radius \(r_{\text{ph}}\) in Table~\ref{Tab:rph}.
		In summary, the shadow radius \(\mathcal{R}_{\text{sh}}\) exhibits a direct correlation with the photon orbit radius \(r_{\text{ph}}\). Parameters that cause \(r_{\text{ph}}\) to increase (\(m_g\), \(q\)) result in a larger shadow, while those that cause it to decrease (\(\alpha\), \(\eta_1\)) result in a smaller one. 
		
		It is worth emphasizing that the above results illustrate the characteristic influence of each parameter when the remaining quantities are fixed. In the Maxwell–dilaton–dRGT-like massive gravity framework, these parameters are not entirely independent, and their effects can interplay in nontrivial ways. Variations in one parameter may partially compensate or amplify the influence of another, leading to different quantitative outcomes. Therefore, while the present analysis captures the dominant trends for representative parameter choices, the precise behavior of the shadow radius and photon sphere may vary under alternative parameter sets. A more exhaustive exploration of the multi-parameter space could thus reveal richer structures in the black hole’s optical appearance and gravitational lensing properties as a potential research project in the feature studies.

		Nevertheless, the present results clearly demonstrate that the combined effects of the dilaton field, electromagnetic charge, and massive gravity sector play a decisive role in shaping the black hole’s shadow. These interrelated contributions give rise to distinct modifications in the shadow geometry, offering potential observational signatures that can be employed to test and constrain such extended gravity models. The next section is devoted to comparing the theoretical predictions for the shadow radius with available observational data.
		
		\section{Constraints from the EHT Observation of EHT}\label{EHT}
		\label{Sec:EHTconstraints}
		The Event Horizon Telescope (EHT) observations of black hole shadows have placed powerful constraints on alternative theories of gravity and provided deep insight into the structure of spacetime in the strong-field regime. In the context of the Maxwell-dilaton-dRGT-like massive gravity framework, these data offer an opportunity to examine how the combined effects of the dilaton field, electromagnetic coupling, and graviton mass influence the observable shadow. In this section, we aim to estimate the range of admissible model parameters consistent with the EHT results for $Sgr A^*$.
		
		To perform this comparison, an accurate mass-to-distance ratio for $Sgr A^*$ is required. Independent determinations from the Keck~\cite{Keck} and GRAVITY/VLTI~\cite{VLTI} collaborations have provided consistent measurements of both quantities. Another key aspect is the calibration connecting the observed bright emission ring to the theoretical black hole shadow, which establishes how reliably the observed angular size can be used to infer the shadow boundary~\cite{vagnozzi2022horizon}. Combining this calibration with the uncertainties in the mass and distance estimates allows us to constrain the deviation between the observed and predicted shadow diameters.
		
		According to the EHT collaboration~\cite{akiyama2022firstSgrA}, the angular diameter of the $Sgr A^*$ shadow is measured as \(48.7 \pm 7~\mu\mathrm{as}\). Using the averaged Keck and VLTI mass-to-distance priors, the following bounds on the dimensionless shadow radius are obtained~\cite{vagnozzi2022horizon,Virbhadra}
		\begin{equation}
			4.55 \lesssim \frac{\mathcal{R}_{\text{sh}}}{M} \lesssim 5.22 \quad (1\sigma),
		\end{equation}
		and
		\begin{equation}
			4.21 \lesssim \frac{\mathcal{R}_{\text{sh}}}{M} \lesssim 5.56 \quad (2\sigma).
		\end{equation}

\begin{figure}[h]
			\centering
			\includegraphics[width=90mm]{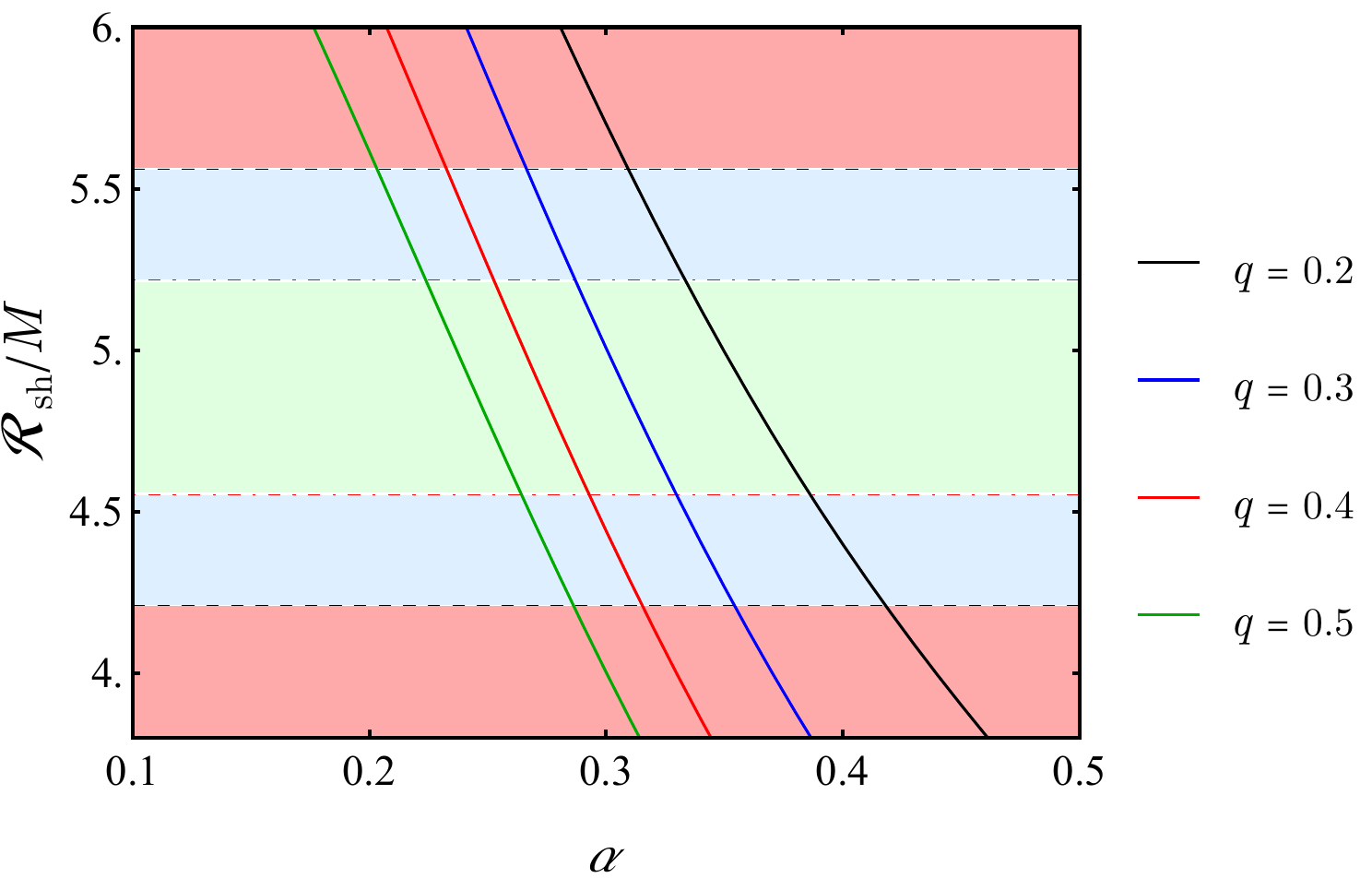}\quad %
			\includegraphics[width=90mm]{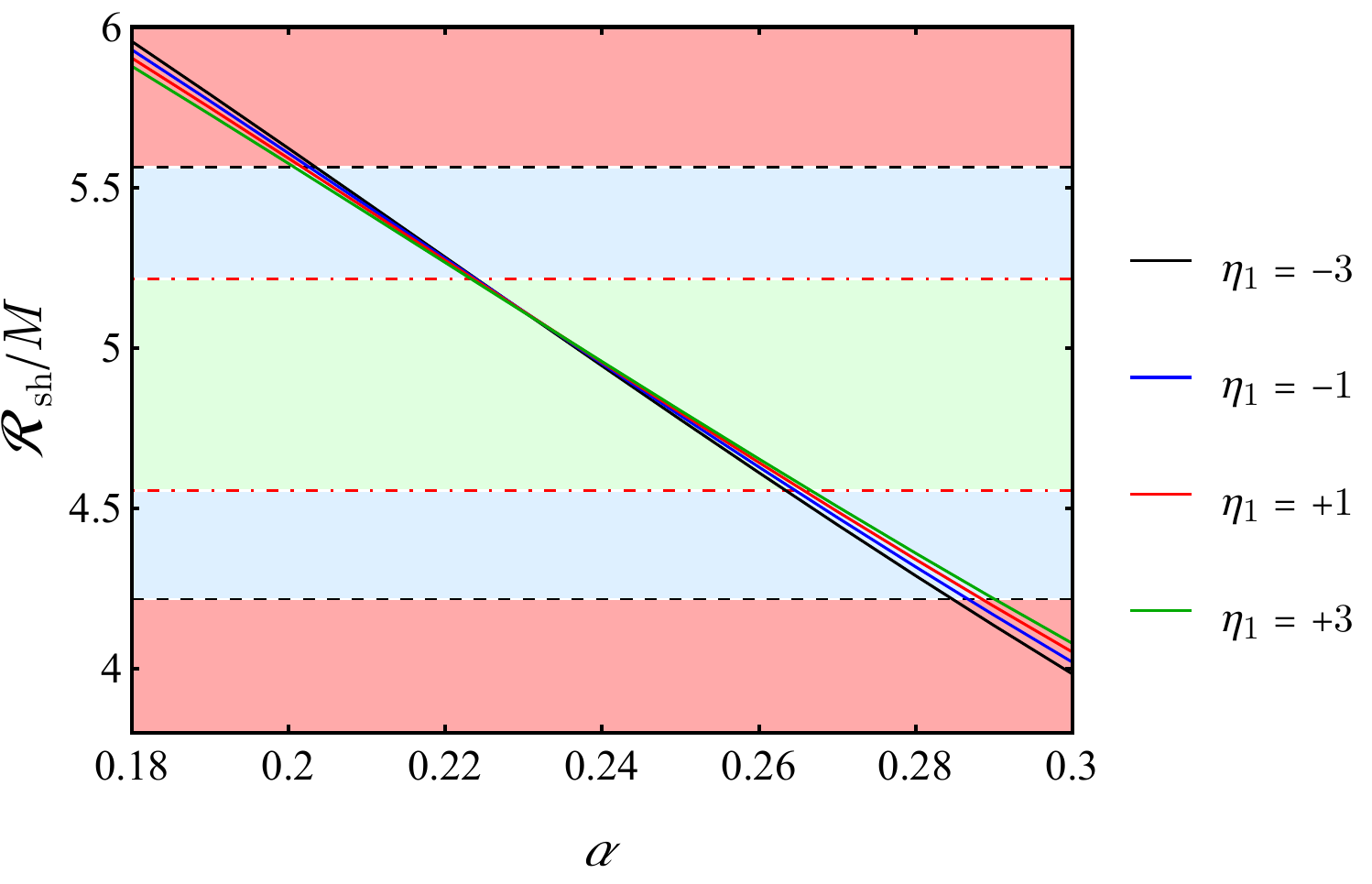}\quad %
			\includegraphics[width=90mm]{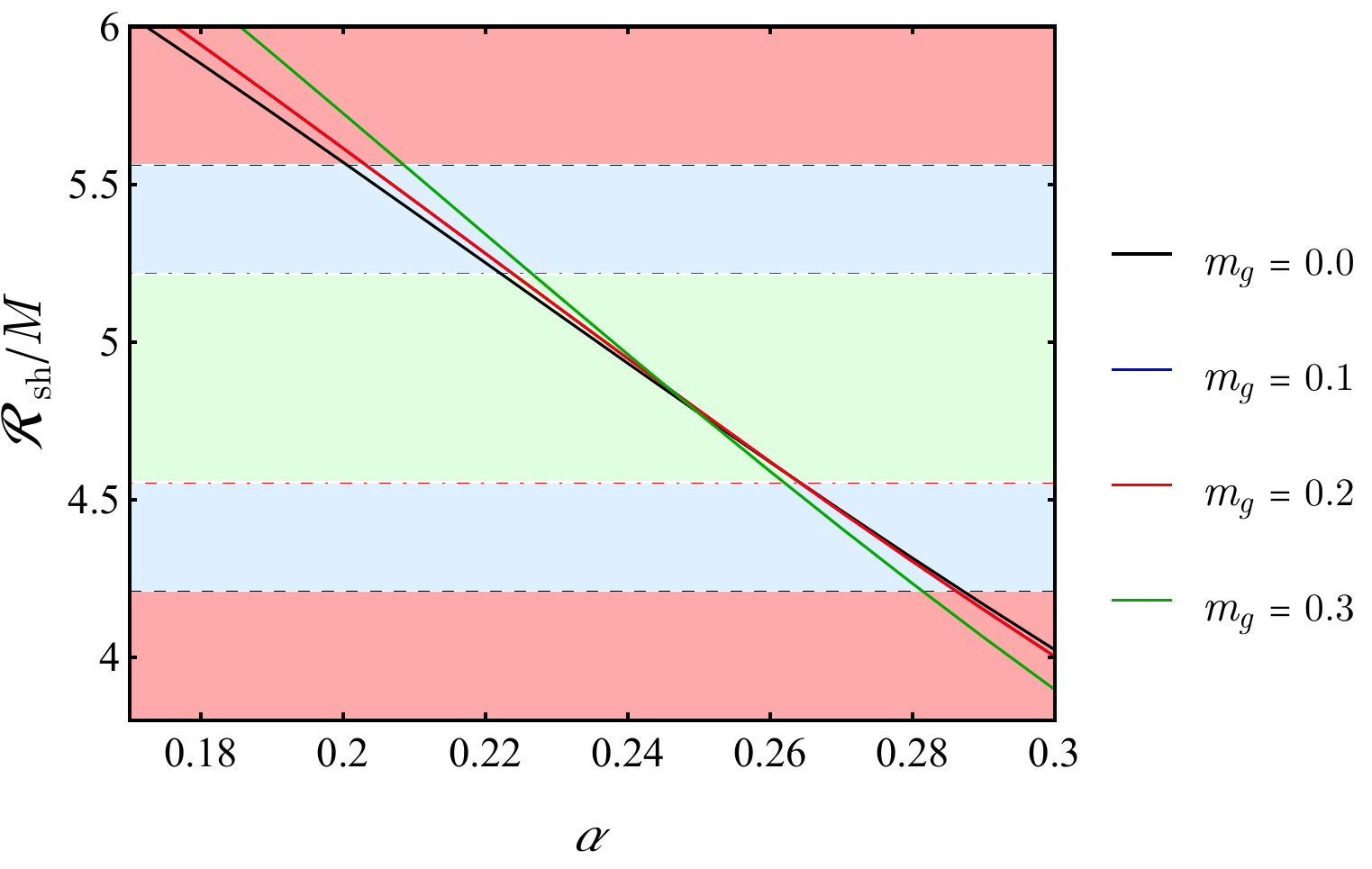}\quad %
			\caption{The shadow radius \(\mathcal{R}_{\text{sh}}\) in the mass unit, as a function of the model parameters in Maxwell-dilaton-dRGT-like massive gravity. The following fixed parameters are used for all plots: \(m_0 = 1\), \(b = 0.2\), \(c_0 = 1\), \(\eta_2 = -0.1\), \(\Lambda = -0.1\), \(\beta = 0.15\), and the observer radius is set at \(r_\text{O} = 10\).}
			\label{fig:Cons}
		\end{figure}

		Figure~\ref{fig:Cons} illustrates the variation of the shadow radius in mass units, \(\mathcal{R}_{\text{sh}}/M\), as a function of the dilaton coupling parameter \(\alpha\) for several representative values of the charge \(q\), the massive gravity parameter \(\eta_1\), and the graviton mass \(m_g\) in the Maxwell-dilaton-dRGT-like massive gravity framework. The shaded green and blue regions represent the \(1\sigma\) and \(2\sigma\) observational intervals derived from the EHT image of Sgr~A\*, corresponding to the measured shadow size range.
		
		For each configuration, the intersection between the theoretical curves and the observational bounds specifies the range of \(\alpha\) values consistent with the EHT data. The allowed interval of \(\alpha\) varies noticeably with the other model parameters. As shown in the upper panel of Fig.~\ref{fig:Cons}, increasing the charge \(q\) slightly narrows the overlap region, shifting the allowed region toward the smaller $\alpha$, indicating that stronger electromagnetic effects reduce the parameter space in which the shadow agrees with observations. In the middle panel of Fig.~\ref{fig:Cons}, larger values of \(\eta_1\) shift the theoretical curves upward, thereby widening the overlap region with the observational bands. This trend implies that the influence of the massive gravity potential (through \(\eta_1\)) permits a broader range of \(\alpha\) values consistent with Sgr~A\*. In contrast, the down panel of Fig.~\ref{fig:Cons} shows that increasing the graviton mass \(m_g\) leads to a downward shift of the theoretical curves, tightening the constraint on \(\alpha\) by increasing its lower bound and decreasing the upper one. This behavior suggests that stronger massive gravity effects suppress the shadow radius, restricting the viable range of the dilaton field.
		
		These results demonstrate how the EHT measurement of $Sgr A^*$’s shadow imposes meaningful constraints on the interplay between the dilaton field, electromagnetic charge, and massive gravity parameters. Within the parameter sets considered, the observationally allowed values of the dilaton field are confined, depending on the black hole parameters. While each constraints are derived for specific parameter choices, they capture the characteristic behavior of the Maxwell-dilaton-dRGT-like massive gravity model and provide a quantitative basis for confronting such theories with future high-precision shadow observations.

	\section{Emission Rate}
	
	\label{Emrate} In this section, we analyze the characteristics of the energy emission rate as a function of photon frequency. Quantum fluctuations near the event horizon of a black hole can give rise to the formation of particle-antiparticle pairs, which may escape the gravitational pull through a quantum tunneling mechanism, a process commonly referred to as Hawking radiation. This mechanism underpins the gradual loss of mass and the eventual evaporation of black holes over time. However, in many theoretical models, the black hole does not fully evaporate; instead, a nonzero remnant mass often remains. This remnant is of considerable interest, as it potentially preserves information that would otherwise be considered lost, thereby contributing to the resolution of the information loss paradox.
	
\begin{figure}[h]
		\centering
		\includegraphics[width=75mm]{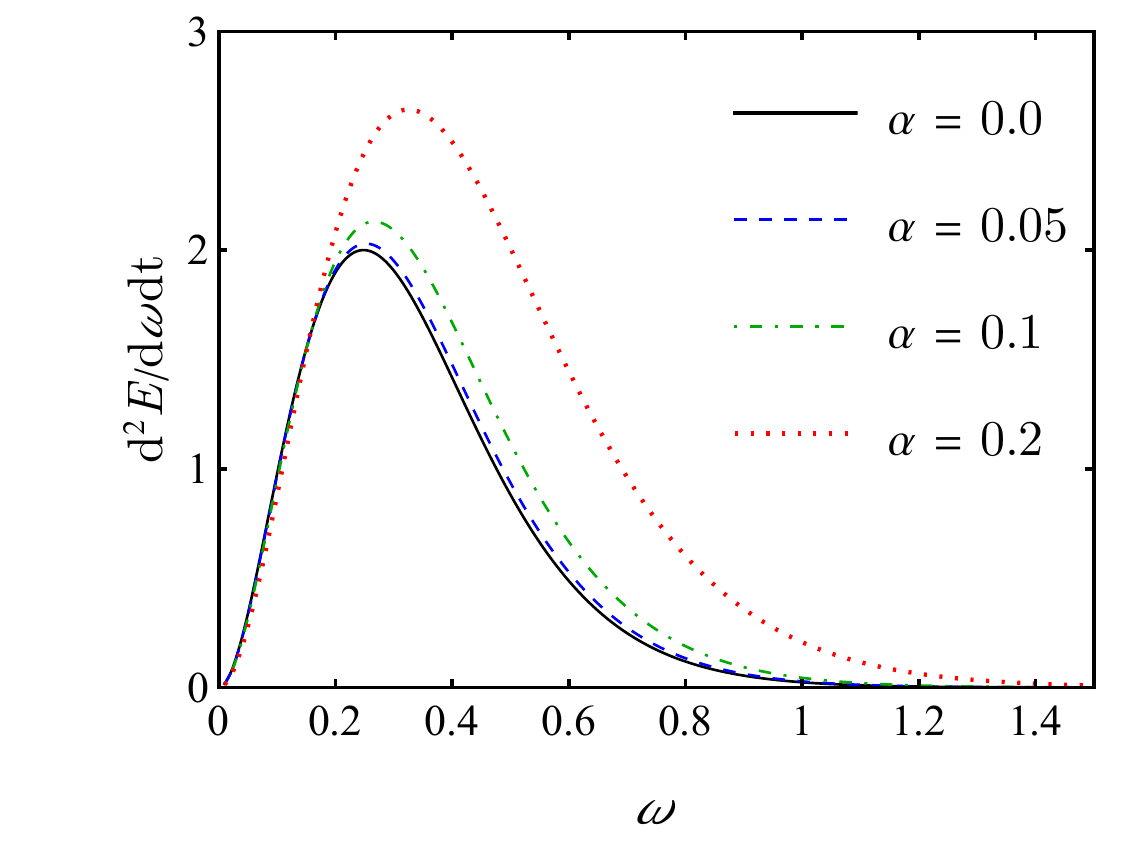}\quad %
		\includegraphics[width=75mm]{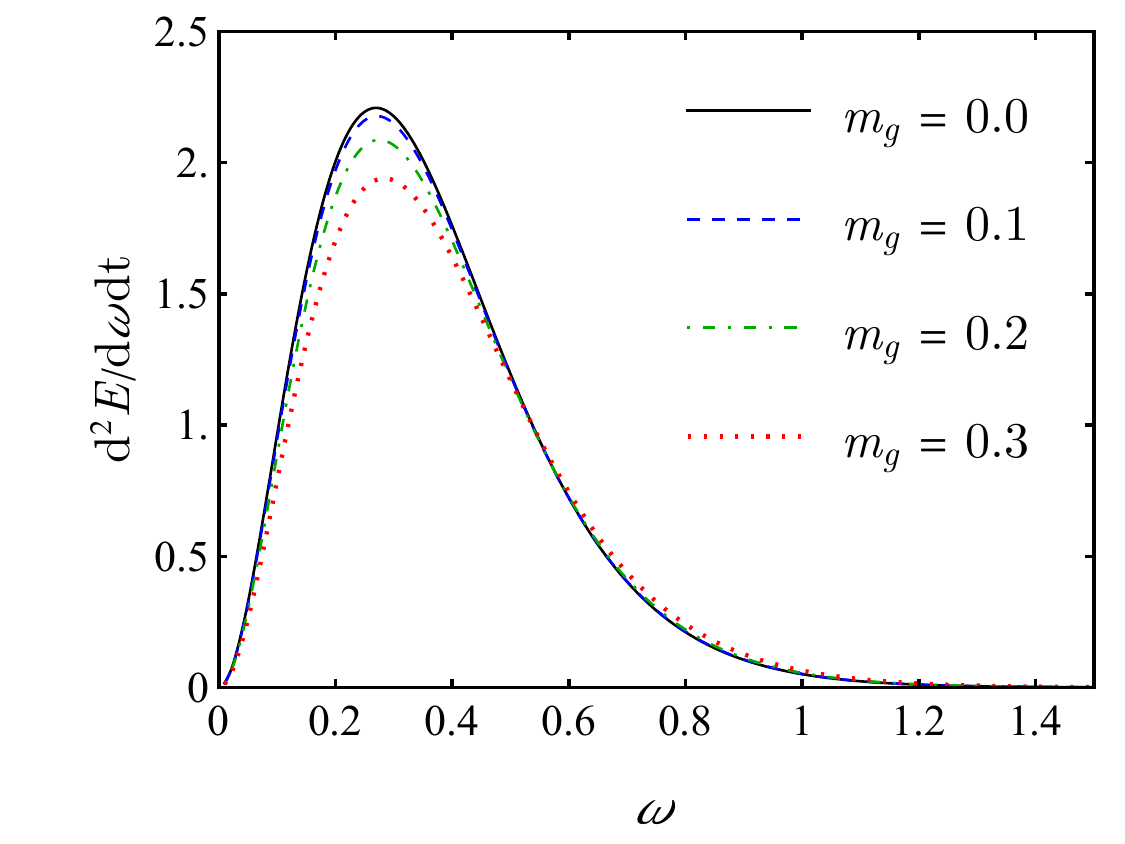}\quad %
		\includegraphics[width=75mm]{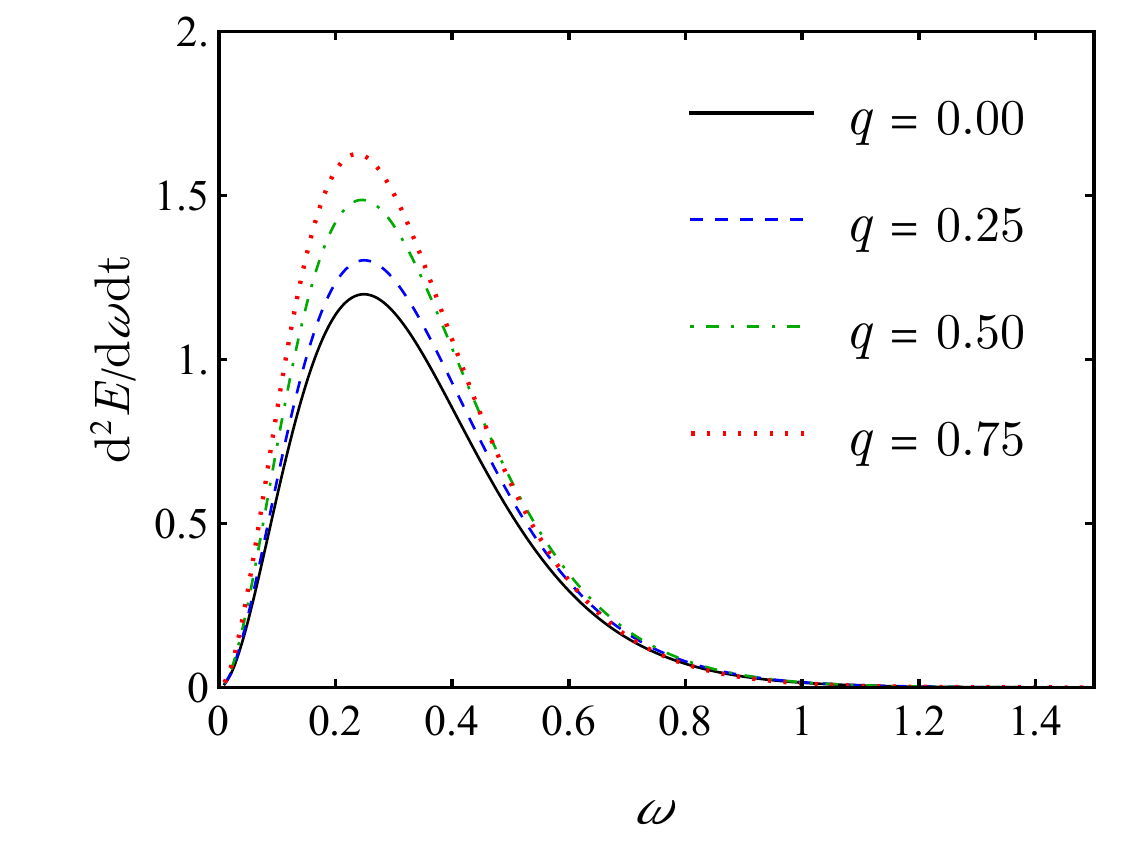}\quad %
		\includegraphics[width=75mm]{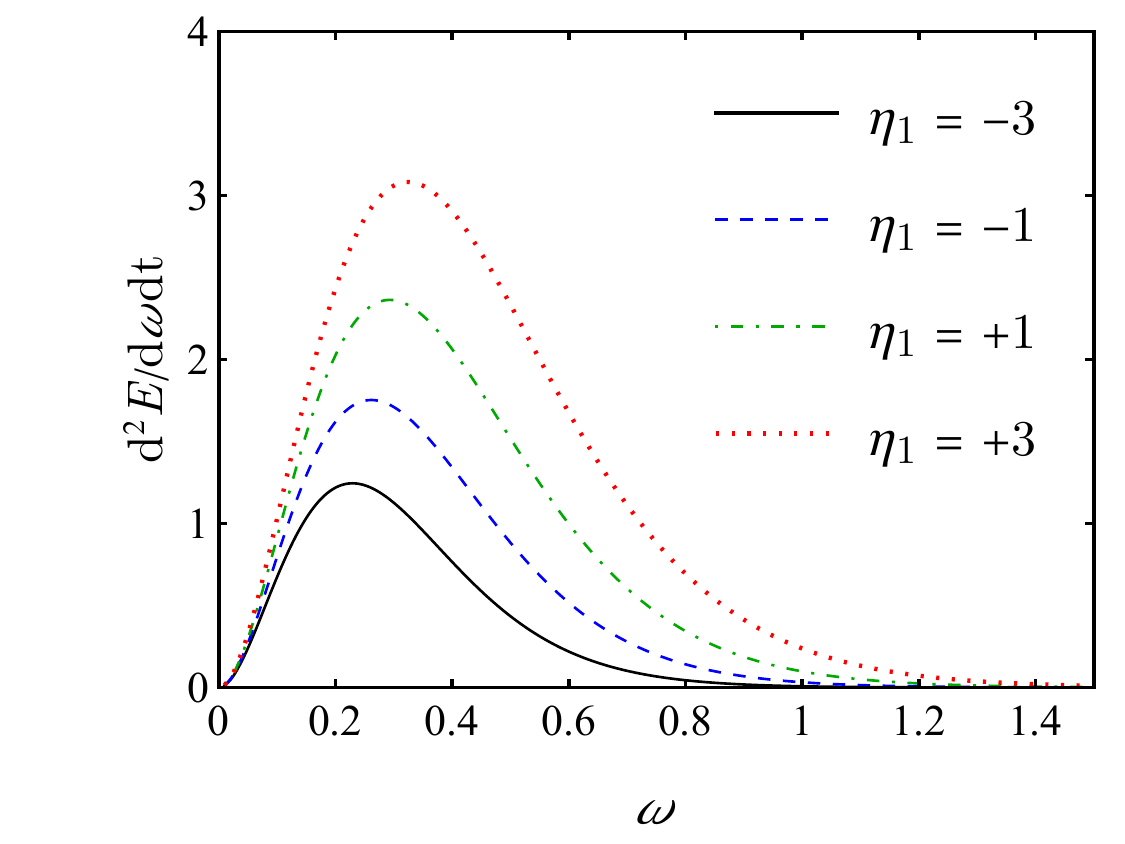}
		\caption{Emission spectrum versus $\protect\omega$ for distinct values of the parameter set for all plots: \(m_0 = 1\), \(b = 0.2\), \(c_0 = 1\), \(\eta_2 = -0.1\), \(\Lambda = -0.1\), \(\beta = 0.15\), and the observer radius is set at \(r_\text{O} = 10\).}
		\label{fig:emission}
	\end{figure}

	Following the framework presented in Refs. \cite{Wei,hendi,ali}, the energy emission rate can be expressed as
	\begin{equation}
		\frac{{d^{2}E}}{{d\omega dt}}=\frac{{2\pi ^{2}\sigma \omega ^{3}}}{{%
				e^{\omega /T}-1}},  \label{emission}
	\end{equation}%
	where $\omega$ denotes the photon frequency, $T$ is the Hawking temperature shown in Eq. \eqref{Temp}. $\sigma$ is the absorption cross-section which approaches a constant value at high frequencies. $\sigma$ is geometrically related to the black hole shadow radius $\sigma_{\mathrm{lim}} \approx \pi \mathcal{R}_{\mathrm{sh}}^2$.	
	In Figs. \ref{fig:emission}, the energy emission rate is plotted as a function of the photon frequency $\omega$ for various values of the parameters. The plots clearly exhibit a peak in the emission rate occurring at a particular frequency, indicating the most probable frequency for radiation emission. Furthermore, as inferred from the analytical expression of the emission rate in Eq. \eqref{emission}, the intensity tends to vanish in both the low-frequency limit $(\omega \rightarrow 0)$ and the
	high-frequency limit $(\omega \rightarrow \infty )$.
	
	The energy emission rate as a function of the photon frequency $\omega$, shown in Figs.~\ref{fig:emission}, exhibits a distinct peak whose position and magnitude depend on the model parameters. In the upper-left panel, for $m_g = 0.1$, $\eta_1 = -1$, and $q = 0.7$, increasing the dilaton field $\alpha$ enhances the emission rate and shifts the peak toward higher frequencies. This indicates that, within the selected parameter range, the dilaton field can deepen the effective potential near the photon sphere, allowing more energetic photons to escape. 
		
		For fixed $m_g = 0.1$, $\eta_1 = -1$, and $\alpha = 0.1$, a larger charge $q$ significantly amplifies the emission peak, while the peak frequency remains nearly unchanged. This behavior may be attributed to the contribution of the electromagnetic field, which increases the overall energy release without substantially altering the effective temperature associated with photon emission.
		
		When $m_g = 0.1$, $\alpha = 0.1$, and $q = 0.7$, increasing $\eta_1$ produces a higher and slightly blue-shifted peak, suggesting that this parameter modulates the massive gravity potential in a way that favors radiation escape from the near-horizon region. In contrast, for $\alpha = 0.1$, $\eta_1 = -1$, and $q = 0.5$, a larger graviton mass $m_g$ leads to a lower and red-shifted peak. This suppression can be interpreted as the massive gravity term increasing the effective curvature scale, thereby confining photons more strongly and reducing the efficiency of emission.
		
		Overall, the trends show that $\alpha$, $q$, and $\eta_1$ tend to enhance high-frequency emission, whereas $m_g$ suppresses it. These results collectively highlight the competing influences of the dilaton, electromagnetic, and massive gravity sectors on the radiation spectrum. It should be emphasized, however, that these interpretations are based on the chosen parameter sets and serve to illustrate the characteristic behavior within this specific regime of the Maxwell-dilaton-dRGT-like massive gravity model.
	

	\section{Conclusions}
	
	\label{sec4}
	
	In this paper, we first reviewed charged black holes in dilaton-dRGT-like
	massive gravity. Then, we analyzed the asymptotic behavior of spacetime by
	adjusting the parameters of the dilaton field ($\alpha$) and the coupling
	constant ($\beta$). Our analysis indicated that: i) for all values of $%
	\alpha $, the spacetime was not asymptotically (A)dS, and it depended on
	both $\Lambda$ and $\alpha$; ii) for $\beta > \frac{-\gamma_{2,-1}}{2\alpha}$%
	, the asymptotic behavior of the spacetime was determined by the
	cosmological constant, the parameters of the dilaton field, the parameters
	of the reference metric, and massive gravity ($\eta_1$), as well as the
	graviton mass ($m_g$); iii) for $\alpha > -\beta$, the asymptotic behavior of the spacetime depended on all parameters of the system. In addition, we
	studied the effects of mass, electrical charge, and the parameters $\alpha$
	and $\beta$ on the metric function to find its roots. We found that these
	black holes could have multiple horizons, depending on the values of the
	various parameters. In other words, our findings revealed that there were
	critical values for the parameters of $m_{0}$ (mass), $q$ (electric charge), 
	$\alpha$, and $\beta$, such that for less or more than these critical
	values, the charged black holes in dilaton-dRGT-like massive gravity
	encounter one, two, and three real roots. This was one of the interesting
	effects of massive gravity and the dilaton field on the roots of the metric
	function.
	
	In the next section, we analyzed the conserved and thermodynamic quantities
	of charged black holes within dilaton-dRGT-like massive gravity. We examined
	how various parameters influenced the thermal stability areas and phase
	transition points by employing both heat capacity and geometrothermodynamics
	approaches. Our investigation into the heat capacity and temperature of
	these black holes revealed four distinct regions: very small and medium
	black holes were found to be non-physical and unstable, while small and
	large black holes met the conditions for physical and thermal stability.
	This indicated that small and large charged black holes in dilaton-dRGT-like
	massive gravity was indeed a physical and stable entity, facilitating a
	phase transition between the two sizes. Additionally, we assessed the
	effects of parameters $\alpha$, $\beta$, $\eta_{1}$, and $\eta_{2}$ on these
	four regions. Our findings include:
	
	i) An increase in the dilaton field parameter ($\alpha$) resulted in a
	reduction of the thermal stability area (see Fig. \ref{fig2} for details).
	
	ii) The thermal stability areas expanded as the value of $\beta$ increased
	(see Fig. \ref{fig3} for details).
	
	iii) An increase in the dRGT-like massive gravity parameter ($\eta_{1}$) led
	to an expansion of the thermal stability area (see Fig. \ref{fig4} for
	details).
	
	iv) The thermal stability areas of these black holes decreased as $\eta_{2}$
	increased (see Fig. \ref{fig5} for details).
	
Geometrothermodynamics provides an alternative method for exploring specific thermodynamic properties of black holes, particularly the critical points of phase transitions, including the divergence points of heat capacity. To achieve this, we analyzed four thermodynamic metrics Weinhold (Fig. \ref{fig6}), Ruppeiner (Fig. \ref{fig7}), Quevedo (Fig. \ref{fig8}), and HPEM (Fig. \ref{fig9}) to identify which metric accurately reflects the divergence and zero points of heat capacity in charged black holes within the framework of dilaton-dRGT-like massive gravity. Our findings indicated that the HPEM metric consistently corresponds with all divergence and zero points of heat capacity of charged black holes within the framework of dilaton-dRGT-like massive gravity.
	
The influence of Maxwell-dilaton-dRGT-like massive black hole parameters on the photonic radius and shadow of the black hole has been explored in the section \ref{sec:photon}, t. The results highlight that the parameters of the dilaton field, massive gravity, and the graviton mass can measurably affect the photonic orbit. The analysis demonstrates that black hole shadows exhibit rich and distinct behaviors under variations of $\alpha$, $\eta_{1}$, $q$, and $m_{g}$.
	
In section \ref{EHT}, the EHT shadow of $Sgr A^*$ constrains the admissible range of the dilaton coupling and reveals how electromagnetic and massive-gravity effects jointly shape the observable geometry. The derived limits offer a quantitative reference for testing this extended gravity framework against future shadow observations.
	
In the section \ref{Emrate}, our analysis of the energy emission rate reveals that the variations in the dilaton field $\alpha$, charge $q$, and massive gravity parameters $\eta_1$ and $m_g$ distinctly influence both the intensity and spectral distribution of the energy emission. The enhancement of the emission peak with increasing $\alpha$, $q$, and $\eta_1$, and its suppression by larger $m_g$, collectively illustrate how the interplay among scalar, electromagnetic, and massive gravity sectors governs the radiative features of the black hole. Although these results are obtained for specific parameter choices, they consistently demonstrate the characteristic behavior of the Maxwell-dilaton-dRGT-like massive gravity model and provide a useful reference for interpreting possible observational or numerical extensions in similar frameworks.
	
Black holes are never isolated; they engage with their surroundings and influence the background environment. Following these interactions, black holes emit gravitational waves characterized by quasinormal modes with distinct frequencies. These modes have a non-vanishing imaginary component and encapsulate all information regarding the relaxation of black holes after a perturbation. It is crucial to recognize that quasinormal frequencies are determined by both the geometry of the black hole and the nature of the perturbation applied to the background whether scalar, vector, tensor, or fermionic, and they remain independent of initial conditions. During the ring-down phase of a black hole merger, a uniquely perturbed object emerges, resulting in damped oscillations in the geometry of spacetime due to the emission of gravitational waves. In this context, quasinormal modes (QNMs) are vital for understanding the underlying physics of gravitational waves, with black hole perturbation theory being essential for this comprehension. Several references, such as \cite{QN1,QN2,QN3,QN4}, provide valuable insights into this field, while more recent studies exploring various methods and backgrounds can be found in \cite{QN5,QN6,QN7,QN8,QN9,QN10,QN11,QN12,QN13,QN14,QN15,QN16} and related references. Gravitational wave astronomy serves as a powerful tool for testing gravity under extreme conditions. Although the literature on the subject is extensive, comprehensive reviews can be found in \cite{QN17,QN18,QN19}. The study of QNMs is significant not only for constraining black hole parameters but also for its implications for area quantization. Therefore, further investigation into the QN spectrum of black holes in dilaton-dRGT-like-massive gravity is warranted for future research.

The investigation of particle dynamics in the immediate vicinity of black holes is fundamentally essential for probing their geometric and physical characteristics. Over the years, extensive research has been dedicated to exploring the trajectories of both massive and massless particles across various parameterized black hole geometries \cite{Particle1,Particle2,Particle3,Particle4,Particle5,Particle6,Particle7,Particle8}. Consequently, exploring particle dynamics in the context of a black hole embedded within dilaton-dRGT-like massive gravity presents a compelling avenue for future investigation, potentially yielding deeper insights into the geometric and physical properties of this specific system.
	
\begin{acknowledgements}
The authors express gratitude to the esteemed referee for the valuable comments. This research is supported by the research grant of the University of Mazandaran (33/60679).
\end{acknowledgements}

\end{document}